\documentclass[a4paper,12pt,titlepage]{book}
\usepackage[utf8]{inputenc}

\usepackage{amssymb}
\usepackage{eurosym}

\usepackage{amsmath}

\newcommand{\spara}[1]{\smallskip\noindent{\bf #1}}

\usepackage{graphicx}
\graphicspath{{figs/}}

\usepackage{rotating}

\usepackage{enumitem}

\usepackage{fancyhdr}

\usepackage[breaklinks=true]{hyperref}

\usepackage{epigraph}
\usepackage[square,sort,comma,numbers]{natbib}
\setcitestyle{semicolon}		% Separate citations with semicolon: [Author, Year; Author, Year]

\usepackage{listings}
\usepackage{color,soul}

\usepackage{url}

\usepackage[stable]{footmisc}

\usepackage{CJK}

\usepackage{xspace}

\newcommand{\thesistitle}{An effective end-user development approach through domain-specific mashups for Research Impact Evaluation}

\newcommand{\thesisauthor}{Muhammad Imran}
\newcommand{\thesisadvisor}{Prof. Maurizio Marchese}
\newcommand{\thesiscoadvisor}{Prof. Fabio Casati}
\newcommand{\thesisdate}{March $2013$}

\newcommand{\emphbf}[1]{\textbf{\emph{#1}}}

\newenvironment{definition}[1][Definition]{\begin{trivlist}
\item[\hskip \labelsep {\bfseries #1}]}{\end{trivlist}}

\newcommand{\clearemptydoublepage}{\newpage{\pagestyle{empty}\cleardoublepage}}
\makeatletter
\renewcommand\part{
	\if@openright \cleardoublepage \else \clearpage \fi
	\thispagestyle{empty}
	\if@twocolumn \onecolumn \@tempswatrue \else \@tempswafalse \fi
	\null\vfil \secdef\@part\@spart
}
\makeatother

\makeindex
\begin{document}

% Active CJK: bsmi = trad. Chinese, gbsn = simp. Chinese, min = Japanese, mj = Korean.
\begin{CJK*}{UTF8}{min} 

% Title page.
% !TeX root = ../phd-thesis.tex
%%% Title page. %%%
\pagestyle{plain}\newpage\clearemptydoublepage\thispagestyle{empty}
\begin{center}
	% Header and horizontal line.
	\textbf{\large PhD Dissertation}\\\ \hrulefill\\\
	
	% Unitn / DISI / ICT School logo and titles.
	\begin{figure}[h!]\centerline{\includegraphics[width=0.7\textwidth]{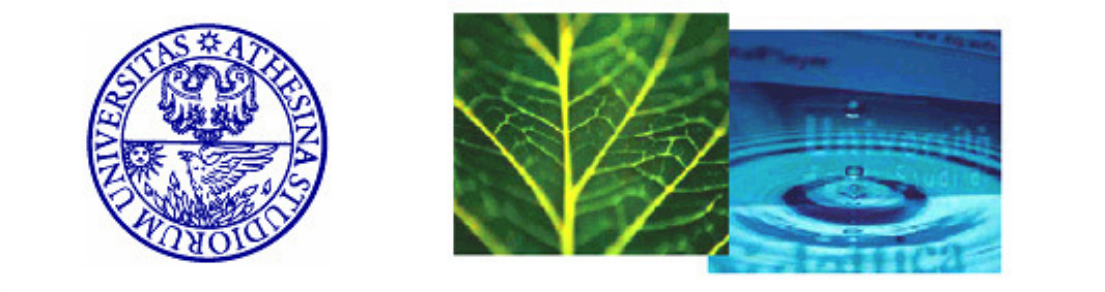}}\end{figure}
	\textbf{\large{International Doctorate School in Information and\\Communication Technologies}}\\
	\vspace{0.3 cm}\LARGE{DISI - University of Trento}\\
	
	% Thesis title.
	\vspace{0.3 cm}\LARGE{\textsc{\thesistitle}}\\\vspace{0.8 cm}

	% Author name.
	\begin{tabular}{l}\Large{\thesisauthor}\\\end{tabular}

	% Advisor name (repeat for co-advisor if needed).
	\begin{flushleft}\begin{tabular}{l}
		\large{Advisor:}\\
		\large{\thesisadvisor}\\
		\large{Universit\`{a} degli Studi di Trento}\\\\
	\end{tabular}\end{flushleft}

	\begin{flushleft}\begin{tabular}{l}
		\large{Co-Advisor:}\\
		\large{\thesiscoadvisor}\\
		\large{Universit\`{a} degli Studi di Trento}\\
	\end{tabular}\end{flushleft}
	
	% Horizontal rule and date of publication.
	\vspace{1.0cm}\hrulefill\\\normalsize{\thesisdate}
\end{center}

% Abstract, keywords and acknowledgments.
% !TeX root = ../phd-thesis-arxiv.tex
%%% Abstract page. %%%
\newpage\clearemptydoublepage\thispagestyle{empty}\large
{\bf \Huge Abstract}

% The abstract.
\vspace{2cm}\noindent{\itshape
Over the last decade, there has been growing interest in the assessment of the performance of researchers, research groups, universities and even countries. The assessment of productivity is an instrument to select and promote personnel, assign research grants and measure the results of research projects. One particular assessment approach is bibliometrics i.e., the quantitative analysis of scientific publications through citation and content analysis. However, there is little consensus today on how research evaluation should be performed, and it is commonly acknowledged that the quantitative metrics available today are largely unsatisfactory. The process is very often highly subjective, and there are no universally accepted criteria.

A number of different scientific data sources available on the Web (e.g., DBLP, Microsoft Academic Search, Google Scholar) that are used for such analysis purposes. Taking data from these diverse sources, performing the analysis and visualizing results in different ways is not a trivial and straight forward task. Moreover, the data taken from these sources cannot be used as it is due to the problem of name disambiguation, where many researchers share identical names or an author different name variations appear in the data. We believe that the personalization of the evaluation processes is a key element for the appropriate use and practical success of these research impact evaluation tasks. Moreover, people involved in such evaluation processes are not always IT experts and hence not capable to crawl data sources, merge them and compute the needed evaluation procedures.

The recent emergence of mashup tools has refueled research on end-user development, i.e., on enabling end-users without programming skills to produce their own applications. Yet, similar to what happened with analogous promises in web service composition and business process management, research has mostly focused on technology and, as a consequence, has failed its objective. Plain technology (e.g., SOAP/WSDL web services) or simple modeling languages (e.g., Yahoo! Pipes) do not convey enough meaning to non-programmers. We believe that the heart of the problem is that it is impractical to design tools that are generic enough to cover a wide range of application domains, powerful enough to enable the specification of non-trivial logic, and simple enough to be actually accessible to non-programmers. At some point, we need to give up something. In our view, this something is generality since reducing expressive power would mean supporting only the development of toy applications, which is useless, while simplicity is our major aim.

This thesis presents a novel approach for an effective end-user development, specifically for non-programmers. That is, we introduce a domain-specific approach to mashups that ``speaks the language of users", i.e., that is aware of the terminology, concepts, rules, and conventions (the domain) the user is comfortable with. We show what developing a domain-specific mashup platform means, which role the mashup meta-model and the domain model play and how these can be merged into a domain-specific mashup meta-model. We illustrate the approach by implementing a generic mashup platform, whose capabilities are based on our proposed mashup meta-model. 
%Further, we illustrate how the generic mashup platform can be tailored for a specific domain, which is achieved through the development of ResEval Mash tool that is specifically developed for the research evaluation domain. 
Moreover, the thesis proposed an architectural design for mashup platforms, specifically it presents a novel approach for data-intensive mashup-based web applications, which proved to be a substantial contribution. The proposed approach is suitable for those applications, which deal with large amounts of data that travel between client and server.

%For the evaluation of our work and to determine the effectiveness and usability of our mashup tool, we performed two separate user studies. The results of the user studies confirm that domain-specific mashup tools indeed lower the entry barrier for non-technical users in mashup development. The methodology presented in this thesis is generic and can be applied for other domains. Moreover, following the methodological approach the developed mashup platform is also generic, that is, it can be tailored for other domains.
%-----------------------------------------

%Moreover this thesis provides a new approach to solve the problem of name disambiguation that is a key aspect to be considered in order to increase the quality of data. In this thesis we try to solve the problem from a different perspective and introduced an unsupervised, heuristic-based name disambiguation method that also consider the user interactions. To validate the algorithm we performed experiments on a set of highly ambiguous researchers names. The results show high precision and recall that proves the viability of the algorithm.
}

% Keywords.
\vspace{0.3cm}\noindent{\bf Keywords}\noindent
[End-user development, Domain-specific mashups, Research evaluation]

%%% Acknowledgements page. %%%
\clearemptydoublepage
{\bf \Large\noindent Acknowledgements}
\vspace{7.5mm}

%To be added

This thesis would not have been possible without the support of many people, whom I want to acknowledge in this section. First of all, thanking God for giving me the amazing opportunity of coming to Trento to pursue my PhD degree.

I would like to express my sincere gratitude to my supervisors Prof. Maurizio Marchese and Prof. Fabio Casati for their valuable guidance, support and constructive comments throughout the journey toward my PhD.

I would also like to express my sincere gratitude to Dr. Florian Daniel for his immeasurable attentive guidance, valuable insights and technical advice throughout my PhD. Thank you Florian, this thesis would not have been possible without your support. I thank my fellows (Soudip roy chowdhury, Stefano Soi) and friends (Zeeshan Munir, Musawar Saeed, Talha Rehman), who have been with their kind behavior contributed to this work directly or indirectly.

This PhD is also the result of much love, encouragement and prayers from my parents and family. Especially my dearest dad, who has been a great source of support and encouragement for me throughout my life. He is truly a great father and a kind person. Dad \& mom, I owe you everything I have. Finally, I want to thank my partner in life, my dear wife. Her constant support and love got me through this process. Thank you all!!

%To my advisors, Prof. Maurizio Marchese and Prof. Fabio Casati, I express my sincere gratitude to both of you, for your guidance, for your support and for your constructive comments throughout the journey toward my PhD. I am grateful for the insight and advice they have given me over the past four years. 

%In addition, I express my sincere gratitude to Florian Daniel for his immeasurable attentive guidance, valuable insights and technical advice throughout my PhD.  I thank my fellows (Soudip Roy Chuwdary, Stefano Soi) and friends (Zeeshan Munir, Musawar Saeed, Talha Rehman) who have been with their kind behavior contributed to this work directly or indirectly.

%This PhD is also the result of much love and encouragement from family. My dad has been a great source of support and encouragement for me through every obstacle I’ve faced during PhD journey and general in life. He is truly a great father and a kind man. He led by example and taught me the same. Finally, I want to thank my partner in life. Her constant support and love got me through this process.

\begin{flushright}
\emph{Muhammad Imran}
\end{flushright}

\vspace{8mm}\hrule\vspace{1.5mm}

%\noindent{\itshape The work compiled in this thesis has been partially supported by the grant number 1234567890 ``Put your grant name here'' etc.}

% Tables of contents and layout settings for the main contents.
% !Tex root = ../phd-thesis.tex
%%% Table of contents and lists. %%%
% Use roman numbers for these pages.
\pagenumbering{roman}

% Insert the table of contents in the document.
\tableofcontents
\clearemptydoublepage

% Build a list of tables in the document.
\listoftables
\clearemptydoublepage

% Build a list of figures in the document.
\listoffigures
\clearemptydoublepage

%%% Layout settings for the main contents. %%%

% Uses different headers and page styles for the main contents.
\pagestyle{fancy}
\renewcommand{\chaptermark}[1]{\markboth{#1}{}}
\renewcommand{\sectionmark}[1]{\markright{#1}{}}
\fancyhf{}
\fancyhead[LE,RO]{\thepage}
\fancyhead[RE]{\textit{\nouppercase{\leftmark}}}
\fancyhead[LO]{\textit{\nouppercase{\rightmark}}}
\fancypagestyle{plain}{\fancyhf{}\renewcommand{\headrulewidth}{0pt}\renewcommand{\footrulewidth}{0pt}}

% Starts using arabic numbers and reset the counter to one.
\pagenumbering{arabic}
\setcounter{page}{1}

%%% Main contents of the thesis. %%%
% To better organize the LaTeX source of the thesis, chapters have been separated into
% different .tex files, all inside the chapters/ sub-folder. Feel free to rename them,
% changing their name also in the \input commands below.

% Outline: temporary
%\input{chapters/outline-temp.tex}
%\newpage

% Chapter 1: Introduction.
% !Tex root = ../phd-thesis-arxiv.tex
%%% Chapter start. %%%
\chapter{Introduction}
\label{sec-intro}
%Before presenting the contributions of this thesis, first we briefly introduce the field of research evaluation, its problems and challenges.
The concepts of \emph{scientometrics} (i.e., the science of measuring and analyzing science) and \emph{informetrics} (i.e., the study of the quantitative aspects of information in any form) \cite{hood2001literature}\cite{rip1984co} are increasingly popular. More specifically, among the other fields that informetrics encompasses, the field of \emph{bibliometrics}, which deals with the quantitative analysis of disseminated information of all forms, has received considerable interest over the last few years. The quantitative analysis of scientific and technological information, under bibliometric field, typically use citation and content analysis techniques. The ultimate goal of such an analysis is to determine the impact of a research work that then contributed to productivity and the impact of researchers (i.e., who actually conduct the research work). Bibliometrics has changed out the way the research assessment practices were following, and as it is now bibliometrics methods are widely being used to evaluate research groups, individual research's, departments, universities and many more.

However, evaluating someone's research output quality is a notoriously challenging problem which, so far, has no well accepted solution. The field of research is a competitive struggle for a researcher. These researchers throughout their career are evaluated on the basis of their research work, especially the disseminated work, which could be of different forms. For example, to name a few, among traditional quantitative indicators include \emph{journals publications count}, or \emph{top tier conference publications count} etc., and among \emph{citation-based} methods include, \emph{journal impact factor}, \emph{h-index} value, or \emph{g-index} value etc.\footnote{A more detailed presentation and discussion of such indicators will be given in chapter \ref{sec-stateoftheart} } Often times, the choice of an evaluation criterion depends on the purpose behind that evaluation practice.

Over the last few years, research impact evaluation received a substantial focus as the amount of contribution to science is increasing heavily, and the competition becomes  tougher among researchers, and at large extent among research groups, departments, universities as well as research institutions. As the research landscape evolves, assessing the impact of researchers and their disseminated research outputs is in high demand for a variety of reasons, such as the self-assessment of researchers, evaluation of faculties or universities, faculty recruitment and promotion, funding, awards \citep{InfoEuroRpt08} as well as to support the search for attractive content within an ocean of scientific knowledge. An evaluation task, which determines the impact and the productivity of researchers, requires the selection one or more information sources, appropriate evaluation indicators, and an uncontroversial evaluation procedure. To this end, a vast collection of such evaluation indicators, information sources and procedures are becoming available, which make the evaluation exercise more subjective. In the next section,  we present diversities along all the above mentioned dimensions.

%\section{Multi-dimensionality of Research Evaluation field: A Motivational Overview}
\section{Research Evaluation: A Multi-dimensional Field}

Research productivity evaluation is a broad endeavor. Among the other goals, the fundamental and the important one is to assess the return of investment in scientific research in the form of quality output. As scientific research heavily funded by the funding bodies, governments and institutions around the world, to establish a consensus about the success or failure of a research project requires making evaluation procedures based on those enriched indicators that can monitor both the productivity of their public money and the quality/impact of research, in order to establish policies for future investments.

Mostly, the evaluators (i.e., university management, funding organizations etc.) produce a new or alter an existing evaluation procedure or its sub-elements (e.g., h-index, g-index etc). The alteration takes place in the form of customization of an indicator tailor it for fulfilling demands in-hand. Moreover, when it comes to the selection of a data source, one may want to use a private data source, one could consider blog posts, keynotes and the like to be used as a performance indicator beside the traditions dissemination activities. Mainly, we observed that an evaluation procedure comprised of three basic, but diverse elements. These are as follows:

\begin {itemize}

\item The selection of one or more appropriate \emph{information sources}. These are the sources which fulfill data requirements (e.g., digital libraries, scholarly search engines). Recently, the presence of a large number of such information sources has provided an opportunity to choose one source over the others.

\item Second, the selection of a set of indicators. These are the smallest units in an evaluation procedure, which hold the logic to determine one particular impact factor. For instance, h-index is a citation based metric.

\item Finally, the formation of an overall procedure, which comprised of both, the information sources and the metrics that collectively determine research impact of researchers. A procedure may also include a customized version of a metric or a private data-source.

\end{itemize}

In the following sub-sections,  we elaborate each of these aspects in more detail.
\subsection{Diverse Information Sources}
An important dimension in the research impact evaluation domain lies in the exponential growth of freely available scientific/scholarly digital content. Bibliographic information sources (aka, digital libraries) maintain and provide bibliographic information. The information sources (e.g. Web of Science (WoS), Scopus, DBLP, Google Scholar etc.)\footnote{Each one of these information sources will be described in detail in chapter \ref{sec-stateoftheart}} as well as information production sources (e.g., authors, journals, books, articles etc.) are growing day by day. Moreover, universities and research institutes also maintain local repositories, which are then used by researchers to keep record of their dissemination activities.

Information integration is an important aspect in the research impact evaluation, which is to collect data from different sources and to apply merging techniques. For example,  several authors can be merged in many ways, like (1) taking an author's papers' information from one source and getting citation information from another (2) comparing two authors with data coming from different sources (3) using one's own private data source in comparison with other sources. %Yet, with the increase of bibliographic sources, the very problem of finding experts or high-profile people in some specific area or do a search for the "best" paper in some topic across different communities, is still a challenging endeavor.

Today, the presence of so many digital data sources overcomes the problem of data availability. On one side,  the excess of data and the data sources is a constructive development, but on the other side it becomes more challenging to decide the selection of one data source over the others. For instance, it is commonly accepted that DBLP data source is a good choice for computer science field in terms of its completeness. It provides a list of published articles for a researcher, but on the other hand it does not provide citation data, which then forces to include other citation sources.

\subsection{Diverse Evaluation Indicators}

In parallel with the growth of scholarly information sources and scholarly literature, people have established richer assessment indicators and metrics than before. These metrics not only incorporate traditional quantitative factors such as publication count or citation count, but also consider various other aspects such as researcher academic age, researcher positions, normalization. To name a few of these bibliographical research quality indicators that are considered to be well established and well-known in different communities include h-index, g-index, citation count, ar-Index etc.

Over the years, these indicators have received a tremendous success, even though different communities prefer to use customized versions of them. These customizations often varies from community to community and often based on a community trends, normalizations and many other factors. The point here is that, after so many efforts from different communities, it is still not guaranteed that a single metric can reflect the in-house demands of an evaluation committee. We also believe that with so many rapid developments in evaluation indicators, it will be extremely helpful to provide a way for research impact evaluation that could provide flexibility and customization support as well as the freedom of expressiveness to the evaluators.

\subsection{Diverse Evaluation Procedures}
As research landscape evolves, universities and research institutions start developing their personal research assessment procedures to meet specific local requirements. As of today, the availability of variety of information sources and also the assessment indicators, on one side gives more freedom to evaluators to choose among several options, but on the other side overall evaluation procedures become more subjective. These evaluation procedures often differ from traditional ones. For example, factors such as customization of the definitions of the traditional metrics such as h-index to contemporary h-index, inclusion of public as well as local private data sources, strict data filtering checks collectively makes an evaluation procedure tailored yet complex. Indeed, software developers cannot anticipate these customizations therefore not able to provide a largely well-accepted solution.

We have gathered a number of such evaluation procedures, which we describe in chapter \ref{sec:scenarios} in more detail to understand their insights. These specific, customized evaluation procedures demand expertise and skills in various ICT-related technical areas that those assessors lack. For example, a typical set of tasks required by these procedures include; fetching a list of publications from a source, applying cleaning process (i.e., to exclude publications which do not belong the queried researcher) and then to send the filtered list for a metric computation and in the end visualizations of results. In the following sections,  we describe in detail all the problems and challenges in this area and state our objectives.

\section{Problems, Challenges and Objectives}

Despite the fact that, the researchers must be evaluated on the basis of their research work; however, there is little consensus today on how an evaluation procedure should be designed and performed, and it is commonly acknowledged that the quantitative metrics available today are largely unsatisfactory. Indeed, today people judge research contributions mainly through publication in venues of interest and through citation-based metrics (such as the h-index), which attempt to measure research impact. However, there are different opinions on how citation statistics should be used, and they have well-known flaws. For instance,  \citep{chapman1989assessing} pointed out shortcomings, biases, and limitations of citation analysis. In another work \citep{seglen1997impact}, authors criticize the use of journal impact factor for evaluating research.

Furthermore, current metrics are limited to papers as the unit of disseminated scientific knowledge, while today there are many other artifacts that do contribute to the Science, such as \emph{blogs}, \emph{datasets}, \emph{experiments}, or even \emph{reviews}, but that are not considered in research evaluation. Besides the flaws of current metrics, the fact remains that people have - and we believe will always have - different opinions on which criteria are more effective than others, also depending on the task at hand (that is, the reason why they are conducting the evaluation). For example, in our department, the evaluation criteria for researchers are defined in a detailed document of 10 pages full of formulas and are mostly based on publications in venues that considered important in the particular community and are normalized following a particular agreed criteria. For instance, other institutions use citation counts normalized by the community to which the authors belong and then grouped by research programs to evaluate each research group, not individuals. Examples are numerous and, much like in the soccer world cup, everybody has an opinion on how it should be done.

Not only individuals may choose different metrics, but also different sources (e.g., Google Scholar vs. Scopus), different normalization criteria (e.g., normalizing the value of metrics with respect to averages in a given community), different ways to measure individual contributions (e.g., dividing metrics by the number of authors), or different ways to compare (e.g., compare a candidate with the group that wants to hire them to determine the autonomy and diversity of the candidate from the group), with different aggregation functions (e.g., aggregated h-index of a scientists co-authors, aggregated citation count, etc.).

We believe that this kind of personalization of the assessment processes (as well as many other personalization of the evaluation process, like for instance, the need of normalizing a traditional metric for a specific community) is a key element for the appropriate use and practical success of the various evaluation tasks. Moreover, people involved in such evaluation processes, most of the time are not IT experts, and not capable of building appropriate software for crawling data sources, automatically parsing relevant information, merging data and computing the required personalized metrics. Therefore, in order to empower the interested end-users, we need to design an appropriate and possibly easy-to-use IT platform, which could make life easier of those domain-experts who do not expert in IT. Indeed, supporting custom metrics for research evaluation is a non-trivial issue and requires addressing interesting research questions like:

\begin{itemize}

\item What is the set of key features that may enable a user to express its own evaluation metrics, i.e., what is the expressive power needed to do so? For instance, assessing the independence of a set of young researchers requires fetching all publications by the researchers, cleaning out papers that have been co-authored by the researchers' PhD supervisor, computing their h-index metrics, and ranking them according to their h-index.

\item How to enable less technical end-users to perform both easy and more complex data integration tasks? We have seen that being able to access an evaluation body (e.g., a set of papers) that is as complete as possible is at least as important as expressing custom metrics over the evaluation body. For example, fetching all publications of the young researchers may imply fetching data from Google Scholar, DBLP, and Scopus as well as fusing the obtained data and cleaning it.

\item Which is the best paradigm or formalism that may allow users to model/express their custom evaluation metrics? A metric may, for example, be expressed in text form via a dedicated domain-specific language, or modeled visually by means of suitable graphical modeling constructs, composed with the help of a guided wizard, and so on.

\item What type of software support does the computation of custom evaluation metrics need? Depending on the logic needed, the actual computation of a metric may be achieved via generated code, a dedicated evaluation engine, a query engine, or similar.

\end{itemize}

One of the most important issues that need especial consideration while addressing the problem is the kind of target end-users. We intend non-IT experts (i.e., non-programmers) as our end-users, who will get benefited from our research work. In following we introduce our proposed solution for all the aforementioned problems. It must be well-understood that throughout the different stages of our work,  we always refer and give examples from the selected domain to convey understanding whenever needed. However, this does not mean that the proposed solution is only valid for the selected domain. Instead, we aim at to keep separate those aspects that purely based on the chosen domain from those of generic type. In essence, we first aim at proposing a generic approach, and a methodology that then given a set of domain-specific aspects we show how to adapt it for that particular domain.
%##################### Intro From Journal paper

\section{Solution Overview and Contributions}

%If we carefully examine all the described scenarios and their requirements, we clearly see that the ingredients used in an evaluation procedure are entirely based on concepts, processes that are of the research evaluation domain. In most of the cases evaluation procedures 
\subsection{Overview}
After about two decades of research in \emph{workflow management} and more or less one decade of web service composition, two research streams whose initial ambitious goal was to enable non-technical users to design processes or compose services with little or no help from developers, we are still in a situation in which these forms of process modeling and execution technologies can only be mastered by specifically trained developers. One of the best examples of this situation is probably the recent standardization of Version 2.0 of the Business Process Modeling Notation (BPMN) \citep{model2011notation}, which brings together the two worlds of BPM and service composition, but that also has become much more like a programming language and less like a modeling instrument targeted at non-programmers (as the size of the documentation also indicates). As a result, people that are not fully familiar with the modeling notation are reluctant to use it since they know that they will not be able to draw a correct and consistent process model.

While this is a concrete issue in business process modeling and service composition, it is even more so in a relatively new, yet highly-related area: \emph{web mashups}. The recent emergence of mashup tools has refueled research on end-user development, i.e., on enabling end-users without programming skills to compose their own applications.

\emphbf{Mashups} are typically simple web applications (most of the times consisting of just one single page) that, rather than being coded from scratch, are developed by integrating and reusing available data, functionalities, or pieces of user interfaces accessible over the Web. For instance, \url{housingmaps.com} integrates housing offers from Craigslist with a Google map adding value to the two individual applications. Likewise, \emphbf{Mashup tools}, i.e., online development and runtime environments for mashups, ambitiously aim at enabling non-programmers (regular web users) to develop their own applications, sometimes even \emph{situational} applications developed ad hoc for a specific immediate need \citep{imranleveraging}.

However, we think that doing so is even harder than enabling non-programmers to model an own process or service composition, because developing full applications is simply complex. While the component-based reuse approach is certainly lowering part of the complexity, developing an own application, however, also means dealing with data integration, application logic, and content presentation issues, all aspects the common web user is not even aware of. Yet, similar to what happened in web service composition, the mashup platforms developed so far tend to expose too much functionality and too many technicalities so that they are powerful and flexible but suitable only for programmers. Alternatively, they only allow compositions that are so simple to be of little use for most practical applications.

For example, mashup tools typically come with \emph{SOAP services}, \emph{RSS feeds}, \emph{UI widgets}, and the like. Non-programmers do not understand what they can do with these kinds of compositional elements \citep{NamounICWEWS2010,NamounECOWS2010}. We experienced this with mashup tools in our own group, mashArt \citep{DanielER09} and MarcoFlow  \citep{DanielBPM10}, which we believe to be simpler and more usable than many composition tools, but that still failed in being suitable for non-programmers \citep{mehandjiev2011empowering}.

Yet, being amenable to non-programmers is increasingly important as the opportunity given by the wider and wider range of available online applications and the increased flexibility that is required in both businesses and personal life management raise the need for situational (one-use or short-lifespan) applications that cannot be developed or maintained with the traditional requirement elicitation and software development processes.

We believe that \emphbf{the heart of the problem} is that it is impractical to design tools that are \emph{generic enough} to cover a wide range of application domains, \emph{powerful enough} to enable the specification of non-trivial logic, and \emph{simple enough} to be actually accessible to non-programmers. At some point, we need to give up something. In our view, this something is generality, since reducing expressive power would mean supporting only the development of toy applications, which is useless, while simplicity is our major aim. Giving up generality in practice means narrowing the focus of a design tool to a well-defined \emph{domain} and tailoring the tool's development paradigm, models, language, and components to the specific needs of that domain only.

\subsection{Contributions}
This chapter presented an introduction of the reference domain and the problems and challenges faced by the users. However, a more detailed discussion and requirements that are of domain-specific type will be presented in chapters \ref{sec-stateoftheart}, \ref{sec:scenarios}. Moreover, the requirements those are related to the end-users (i.e., non-programmers) will be presented in chapter \ref{ch:eud-soa}. In following we summarize contributions of this thesis. 

\begin{enumerate}

\item First of all, we present the novel idea of \emphbf{domain-specific mashups} and describe what they are composed of, how they can be developed, how they can be extended for the specificity of any particular application context, and how they can be used by non-programmers to develop complex mashup logics within the boundaries of one domain.

\item We detail and exemplify all \emphbf{design artifacts} that are necessary to implement a domain-specific mashup tool, in order to provide expert developers with tools they can reuse in their own developments. 

\item We show what developing a domain-specific mashup tool means, which role the \emphbf{mashup meta-model} and the \emphbf{domain concept model}, the \emphbf{domain syntax model} play and how these can be merged into a \emphbf{domain-specific mashup meta-model}.

\item We describe a \emphbf{methodology} for the development of domain-specific mashup tools, defining the necessary concepts and design artifacts. As we will see, one of the most challenging aspects is to determine what is a domain, how it can be described, and how it can both constrain a mashup tool (to the specific purpose of achieving simplicity of use) and ease development. The methodology targets expert developers, who implement mashup tools.

\item We apply the methodology in the context of a \emphbf{mashup platform} that supports the development of domain-specific mashup tools. To achieve this, we present a baseline platform, which is then used to develop and tailor a mashup tool to support a domain most scientists are acquainted with, i.e., research evaluation. This mashup platform targets domain experts (i.e., non-programmers). 

\item In this thesis, we also present an efficient approach for mashup-based web application, those communicate big data between client and server. The proposed approach prevents heavy data communication using suitable communication-pattern (i.e., among the four proposed patterns) and a server-side cache. 

\item To evaluate our work, we performed twofold validations. First, we performed a \emph{usability and comparative evaluation}, which is to understand end-users preference between a generic versus a domain-specific mashup tool and to learn the right balance a mashup tool should offer in terms of complexity, flexibility, and expressiveness. Second, we performed a user studies in order to assess advance usability aspects of the developed platform and the viability of the respective development methodology.

\end{enumerate}

While we focus on mashups, the techniques and lessons learned in the thesis are general in nature and can easily be applied for other domain sand to other composition or modeling environments, such as web service composition or business process modeling.

\section{Structure of the thesis}

Literature reviews and the aforementioned contributions of this thesis are presented in different chapters as described below:

\begin{itemize}

\item Chapter 2, presents state of the art related to the domain of research evaluation. We present different evaluation indicators, data sources and techniques, which are being used for different evaluation purposes by different communities. We also present the related tools that are currently available for performing research evaluation. 

\item Chapter 3, presents state of the art related to the End-user development. We present different approaches that end-user development based upon. Various programming paradigms especially for the end-user are reported. Moreover, we present mashups approaches, and see how this paradigm can be used for effective end-user development.

\item Chapter 4, describes a few real-life research evaluation procedures, which we have collected from different sources, to devise a set of concrete requirements and in the end we present our analysis in terms of major design-principals that are to facilitate end-users for their development tasks.

\item Chapter 5 states a set of methodological steps. We present the definitions of important concepts, various design artifacts, formalisms, and a detailed methodology for the development of domain-specific mashup tools. We show what role a domain-model, meta-model and a domain-specific meta-model play in the development of a domain-specific mashup tool. 

\item Chapter 6 shows an implementation of a generic mashup tool, its design principals, architecture and shows how and where domain knowledge can be injected for tailoring it to a domain-specific mashup tool.

\item Chapter 7 presents ResEval Mash, a mashup tool that is tailored to the domain of research evaluation. We present how different domain related artifacts are used in the development following the methodological steps presented in the chapter 5.

\item Chapter 8 reports on a few user studies that we conducted to evaluate of our approach, methodology and domain-specific mashup tool. 

\item Chapter 9 concludes the thesis. We present future work, lessons learned specific of the selected domain and of related to the development of mashup tool in general.

\end{itemize}

\newpage

% Chapter 2: State of the art.
% !TeX root = ../phd-thesis-arxiv.tex
%%% Chapter start. %%%
\chapter{Research Impact Evaluation: State of the Art}
\label{sec-stateoftheart}

\section{Overview}
This chapter presents comprehensive insights of the research impact evaluation field. Exploring fundamental questions, like \emph{what is research impact evaluation?}, \emph{why is it needed?}, \emph{how is it performed?} and \emph{who performs it?}, provide us a consolidated base through which we tend to understand various associated aspects of the field. In response to the \emph{how}, we also present different evaluation indicators that are developed over the years and are being used by different communities. Although, these communities have adopted and tailored these indicators to meet their community-specific trends and requirements, even understanding those specific details lead us to a solid understanding. This chapter also reports on the impact evaluation tools that have been developed and used over the years and we explain why these tools failed to support the current practices in research evaluation field. In response to the \emph{who}, we present end-users who perform such evaluation tasks and what are their expertise level with respect to this domain and to the technology.

\section{Multiple Faces of Research Impact Evaluation}

Impact evaluation, in terms of a project, program or policy, assesses the changes that could happen after a particular intervention. In essence, the impact evaluation is a comparison between what happened and what would have happened if we take those interventions aside. In theory, the concept of impact evaluation is slightly different from "outcome monitoring", which is to check on whether targets have been achieved or not. While the field of \emph{research impact evaluation} deals with the growing concerns related to the productivity assessment of a research work, sometimes, both in terms of research inputs and outputs. The research assessment could be of various types, for instance, ranging from the traditional ways (i.e., peer review process which usually performed before dissemination, for an early evaluation) to more sophisticated assessment methods (i.e., using citation-based, content-based indicators; mainly performed after dissemination).  Likewise, the evaluation can be an ongoing process that monitors the progress of work, or it can be a process that evaluates at some certain stages (e.g., midterm evaluation, final-stage evaluation).

From the point of view of an early or pre-dissemination evaluation approach (i.e., peer review), the assessment takes place by the recognized experts in a particular field. In practice, peer review usually performed by experts with general expertise in a specific field, which is largely an accepted way, however, sometimes this particular scrutiny process considered controversial, as according to some others, the evaluation committee should be comprised of specialists of the field rather than a general competence committee. On the other side, the post-dissemination evaluation process, which is the main focus of our discussion, is much more controversial than of pre-dissemination. Over the years, many approaches have been proposed and to some extent fulfill a general set of evaluation requirements. However, despite many efforts, different communities have developed new or tailored exiting evaluation methods for their specific needs. In the last few years, It has been observed that the research spectrum crosses the boundaries, researchers are becoming more collaborative than ever, research groups are formed of experts from different affiliations and different continents. In such a conducive environment for research to grow, the amount of research dissemination to science is rapidly increasing. In parallel to this increase,  the assessment of research outputs has become a crucial issue for a wider range of stakeholders (e.g., funding bodies, universities, research institutions etc.). The field of research impact evaluation primarily focuses on a number of aspects that need to be considered first. For example, amongst many others, the fundamentals are:

\begin{itemize}

\item For whom the evaluation procedure is taking place? A clear vision of a body (e.g., individuals, groups, universities etc.) to be evaluated is a core element before performing further steps.

\item What types of research artifacts to be considered in the evaluation? After the selection of \emph{whom}, the next step is to agree upon what research outputs of the selected unit will be considered in the evaluation.

\item What evaluation methods to adopt? This aspect addresses the most controversial part of the evaluation process i.e., evaluation approach, method, the nature of the process.

\end{itemize}

The first and the fundamental aspect, that must be considered before investigating further into the details, is for whom the evaluation procedure will be performed. That is the selection of an \emph{unit} to be evaluated (i.e., whose research work to be evaluated). The units of assessment include individuals, research groups, departments, universities, research fields and even countries. The complexity of an evaluation procedure is directly proportional to the selected unit. To determine the productivity of an individual researcher is far easier than to determine the productivity of a university where normally hundreds of researchers work. The second noteworthy aspect in the research impact evaluation field is the selection of the types of research outputs to be evaluated. To this end, different disciplines prefer different types of research output to be considered. Usually these types include, to name a few of them, journals, conference and workshop proceedings, book chapters, books, prototypes etc. Amongst the other important aspects, the selection of appropriate assessment indicator is highly important, and to some extent is highly controversial in some cases. Often, one's opinion on an indicator for an assessable unit differs from others as everyone has his own opinion on what criteria/indicator should be used.

%\begin{figure}[t]
%  \centering
%    \includegraphics[width=1.05\columnwidth]{figs/multi-dimensions-reseval.pdf}
%  \caption{A multi-dimensional research assessment matrix, presenting various units of evaluations, purposes, and indicators.}
%  \label{fig:multi-dimension-reseval}
%\end{figure} 

Based on these diversities, in 2010, a multi-dimensional research assessment matrix was published by the Expert Group on the assessment of University Based Research (AUBR) \citep{aubr2012}, operated under European Commission. 
%Figure \ref{fig:multi-dimension-reseval}, depicts the core part of that research assessment matrix, which is not the full version of the actual matrix. If we carefully look at the figure \ref{fig:multi-dimension-reseval}, it can be easily seen that how much diverse the field of research evaluation is. 
The matrix presents five basic units of assessment, obviously one can think of a different one. The matrix also represents a few purposes (i.e., why a particular research work conducted) for each unit to be assessed. Moreover, the matrix also shows a very basic set of bibliometric as well as a few other emerging indicators that can be applied to various assessable units. In essence, the matrix shows a glimpse of the diversity of the field and clearly it is not restricted to only these aspects, one can think of many other trivial as well as non-trivial aspects.

In the field of research impact evaluation, the central role in an assessment procedure holds by the selected assessment indicators. Over the years, many different indicators have been proposed. These include quantitative as well qualitative ones. In a report published by Scopus \footnote{http://www.researchtrends.com/wp-content/uploads/2011/06/Research\_Trends\_Issue23.pdf} in 2011, amongst the others, they only focused on bibliometric indicators. According to the report, bibliometric indicators are divided into three generations. In table \ref{tab:bib-generations}, we show the division of all three types of bibliometrics indicators. The first generation corresponds to a basic set of indicators (e.g., publications count, citations count etc.), which are easily available and can be obtained from various sources. The second generation, which is relatively more advance than the first ones, includes indicators that used to be normalized based on a specific filed to remove the biases and so on. The third and the most non-trivial set of indicators were categorized in this generation that include influence weights, Journal Rank, SCImago and other more sophisticated indicators etc.

\begin{table}[ht!]
\begin{center}
\begin{tabular}{|p{2.24cm}|p{5cm}|p{5cm}|}
\hline
{\bf Type (generation) } & {\bf Description} & {\bf Typical examples}\\
\hline
First & Basic indicators; relatively easy to obtain from sources that have available for decades & Number of publications; number of citations; journal impact metrics\\
\hline
Second & Relative or normalized indicators, correcting for particular biases (e.g., differences in citation practices between subject fields) & Relative or field-normalized citation rates\\
\hline
Third & Based on advance network analysis using parameters such as network centrality & Influence weights; SCImago Journal Rank; 'prestige' indicators \\
\hline
\end{tabular}
\caption{Generations of bibliometric indicators}
\label{tab:bib-generations}
\end{center}
\end{table}

To practically devise an evaluation procedure, it requires making decisions about which \emph{unit} needs to be assessed, for what purposes, on which output dimensions, using which assessment indicator (i.e., a bibliometric or other emerging indicators). Clearly, there is not a single answer to these questions, it is entirely, on one side, based on the purpose of an evaluation, the selected unit to be assessed, and on the other side the selection of appropriate indicators. In our opinion,  the field of research impact evaluation is highly diverse, and the use of one indicator over the others is highly subjective. Even the citation-based approaches can alone raise significant challenges, but a proper use of these can also provide a clear indication of someone's performance. Many studies, for example, according to \citep{van1996advanced}, quantification through citation analysis of past performance can be used to predict future performance. Moreover, in a similar study that is based on several related aspects of citation analysis has been presented in \citep{moed2005citation}, where author presented a detailed analysis of accuracy, theory, and effective use of citation analysis in parallel to its strengths and weaknesses.

\subsection{Quantitative and Qualitative Research Evaluation}

By and large, the impact evaluation approaches can be divided into two basic methods: 1) \emph{quantitative} 2) \emph{qualitative}. Both methods can be distinguished based on the type of evaluation experiments conducted on the data produced by some research work. In general, quantitative methods focus more and deal with real numbers. For instance, count on the number of publications, count on the number of citations, and other indicators that rely on such numbers in one way or the other, like H-Index, G-Index etc. While qualitative methods are more based on the descriptive properties of the data. For examples, evaluation practices those involve aspects like reputation, peer ranking analysis through participatory studies, interviews, and other socially enhanced indicators. Quantitative approaches are typically used and kind of considered standard method. Whereas, qualitative approaches are less common and rarely used. We mostly focus and study bibliometric methods that are quantitative in nature than of qualitative ones.

\subsection{Bibliometrics, Scientometrics and Informetrics}
Often interchangeably used terms: Bibliometrics, Scientometrics and Informetrics, refer to the methods that study various aspects related to the science and information (i.e., the information present in any form). To some extent, there has been confusion for these closely related terminologies. Over time, people have defined these terminologies for the field they belong, but still all definitions show considerable overlap among different terms that they used.

In 1969 \emph{Pritchard} introduced the term \emph{Bibliometric} in his paper \citep{pritched1969bibliometric} as \emph{``the application of mathematical and statistical methods to books and other media of communication"}.  He stressed more on quantitative aspects, like count on the number of articles, publications, citations, books and in general any statistically significant measures of recorded information. The term \emph{Scientometrics} was introduced as a science for analyzing and measuring science through relationships and social structure and also to check the status of an individual within a group \citep{de1986little}.

A field that encompasses both the bibliometrics and scientometrics fields is \emph{Informetrics}. In \citep{bar2008informetrics}, the author defined it as a study of the quantitative aspects of information in any form that include the production, dissemination and use of the information regardless of its form. In the following section, we mainly focus on the bibliometrics based approaches and indicators.

\section{Research Evaluation Through Bibliometrics Approaches}

\label{sec:research_evaluation_soa}

Over the last few years, bibliometric indicators are considered to be a standard and popular way to assess research impact. All significant indicators heavily rely on publication and citation statistics and other, more sophisticated bibliometric techniques. In particular, the concept of citation\citep{garfield1992nobel,garfield1979citation} became a widely used measure of the impact of scientific publications, although problems with citation analysis as a reliable method of assessment and evaluation have been acknowledged throughout the literature\citep{chapman1989assessing}. Indeed, a research work not always gets citations because of its merits, but also for some other reasons such as flaws, drawbacks or mistakes. A number of other indicators have been proposed to balance the shortcomings of citation count and to "tune" them so that they could reflect the real impact of a research work in a more reliable way. As with the increase of scholarly literature, different communities introduced new indicators for the assessment. Although these indicators widely based on citation analysis, but they gained popularity over simple citation indicators like a simple publication or citation count.

Of the many famous indicators, like \emph{h-index} that is proposed by \citep{hirsch2005index} by Jorge Hirsch, considered as a more comprehensive indicator to assess the scientific productivity and the impact of an individual researcher. The \emph{h-Index} is among the recent and most successful indicators over the last few years because it is straightforward to compute based on the citations of a researcher's publications. The \emph{h-index} takes into account both the quantity and the impact of the researcher's contributions. That is why some of the most significant journals\citep{ball2005index} take interests into it. The original definition of the h-index by Hirsch is as:

\begin{definition}\label{def:gindex}
A scientist has index \textit{h} if \textit{h} of his or her $N_p$ papers have at least \textit{h} citations each and the other $(N_p - h)$ papers have $\leq$ \textit{h} citations each.
\end{definition}

The \emph{h-index} has been widely acknowledged because of the good properties it holds, for example in \citep{costas2007h}, authors considered this index as an objective indicator and based on this they stated that it can play a significant role when allocating funds, making decisions about personnel or awarding prizes. In \citep{vanclay2007robustness} highlighted another advantage of the \emph{h-index}, where author reported that the \emph{h-index} does not care much about the low cited papers, which is a good thing that makes this index viable than others. According to them, as the majority of the confusions and errors tend to occur in the lower part of someone's citation record so neglecting that part certainly reduces possible errors.

However, some flaws and drawbacks of the h-index have been identified over time and often different authors have tried to solve those errors by introducing new indicators or its variations. Hirsche in his paper \citep{hirsch2005index}, himself mentioned that due to differences in the productivity of different fields, there are differences in $h$ values. Hence, comparing two researchers based on their h-index values those belong to two different disciplines is not an appropriate comparison. Another disadvantage of the h-index is that, it is used to compare researchers which are at a different level of their career, since h-index depends on the scientist's entire career, but publications and citations increases over time, claimed in \citep{kelly2006h}.

To overcome the shortcomings of the h-index, recently a number of variations of the h-index have been proposed. One of the proposals presented in \citep{van2008generalizing}, where authors considered the h-index is quite arbitrary. From their point of view Hirsche could defined h-index as: "a scientist has h-index of $h$ if $h$ of his $n$ papers have at least $2h$ citations each and the other $n-h$ papers have $\leq2h$ citations each''. That is how they extended the h-index to $h_\alpha$-index, which is formally defined as:

\begin{definition} "A scientist has $h_\alpha$-index of $h_\alpha$ if $h_\alpha$ of his n papers have at least $\alpha . h_\alpha$ citations each and the other $n - h_\alpha$ papers have fewer than $\leq \alpha . h_\alpha$ citations each.'' Where $\alpha \in (0, \infty)$.
\end{definition}

In \citep{jin2006h}, author proposed $A-index$, according to which they proposed to use average of the citations in the Hirsch core \citep{rousseau2006new}. Formally A-index is defined as:

$ A = \frac{1}{h} \sum\limits_{j=1}^{h} cit_j$\\

In the above definition of A-index, $h$ is the h-index value and $cit_j$ is the total citations received by $j-th$ most cited paper. Another problem that is also solved by the \emph{A-index} is that the index increases its value if the most cited papers receive more citations, while in case of h-index, it does not increase if a most cited paper gets more citations. To the best of this side, it is crucial that if an indicator which should indicate quality of a researcher, should consider the performance of top cited papers too. To this end, an indicator which is known as \emph{g-index} was proposed by Egghe \citep{egghe2006theory}. The formal definition of the \emph{g-index} according to Egghe is as follows:

\begin{definition}\label{def:gindex}

A set of papers has a \emph{g-index} $g$ if $g$ is the highest rank such that the top $g$ papers have, together, at least $g^2$ citations. This also means that the top $g + 1$ papers have less than $(g + 1)^2$ cites.

\end{definition}

Egghe's concern with the h-index was, once the h-index is computed, for the highly cited paper it remains insignificant that those receive further citations as new citations do not effect the $h$ value. The consequences of this would impact highly cited researchers, as they may have h-index similar or equal to moderate researchers. However, the g-index also suffers from problems. For instance, if a researcher receives a high number of citations in one paper, but for other papers he gets average citations. The g-index for that researcher would be higher as compared to other scientists with higher average citations in their papers, reported by \citep{alonso2010hg}.

To overcome the limitations of both $h$ and $g -indices$, a new index has been proposed in\citep{alonso2010hg} with the aim to combine the good properties of both indices and to minimize the disadvantages. This index is known as \emph{hg-index}, and is defined as $hg = \sqrt{h.g} $, which is the geometric mean of the $h$ and $g$-index. It is easily understandable that $h \leq hg \leq g$ and that $hg - h \leq g - hg$. Indeed this index is very simple to compute once both $h$ and $g$-index values have been obtained. It has more granularity, which makes it even easier to compare researchers with similar $h$ or $g$-index values.

In \citep{jin2007ar} authors proposed a new index, which is known as $AR$-index. This particular index not only takes into account citations of a researcher and also the publication age. As with the time, the performance of a researcher can increase or decrease, which is an aspect that was ignored before. However, the $AR$-index claims to observe these changes and can increase or decrease with time. The AR-index is formally defined as follows:

$AR = \sqrt{\sum\limits_{j=1}^{h}\frac{cit_j}{a_j}}$.

Where $h$ is the h-index value, $cit_j$ is the total number of citations of the $j$-th most cited paper, $a_j$ is the number of years since the publication of the $j$-th paper. In another work \citep{egghe2008h} in which the authors proposed the idea to give weights to citations. This variation of the h-index is known as $h_w$-index and is defined as follows:

$h_w = \sqrt{\sum\limits_{j=1}^{r_0}cit_j}$.

Where $cit_j$ is the number of citations for the j-th most cited paper, $r_0$ is the largest row index i such that $r_w(i) \leq cit_j$ and $r_w(i) = (\sum\limits_{j=1}{i}\frac{cit_j}{h})$.

In \citep{kosmulski2006new}, author presented the $h^{(2)}$-index. In this work, the authors proposed to give more weight to the most cited papers, as this idea originally been presented in the g-index. Based on this idea, the $h^{(2)}$-index is defined as: "A scientist's $h^{(2)}$-index is defined as the highest natural number such that his $h^{(2)}$ most cited papers received each at least$[h^{(2)}]^{2}$ citations''. This index is easier to compute because it only focuses on highly cited paper. It can be used with data where some uncertainty exists, especially in low cited papers. This index also sufferd by problems identified in \citep{sidiropoulos2007generalized}, where author emphasized that as a small set of papers are needed to compute $h^{(2)}$-index, and since researchers with different number of publication and citation rate, which is not suitable for this type of index. Thus, they proposed the \emph{normalized h-index}, which is defined as: $h^{n} = \frac{h}{N_p}$. Where $h$ is the $h$-index and $N_p$ is the total number of publications of a researcher. This index is also considered more suitable for younger researchers, as they can less productive at the beginning of their career.

In \citep{anderson2008beyond}, author proposed an interesting index, which is called tapered h-index. They propose to incorporate all citations for all papers of a researcher. One of the Shortcomings of the h-index is that it ignores very low cited papers as well as new citations to highly cited papers. However, this index claims to consider complete citation records of a researcher despite a paper has low or high citations. It uses the idea of representing the citations of the papers in a Ferrers graph, where columns represent the partition of the citations among the papers. The largest filled square in Ferrers graph, is called the Durfee square. In another similar approach\citep{ruane2008rational}, authors presented the rational h-index $h_{rat}$-index, which is defined as: $h_{rat} = (h + 1) - \frac{n_c}{2.h + 1}$ where h is the h-index, $n_c$ is the number of citations. Intuitively $h\leq h_{rat} < h + 1$.

There are some other factors that might implicitly influence the interpretation of the results using a citation-based metric. Therefore, the evaluation process may produce incorrect results. One of these factors could be the \emph{self-citation} count. The controversial phenomenon of self-citation is generally believed to create problems for those who would attest to the reliability of citation analysis for evaluative purposes\citep{purvis2006h,van2006comparison}. The inclusion of self-citation in the calculation of citation statistics inflates the research impact of a given artifact, thus taking out self-citations from citation count would be better in quantification of a more realistic research impact.

Mich\`{e}le Lamont's book \citep{lamont2009professors} holds a complete analysis on how evaluation is performed by professors. In the book, she analyzed the complicated details of peer reviews and 12 panels of experts in the humanities and social science, extrapolating subjective criteria for decision-making in each different discipline, giving an interesting overview of possible features that influence reputation of researchers. The Altmetrics Initiative \citep{priem2010alt} goes one step further and aims at using social interactions for proposing new indicators of research impact more related to the reputation of the researchers.

We have presented a number of different metrics that have been proposed and used. We can clearly see that the present literature on research impact evaluation emphasizes that there are so many different criteria, proposals and thoughts for conducting the evaluation and there are different opinions on which criteria are more effective than others (depending on the reason why they are conducting the evaluation). We provide a more detailed critical analysis of all these metrics in the section \ref{sec:analysis}. However, in the next section we present a comprehensive review of the different information sources (i.e., bibliographic databases) and various tools developed support providing evaluation services.

\section{Bibliographic Databases}
\label{bib-dbs}
Bibliographic databases also known as digital libraries maintain and provide bibliographic records such as,  journals, conference proceedings, technical reports, books, patents etc. A bibliographic database can be a multidisciplinary in terms of coverage (i.e., covering various disciplines like computer science, physics etc.) or can be a discipline-specific (i.e., covering one discipline). Of the several bibliographic databases,  a few of them are proprietary, available under licensing, and other are freely available on the Internet. The ones, freely available either offer their services as a scholarly search engine or as a digital library (i.e., a system that store content in digital formats and accessible via computers through an API). In the next section,  we present a few of these bibliographic databases and present services these databases provide. We also report on diversities, completeness, and coverage issues related to these databases.

\subsection{Web of Science}

A decade ago, researchers had essentially a very few bibliographic data sources available, among those  the \emph{Web of Science}\footnote{http://scientific.thomson.com/products/wos/}, which is an online academic citation index provided by Thomson Reuters, was very popular. Web of science provides access over 12,000 journals worldwide, including 150,000 conference proceedings\footnote{ Recorded on Jan 10, 2013}. Web of Science provides coverage of nearly 256 disciplines that include science, social science, arts, humanities etc. Along with the bibliographic data, web of science also provides a few numbers of indicators that can be used for research impact evaluation. The commonly used indicators provided by WOS include: \emph{p-index} (number of articles of an author), \emph{cc-index} (number of citations excluding self-citations), \emph{cpp} (average number of citations per article), productivity (quantity of papers per time-unit). To some extent, these indicators can be used to determine the impact of communities, journals, academic institutes using various aggregations. Another, academic citation indexing search service known as \emph{Web of Knowledge}, is also provided by Thomson Reuters. This wrapper service covers a few disciplines like sciences, social sciences, arts, humanities, that also include a number of journals from the web of science. It provides tools to analyze the bibliographic content over several databases.

Despite all the benefits the web of science and web of knowledge provide, they still have some limitations, and thus become very crucial in some assessments tasks. Among these drawbacks, the limited coverage of these services that only targets, as mentioned above, a few high impact peer-reviewed journals. These journals only represent a fraction of research work that is published. In various disciplines internationally recognized high impact journals are not the only way to disseminate research work, so those cannot take advantage of the Thomson Reuters services. Moreover, the web of science does not provide free access to their data and tools, which can also be considered as a drawback for these kinds of bibliographic database.

\subsection{SciVerse Scopus} 
Recently, many other competitors of the Web of Science emerged that also provide bibliographic data. One of these is \emph{Scopus}\footnote{http://www.info.sciverse.com/scopus}, that maintains bibliographic records including citations, abstracts, journal articles. As of today\footnote{Scopus database status published on their website on Jan 17, 2013}, Scopus claims of having a bibliographic database that contains more than 20,500 peer-reviewed titles from more than 5,000 international publishers. In case of scopus, it only indexes journals, book series, conference proceedings that have an ISSN assigned to them. Scopus does not index an article whose author is not the person behind the presented material such as obituaries or book reviews. Scopus provides various tools that work on their own database and provide value-added services. For instance, citation tracker is a tool that can be used to find highly cited author in a field or hot topic in some subject areas.

Similar to the web of science approach, Scopus is also a paid source of bibliographic type of information. Elsevier that operates Scopus also operates a free service called \emph{Scirus}. It is a science-specific search engine that only works for Computer science field. One can search bibliographic records using this service; however, they do not provide any kind free public API to take advantage of the data they maintain.

\subsection{Microsoft Academic Search}

On the contrary to both Web of science, and Scopus services as mentioned above, the \emph{Microsoft Academic Search}\footnote{http://academic.research.microsoft.com/} is a \emph{free} academic search engine. This search engine is developed by Microsoft Research and it came into being during the recent years. This multidisciplinary search engine covers more than 48 million publications and more than 20 million authors from various domains. The service is free and provides an easy to use interface to query scholarly literature. Moreover, Microsoft Academic Search provides a few basic indicators (e.g., h-index, g-index etc.) for assessment, and it also provides a visual explorer where one can visualize a researcher's co-authors graph or a citation graph.

Another appealing yet highly demanding feature, which is researchers name disambiguation, is also provided by Microsoft Academic Search. This feature to some extent works, but we personally observed that it too does not completely disambiguate many cases. To disambiguate a researcher, it shows a list of authors who share the same names along with their affiliations. From the given list a user can select one among many based on the affiliation. However, the problem still exists and the service does not completely disambiguate more complex cases. In the beginning their data service suffered by the problem of coverage. Until the year 2010, they only covered the computer science field, but quiet recently the coverage has been increased to other disciplines like biology, chemistry, mathematics etc., which makes the service more useful.

\subsection{Google Scholar} 
Likewise the Microsoft Academic Search service, Google also started in 2004 a bibliographic search service named \emph{Google Scholar}\footnote{http://scholar.google.com/}. Google Scholar provides a very simple interface to search bibliographic content over a large set of disciplines from many sources. Google Scholar maintains its database by crawling data from quite a large number of sources. The type of bibliographic data that Google Scholar indexes include peer-reviewed online journals, conference proceedings, books, non-peer reviewed journals, preprints, technical reports, theses etc. Moreover, Google scholar maintains the citation records of scholarly literature. 

It does not guarantee that an article indexed by Google Scholar can be freely available, though a request made through certain universities, institutes those subscribed to various services can access articles freely. Google Scholar claims and apparently considered trusted bibliographic source in terms of its coverage. Moreover Google Scholar seems the most updated scholarly data providers, though nobody knows when and which journals Google scholar crawls. However, the data quality in some cases seems compromised. Google Scholar does not provide the support for name disambiguation problem, that is, for example in the case where two or more authors share the same name \citep{imran2013real}. 

\subsection{DBLP} 
\emph{DBLP} is largely a computer science specific bibliographic database hosted in Germany by the Universitat Trier. As of November 2012 DBLP maintains 2.1 million bibliographic data. DBLP provides a browser-based user interface for performing search over the data and also it allows to download the entire dataset in XML format. Moreover, DBLP offers an API that developers can use to query specific records. The service is free, though as it is today, a disadvantage of this service is that it only covers the computer science field. Moreover, DBLP does not maintain citations references. Despite these flaws, the DBLP service considered a clean and reliable source for bibliographic data.

The above mentioned bibliographic services are just the tip of the iceberg. Over the years, a number of other bibliographic data sources have been emerged. Among these bibliographic databases,  \emph{CiteSeerX}\footnote{http://citeseerx.ist.psu.edu/}, arXive\footnote{http://arxiv.org/}, \emph{Association for Computing Machinery (ACM)}\footnote{http://www.acm.org/}, \emph{GoPubMed}\footnote{http://www.gopubmed.org/}, \emph{Science.gov}\footnote{http://science.gov/}, \emph{SpringerLink}\footnote{http://www.springer.com/} are the popular ones. 

The proliferation of data sources makes it evident that the scholarly data and the data providers are numerous, however, the main problem for non-experts users is the lack of technical expertise that are required to use these sources to crawl, call API etc. For simple scenarios, for instance, to get a list of publications of a researcher seems reasonable and can be performed manually. However, tasks such as to get all the publications and citations of all the researchers of a university poses serious challenges that cannot be performed manually as it requires huge human efforts. Thus, an easy-to-use, flexible and as much as automated software support is required that could perform such complex tasks. Recently, a number of such tools have emerged. In the next section we report on these tools that provide the research evaluation services based on the different data sources mentioned in this section.

\section{Research Impact Evaluation Tools}
\subsection{Publish or Perish} 
Based on the existing bibliographic data sources, new tools are beginning to be available to support people in their research evaluation analysis. Such a tool named \emph{Publish or Perish} was developed by \citep{harzing2007publish}. The tool is freely available to download on the Internet. It is a desktop software that crawls Google Scholar pages for a given query and then analyses the data for further computation of citation based metrics. It provides a few numbers of famous metrics like h-index, g-index, zhang's e-index and a few more. A user can filter out publications of his/her interest from a given list of publications that the tool actually crawls. To some extent, this approach is useful for someone who intends to perform analysis of his own data, because it's easy to determine what publication data belong to him. But the very approach does not work in those cases where users want to search other researchers as it is less likely and hard to remember about someone's else complete publication details. Among the other weaknesses that this tool has, include, (1) its reliance on only one information source i.e., Google Scholar; (2) the need for manual cleaning of the obtained data (for example for author disambiguation and self-citations among others); (3) the lack of Application Programming Interface (API) over which other applications or web services could use their services; (4) the tool does not provide a way to call a third party API, a feature which is useful if provided. Moreover, a user cannot customize or provide a new user-defined evaluation procedure.

\subsection{Scholarometer} 
A different approach is provided by \emph{Scholarometer} \citep{hoang2010crowdsourcing}, which is a kind of social tool that is used for citation analysis and also for the evaluation of the impact of an author's research work. It is a browser-based free add-on for Firefox and Chrome that provides a smart interface for fetching data from Google Scholar. However, the service requires users to tag their queries with one or more discipline names from a predefined list of disciplines. This generates annotations that go into a centralized database, which collects statistics about the various disciplines, such as average number of citations per paper, average number of papers per authors, etc. The impact measures are then dynamically recalculated based on the user's manipulations. Scholarometer has a server where information about the queries performed and their results are stored. However, it does not offer an API to retrieve or use this information. This tool also only depends on Google Scholar data, and no other data providers can be injected or used or linked with it. Moreover, the functionality to add or to customize existing evaluation indicators is not provided, so it is not suitable for those users who want to implement a very specific evaluation procedure. The use of predefined disciplines makes this tool more restricted to only tool provider's chosen fields, no provision is provided to introduce new disciplines though. 

\subsection{ResEval} 
Over the time, information sources and evaluation enabler tools are becoming available but they still have many shortcomings. For example they differ in data coverage, data quality as the same case for Scholarometer. Moreover, these tools are data-source specific and cannot be extended to use other data sources. Moreover, personalization of metrics, an important feature for the diverse field of research evaluation, is still missing. 

With an aim to overcome the above mentioned deficiencies of the existing solutions, we introduced our own tool for the research evaluation purposes as a part of LiquidPub project \citep{baez2011d5}. Lessons learned from the existing experiences, in our own tool ResEval \citep{imran2010resevalRE}, we focused on the computation of more informative citation based measures. The tool focuses on providing an open and resource-oriented research impact ways and stresses the customization of existing evaluation procedures, such as the h-index and g-index measures. ResEval provided the provision to introduce new customized evaluation procedures in the form of web services. Likewise, new data sources can also be added with the help of web services, which actually encompasses the logic of calling a data source API or crawling data from its web pages. That data then can be used to leverage various metrics provided by the tool. 

By and large, the functionalities that ResEval provided mainly targeted only the experience developers as the implementation of new web services, crawling data from web pages, performing filtering, aggregating results etc. are all aspects that an experienced developer is capable to perform \citep{imran2010resevalecss}. That is the reason, the tool failed to achieve its objective as no end-user (non-technical user) support was provided, which is the main requirement of this field. The lessons learned from other and our own tool motivated us to think about a solution that stays in the boundaries of an end-user's expertises. 

\subsection{Research Gate}
\emph{Research Gate}, is a new and a different kind of entry in the list of already existing tools. The tool is not built on the same theme as other tools aimed at, however, it aims at providing a social networking platform for scientists and researchers. It is more towards finding collaborations, sharing papers, asking and answering questions than performing research evaluation. Although, we believe that in near future new and advanced research evaluation methods will be used instead of the traditional ones. These methods could be based on social reputation of a researcher that the researcher might gain based on his/her social interaction in the form of valuable shares of scientific papers, datasets, experiments, and likewise answering peers' questions and the like. 

To the best of our knowledge, there are not so many other tools left that are built for the purpose of research evaluation for a broader audience. However, there are efforts within different communities and those only addresses the specific problems of a specific community. The lack of a general purpose, flexible, yet end-user oriented tool left a huge gap for the growing community of researchers, which is why complex research evaluation tasks still pose challenges for non-technical users and these challenges still have not been addressed yet by the existing solutions.  

\section{Analysis and Discussion} \label{sec:analysis}
This section presents a critical analysis of all aforementioned bibliographic indicators, data sources, and impact evaluation tools. We have presented different indicators that have been developed and used for the assessment purposes over the years. We also noticed that these indicators evolved over time, and scientific communities have adopted these indicators in one or the other way (e.g., a customized version of an indicator). However, we have not found any consensus on a commonly accepted indicators, and that proves the fact that the field of research impact evaluation is a diverse field, where everyone has its own interpretation of what an evaluation procedure should be. To further support the justification for this fact, in following we present studies that have been conducted and showed the same claim as we do.

In \citep{costas2007h}, authors analyzed the relationship of the well-known \emph{h-index} with other bibliometric indicators. Their analysis was based on a set of publications downloaded from the Web of Science(1994-2004) for Spanish CSIC scientists in Natural Resources, where the actual impact assessment conducted through the h-index. Their claim was to give more weight to those researchers who do not produce a high number of publications but who achieve a very significant impact. As the h-index considers both quantity and impact of publications, however, a researcher's maximum h-index value cannot exceed his publication count. They emphasized the use of diverse indicators for the better productivity assessment instead of just h-index, moreover they noticed that widespread use of a single index (e.g., h-index) might influence their publication behavior. Several other different bibliometric indicators have been analyzed to distinguish between researchers. For example in \citep{lehmann2008quantitative}, author analyzed h-index with other indicators using Bayesian statistics, in order to confirm which indicator performs better with respect to publication data. They concluded that, in order to achieve long term scientific productivity of a researcher, most indicators require minimum 50 publications as input.

It is widely accepted that some indicators show a strong bias towards some scientific fields. For instance, in case of h-index, when it is used to compare researchers from different fields tends to create problems, as also identified in a related study conducted by \citep{van2008testing}, where they analyzed the level of a researcher with the academic reward system in the Netherlands. They compared the h-index with other different bibliometric indicators in different fields. They concluded that comparing scientists from different fields using the h-index is not appropriate. Another interesting analysis has been conducted by \citep{costas2008g} among different types of scientists such as, low producers, big producers, selective scientists\footnote{Those researchers who do not produce a very high number of documents but who do attain a high impact} and top scientists in the Natural Resources field at Spanish CSIC. Their analysis was based on the g-index and h-index. They found that these indicators clearly distinguish between low producers and top scientists. However, in the case of selective scientists and big producers, these indicators do not perform well. Their results show that g-index is more sensitive than the h-index. Therefore, this research work shows that both indicators do not replace each other, and both have their own advantages and disadvantages. Another similar conclusion deduced in \citep{schreiber2008empirical}. They analyzed 26 practical cases of physicists from the Institute of Physics from Chemnitz University of Technology.

Some studies have been conducted regarding most criticized aspects of these indicators, which is the possible influence of self-citation. The inclusion of self-citation in the computation of citation-based indicators inflates the reflection of research impact of a scientist. In \citep{schreiber2007self}, author presented the results conducted on several bibliography datasets. They showed that self-citations do have an impact on the h-index, particularly in the case of young researchers. They proposed to discern self-citations while checking the impact. Mainly various scientific communities have a consensus on the exclusion of the self-citations before performing research evaluation tasks.

The correct usage of the indicators has been the primary concern of many studies and even in Hirsch's h-index proposal, he presented that the h-index, when applied to compare scientists from different communities is not appropriate. Factor such as normalization varies based on different fields, thus reference practices and traditions in different fields should also be considered.

A particularly interesting aspect in the computation of these indicators is the data sources used to fulfill data requirements. Until a few years ago there was essentially only a very few data sources available (e.g., ISI Web of Science, Scopus etc.) to compute various indicators. However, this number has increased during the recent years and now a number of different alternatives have become available as also presented in the section \ref{bib-dbs}. Some of these sources only cover single discipline, like Chemical Abstract produced by the American Chemical Society, MatchSciNet by American Mathematical Society etc. On the other hand,  a number of multidisciplinary data sources have emerged, like Google Scholar, Scopus, CiteSeer. These sources have been used in many studies and also for the scientific evaluation purposes as compared to discipline-oriented sources.

In a study \citep{sterne2009multiple}, author analyzed three main data sources (Google Scholar, Scopus and Web of Science). The study focussed on the analysis of pros and cons of these three largest, cited-reference-enhanced, multidisciplinary databases. They proposed that, some of the aspects to determine the h-index need scrutiny because they believe that content from reference databases can influence the h-index values due to problems such as completeness of data, the scope of data source and coverage. In another study \citep{meho2008citation}, authors examined the citation counts, ranking by citation and h-index values for top 22 researchers belongs to human-computer interaction (HCI) field. They used Scopus and Web of Science as data sources. Their results show that Scopus provides more coverage in this field as compared to Web of Science. They found significant differences in the value of the h-index, where Scopus performs much better which is near to the actual case.

In our literature review, we observed that the usage of bibliometric indicators in different perspective is highly subjective. We noticed that a number of studies showed their concerns about data sources problems in terms of completeness, coverage and their scope. A number of studies have been conducted regarding most sensitive issues about the use of proper indicators. Moreover, we also found that their usage is highly variable aspect across different scientific communities. Research executives, institutes, and communities have different assessment requirements hence it is hard to say that a single indicator would be truly effective. Some studies proposed to use one indicator, and on the other hand some propose to use its variation or they recommend using other indicators. Moreover, the issues related to the comparison of researchers, research groups and institutes those belong to different community have not been addressed yet and rely on a single indicator is not a recommended practice.

We have also noticed that, all the currently available tools lack, in our view, some key features, mainly: (1) completeness of data, (2) flexibility and personalization features (3) languages to support users' defined evaluation procedures, queries and metrics and (4) data processing features. The possibilities to define customized metrics is an essential feature in order to have a personalized access to the information, e.g., one might want to exclude self-citation from the h-index value of a researcher or see how an index could change excluding citations coming from the top co-authors \citep{parra2011scientific}. To this end, in this thesis, we propose an approach to tackle these challenges, which we believe mainly the reason that this field is highly diverse. Thus, providing ingredients to be used in research evaluation procedures will be more beneficial than to restrict users to a fixed set features. Moreover, the people responsible for performing these tasks often lack technical skills which is also a main setback for the current solutions as they do not aim at these non-technical users. To this end, in the next chapter we explore techniques that could enable these users to easily and effectively involve in such complex and technical tasks.

\newpage

% Chapter 3: First contribution of the thesis.
% !TeX root = ../phd-thesis-arxiv.tex

\chapter{End-user Development \& Mashups: State of the Art}
\label{ch:eud-soa}

\section{Overview}
By and large, in the current era, most people are familiar with the use of computers, at least with the basic functionalities and user-experience that computers provide. These computer users include engineers, teachers \citep{wiedenbeck2005facilitators}, doctors, salesmen, scientists \citep{segal2007some}, managers, and children \citep{petre2007children}. Based on a survey conducted by the U.S. Bureau of Labor and Statistics, Boehm et al. in his paper \citep{boehm1995cost}, predicted that in 2005 there would be 55 million such end-users (i.e., computer users using spreadsheets, databases, writing formulas, and queries for their daily work requirements). In another work \citep{scaffidi2005estimating}, which was also based on a survey conducted in 2005 by the U.S. Bureau of Labor and Statistics, reported that these end-users population already increased to 80 million. Moreover, in the same work, based on the rate of increase from 1995 to 2005, they also predicted that this number will be 90 million in 2012.

The nature of work that many of these users involved - vary - rapidly on the basis of months or even days. Thus, the requirements for more intuitive, easy-to-use and flexible enabling development environments increased as with the growth of end-users. Despite many efforts, it is still a challenging endeavor for the end-users to develop or modify applications that support and fulfill their goals. As this process requires considerable expertise in programming languages that these users lack. On the other hand, traditional requirement elicitation methods and computer programmers simply cannot anticipate and meet all of these requirements. %because of their limited domain knowledge and mostly traditional requirement elicitation methods cannot anticipate new requirements. 

End-user development (EUD) is a way to solve this problem. EUD helps to empower less skilled users in such a way that they can easily and effectively be involved in development processes so to develop and tailor applications by their own. More specifically, EUD provides different techniques, methods, and tools that allow users to easily cope with the new requirements within the boundaries of a particular user's expertise \citep{lieberman2006end}. Over the time, different EUD techniques emerged that target different classes of end-users having different expertises \citep{lieberman2001your}, \citep{burnett2001far}, \citep{little2007koala}, \citep{pane2002using}, \citep{repenning2006agentcubes}. 

In this chapter, we present state of the art methods that have been proposed in the field of end-user development and we also present an analysis of the major techniques, methods, and tools used for this purpose. We also discuss major paradigms those considered as a fundamental base for EUD. Moreover, this chapter introduces the newly emerging field of \emph{Mashups}, especially in the context of EUD along with various developments in mashups field that have been proposed. In the end we discuss on how mashups can be better choice for less-skilled users.

\section{End-user Development}

The term \emph{end-user} typically refers and uses for a user of computer applications. The user in this context considered a non-technical or less skilled and a non-programmer. The intentions of these users are to use the computer applications to fulfill their daily life work requirements. While the term \emph{end-user development} refers to, when an end-user, who is not an expert on conventional computer programming languages, writes computer programs using either declarative or imperative programming techniques\footnote{More details on declarative and imperative techniques will be presented in the next section}. Thus, end-user development, for these kinds of users (i.e., end-user), provides enabling techniques, method and tools that facilitate them to configure, tailor, modify or write new computer programs. Among various forms of end-user development, to name a few, include use of spreadsheets, writing database queries, configuring software programs, visual programming, use of Wikis etc.

Early efforts in the field of EUD were focused around the concepts like customization, parameterization of software programs and some other on tailoring and writing small scripts \citep{trigg1994implementation}, \citep{eagan2008buzz}. These enabling techniques allow end-users, for example, to write scripts in the form of macros for MS Word using Visual Basic syntax, or to perform complex computations or data processing with the help of spreadsheets (e.g., MS Excel), or configuring a software settings using different parameters (e.g., use of various graphical settings). With the passage of time and in parallel the increase in more complex users' requirements made some of these technologies (e.g., writing scripts or macros), due to their richer technical usage demands, off-track and out of non-programmers technical expertise domain and others (e.g., use of spreadsheets) become simply useless for performing non-trivial tasks.   

However, new ways emerged and among those, for instance, \emph{programming by example} also known as \emph{programming by demonstration}, to some extent, reduces the efforts a user needed to learn traditional programming abstractions \citep{lieberman2001your}. In this approach a computer program records the user's action and after generalizing those set of actions it performs the same actions (not necessarily exactly same) in some other similar situations. With the passage of time, the presence of the Internet, especially with the growth of newly emerged Web 2.0 technologies, made it possible to provide a common platform for everyone to produce and consume resources at any time, in any form and from anywhere. For example, among these resources, open data access, Web Services, Online APIs, feeds (i.e., RSS/ATOM feeds) are the most popular. Although the requirement for more intuitive development environments and design support for end-users clearly emerge from research on end-user development, for example for web services \citep{NamounICWEWS2010,NamounECOWS2010}, not many tools and frameworks are yet available to satisfy this need. From a conceptual point of view, there are currently two main approaches to enable less skilled users to develop programs, which are \emph{simplifying} development practices and enabling \emph{reusability}. That is, in general development can be eased either by \emph{simplifying} it (e.g., limiting the expressive power of a programming language) or by \emph{reusing} knowledge (e.g., copying and pasting from existing algorithms). 

Among the \emph{simplification} approaches, the workflow and Business Process Management (BPM) community was one of the first to propose that the abstraction of business processes into tasks and control flows would allow also less skilled users to define their own processes. Yet, according to our opinion, this approach achieved little success and modeling still requires training and knowledge. The advent of the \emph{Service-Oriented Architecture} (SOA) substituted tasks with services, yet the composition is still a challenging task even for expert developers \citep{NamounECOWS2010} \citep{NamounICWEWS2010}. The \emph{reuse} approach is implemented by program libraries, services, or templates (such as generics in Java or process templates in workflows). It provides building blocks that can be composed to achieve a goal, or the entire composition (the algorithm -– possibly made generic if templates are used), which may or may not suit a developer's needs. 

In recent years, several research projects such as Search Computing\footnote{http://www.search-computing.it/} \citep{ceri2010search}, mashArt \citep{daniel2009hosted}, FAST\footnote{http://fast-fp7project.morfeo-project.org} \citep{hoyer2009fast} and even our own old tool ResEval \citep{imran2010reseval} spent substantial effort towards empowering end-users ( as for some of these tools refer end-users sometimes as expert users, to distinguish them from generic, completely unskilled users), with tools and methods for software development. In the following we look at this field from a different perspective and we elaborate on which paradigms and ingredients best aid end-users in performing development tasks, and most notably formulating complex tasks. We also discuss various dimensions of end-user programming, including vertical versus horizontal language definition, declarative versus imperative approaches.  

\section{Enabling Practices and Techniques}

Enabling end-users to develop own applications or compose application programs by combing together the different pieces available online in the form of public web services, APIs or data in various forms, requires simplifying current end-user development practices. To this end, a variety of approaches may help simplify the end-user development, as also discussed a few of these approaches in the previous section. However, in this section we discuss in detail the most important ones, in order to use them in the next section to analyze these approaches that partly aim at supporting end-users for composing complex applications.

\subsection{Simple Programming Models} 
The first issue is to understand which programming paradigms are best suited for end-user programming. The solution to this issue can take inspiration from existing experiences in the orchestration and mashup languages which are targeted at process automation and at relatively inexperienced users. Although they have not been that successful in reaching out to non-IT experts, as yet. The aim is to find programming abstractions that are simple enough to appeal to domain experts and at the same time complex enough to implement enterprise procedures and Web application logic. 

For instance, some mashup approaches heavily rely on connections between components, which is for instance, the case of Yahoo! Pipes and IBM Damia \citep{altinel2007damia}, and therefore are inherently imperative; other solutions completely disregard this aspect and only focus on the components and their pre- and post-conditions for automatically matching them, according to a declarative philosophy like the one adopted in choreographies. For instance, as also stated in the FAST European project \citep{hoyer2009fast}.

\subsection{Domain-specific Modeling.} 
The idea of focusing on a particular domain and exploiting its specificities to create more effective and simpler development environments is supported by a large number of research works \citep{ledeczi2001} \citep{costabile2004} \citep{mernik2005} \citep{france2005}. Mainly these areas are related to Domain Specific Modeling (DSM) and Domain Specific Language (DSL).% and Domain Driven Design (DDD).

In DSM, domain concepts, rules, and semantics are represented by one or more models, which are then translated into executable code. Managing these models can be a complex task that is typically suited only to programmers but that, however, increases users' productivity. This is possible thanks to the provision of domain-specific programming instruments that abstract from low-level programming details and powerful code generators that "implement" on behalf of the modeler. Studies using different DSM tools (e.g., the commercial MetaEdit+ tool and academic solution MIC \citep{ledeczi2001}) have shown that developers' productivity can be increased up to an order of magnitude.

\subsection{Domain-specific Languages (DSLs)} 
Simple programming models are not enough. Typically, end-users simply do not understand what they can do with a given development tool, a problem that is basically due to the fact that the development tools does not speak the language of the user and, hence, programming constructs do not have any meaning to the user. Domain-specific languages aim at adding domain terminology to the programming model, in order to give constructs domain meaning.

In the DSL context, although we can find solutions targeting end-users (e.g., Excel macros) and medium skilled users (e.g., MatLab), most of the current DSLs target expert developers (e.g., Swashup \citep{maximilien2007}). Also here the introduction of the "domain" raises the abstraction level, but the typical textual nature of these languages makes them less intuitive and harder to manage and less suitable for end-users compared to visual approaches. A number of benefits and limits of the DSM and DSL approaches are summarized in \citep{france2005} and \citep{mernik2005}.

In some fields, such as database design, domain-specific languages are a consolidated practice: declarative visual languages like the ER model are well accepted in the field. Other, more imperative approaches, like WebML \citep{ceri2000web}, address developers that are willing to embrace conceptual modeling. Business people, on the other hand, are well aware of workflow modeling practices and are able to work with formalisms like BPMN, completely ignoring what happens behind the scenes both in terms of technological platform and of transformations applied to get to a running application. Another example in this category is Taverna \citep{kuhn2010cdk}, a workflow management system well known in the biosciences field. As DSL approach is more closely related to our proposed solution so we present a more precise classification of DSLs in Section \ref{sec:dsl-assessmnet}.

\subsection{Web Service Composition.}
BPEL (Business Process Execution Language) \citep{bpel20} is currently one of the most used solutions for web service composition, and it is supported by many commercial and free tools. BPEL provides powerful features addressing service composition and orchestration but no support is provided for UI integration. This shortcoming is partly addressed by the BPEL4People \citep{bpel4people} and WS-HumanTask \citep{wshumantask} specifications, which aim at introducing also human actors into service compositions. Yet, the specifications focus on the coordination logic only and do not support the design of the UIs for task execution. In the MarcoFlow project \citep{DanielBPM10}, they provide a solution that bridges the gap between service and UI integration, but the approach, however, is still complex and only suited for expert programmers.

\subsection{Intuitive Interaction Paradigms} 
The user interfaces of development tools may not be a complex theoretical issue, but acceptance of programming paradigms can be highly influenced by this aspect too. The user interface comprises, for instance, the selection of the right graphical or textual development metaphor so as to provide users with intelligible constructs and instruments. It is worth investigating and abstracting the different kinds of actions and interactions the user can have with a development environment (e.g., selecting a component, writing an instruction, connecting two components), to then identify the best mix of interactions that should be provided to the developer. 

\subsection{Reuse of Development Knowledge} 

Finally, even if a tool speaks the language of the user, it may still happen that the user does not speak the language of the tool, meaning that he/she still lacks the necessary basic development knowledge in order to use the tool profitably. Such a problem is typically solved by asking more expert users (e.g., colleagues or developers) for help – if such is available. The challenge is how to reuse or support the reuse of development knowledge from more expert users in an automated fashion inside a tool, e.g., via recommendations of knowledge \citep{roy2011wisdom}. 

Recommendations can be provided based on several kinds of information, including components, program specifications, program execution data, test cases, simulation data, and possibly mockup versions of components and program fragments used for rapid prototyping. Information may or may not be tagged with semantic annotations. When present, the annotations can be used to provide better/more accurate measures of similarity and relevance. In a general sense, the approach we envision is an alternative to design patterns for exploiting the expertise of good developers, thus allowing reuse of significant designs.

Programming, testing, and prototyping experiences of peers or of more experienced developers may support the entire development lifecycle. If knowledge is harvested and summarized from peers (e.g., by analyzing their mashup definitions), this opens the door to what we can call "implicit collaborative programming" or "crowd programming", where users, while going through a software engineering lifecycle for implementing procedures of their own interest, create knowledge that can be shared and leveraged by other domain experts for their own work.

\section{Domain-Specific Languages: Discussion}
\label{sec:dsl-assessmnet}
We have seen that Domain-Specific Languages (DSLs), i.e., design and/or development languages that are designed to address the needs of a specific application domain, are important to provide the end-user with familiar concepts, terminology and metaphors. That is, DSLs are particularly useful because they are tailored to the requirements of the domain, both in terms of semantics and expressive power (and thus do not enforce end-users to study more comprehensive general-purpose languages) and of notation and syntax (and thus provide appropriate abstractions and primitives based on the domain). In following we highlight a few possible classifications of these languages, which can become handy for EUD. In particular, we describe the dimensions of focus, style and notation. 

The focus of a DSL can be either vertical or horizontal. Vertical DSLs aim at a specific industry or field. Examples of vertical DSLs may include: configuration languages for home automation systems, modeling languages for biological experiments, analysis languages for financial applications, and so on. On the other side, horizontal DSLs have a broader applicability and their technical and broad nature allows for concepts that apply across a large group of applications. Examples of horizontal DSLs include SQL, Flex , WebML , and many others.

The style of a DSL can be either declarative or imperative. Declarative DSLs adopt a specification paradigm that expresses the logic of a computation without describing its control flow. In other words, the language defines what the program should accomplish, rather than describing how to accomplish it. Imperative DSLs instead specifically require defining an executable algorithm that states the steps and control flow that needs to be followed to successfully complete a job.

The notation of a DSL can be either graphical or textual. The graphical DSLs (also known as Domain Specific Modeling Languages, DSML) imply that the outcomes of the development are visual models and the development primitives are graphical items such as blocks, arrows and edges, containers, symbols, and so on. The textual DSLs comprise several categories, including XML-based notations, structured text notations, textual configuration files, and so on.

Despite the various experiences in DSL design and application, there is no general assessment on the preferences of the developers for one or the other kind of language depending on the user profile. However, typically languages oriented to the end-users tend to be more visual and declarative, while the ones for developers are often textual and imperative.

\section{Mashups from an End-User Development Prospective}
\subsection{Web 2.0 \& Enabling Technologies}
During the last decade, the advent of Web 2.0 has been drastically and successfully proved as an enabling environment for normal web users to enable them to involve into the creation and consumption of Web resources of various types, like blogs, Wikis, Social Media etc. In respect to Web 1.0 which was known as "web as information source", web 2.0 is called "web as participation platform". Of the major key features of Web 2.0 from EUD point of view include "rich user experience", "user as a contributor", and "user participation". 

Among relevant Web 2.0 technologies, the Service-Oriented Architecture (SOA) field emerged as a paradigm in software development. The emerging visions of an Internet of Services (IoS) and a Web Service Ecosystems \citep{barros2005move} \citep{barros2006rise} supported SOA and have shown much potential in the field. However, the major focus of these technologies remained on the technical level of a service to service based interactions systems \citep{schroth2007brave} and a little on service to user (i.e., non-programmer) communication. Due to the high technical complexity of the relevant standards (e.g., WSDL, SOAP, UDDI, REST), we think that doing so is even harder than enabling non-programmers to model an own process or service composition, because developing full applications is simply complex as it require a lot of programming knowledge to deal with data, application and presentation issues

In parallel to Web 2.0 technologies evolution, the Web mashup \citep{YuIC08} phenomenon emerged, which provided easier ways to glue these services and data together \citep{hartmann2006hacking} and claiming to enable also non-programmers to use and mash pre-built components that provide an abstraction of complex programming concepts. Before investigating further on mashups and to set the context, lets just introduce the terminologies that are mostly used. Typically the term \emph{mashup} refers to those web applications that, rather than being developed from scratch, are developed using various available data, functionalities or user interfaces over the Web. While, \emph{Mashup tools}, provide development and runtime environments for the composition and execution of applications (i.e., mashup applications) to non-programmers to enable them to create their own situational applications \citep{imran2012systematic}.

Based on the Web 2.0 philosophy, a new type of mashup-based approach emerged, which is known as Enterprise Mashups \citep{hoyer2008enterprise}. As it is more adapted and evolved in large companies where more rapid requirements require employees to be dealt with more sophisticated information technologies. Within an organization the key components of Enterprise Mashups include "resources", "widgets" and "mashups", that deal with data (i.e., actual content), application logic (i.e., implementing actual business logic) and mashup application (i.e., assembling together a collection of widgets) respectively. In \citep{abiteboul_modeling_2008} author introduces mashup concepts and present a mashup model for syntactically composing mashups. In this model, a mashup is defined as a network of \emph{mashlets}. These mashlets are the main mashup components and consist of a set of relations, e.g. internal relations, I/O relations and web service relations. They can be GUI-based and can be organized in a hierarchical way, i.e., complex mashlets can contain simpler ones. In this model mashlets are defined by means of rules that state which the input, the output and the possible services calls are. The authors also explain the necessity of allowing the user to query and update the data dynamically in the mashup, as well as to add, update or remove mashlets at run time.

As mashups aim to bring together the benefits of both \emph{simplification} and \emph{component reuse}. We believe that, in order to make application development from programmers-centric to end-user centric, we need to achieve simplicity from both ends (i.e., from the technology as well as from end-user ends). While the component-based reuse approach is certainly lowering part of the complexity, developing an own application, however, also means dealing with data integration, application logic, and content presentation issues, all aspects the common web user is not even aware of.  However, in the case of domain-specific mashup environments, as also in our case, we aim to push simplification even further compared to generic mashup platforms by limiting the environment (and, hence, its expressive power) to the needs of a single, well-defined domain only. Reuse is supported in the form of reusable domain activities, which can be mashed up.

\subsection{Tool-Assisted Mashup Development}
In this section we present and review a number of representative mashup tools, and evaluate them based on those main aspects we consider fundamental for addressing real-life end-user needs. Of the main assessment aspects, the support for the integration of data, services and user interface is fundamental. This functionality is known as \emph{universal integration}. Moreover, we also present our analysis based on the requirements we gathered during our domain analysis, those best suited for end-users, which are \emph{intuitiveness of UI, modeling constructs, execution paradigm}, and data-mappings.

%\begin{figure}
%  \centering
%    \includegraphics[width=\columnwidth]{figs/yahoopipes-ss.pdf}
%  \caption{Yahoo Pipes! Composition Editor showing a compistion pipe}
%  \label{fig:yahoopipe}
%%\vspace{-3mm}
%\end{figure} 
%\smallskip
\spara{Yahoo Pipes!}\footnote{\url{http://pipes.yahoo.com}} is a well-known mashup tool by Yahoo. It provides a number of built-in components and a visual composition editor that allows to design data processing logics. The Yahoo Pipes composition editor offers a set of components and works in a drag-drop fashion, where a user can drop, connect and configure components. Yahoo pipes is quite an attractive with its composition environment, which allows web users to make data centric compositions. The components follow data-flow based approach, as each component waits until data becomes available at its input port. The data-flow based approach is more intuitive for an end-user as compared to a control-flow approach as long as it stays trivial. As in case of Yahoo pipes, it mainly offers a very technical set of components (i.e., modeling constructs) like loops, regular expression, URL builder, RSS feeds etc. that makes composition task complex for a non-technical users. A non-technical user by no means can understand these components and consequently unable to make compositions. Moreover, these programming-related components, which require the basic expertise of programming concepts, may have multiple input and configuration parameters through which connection between two components take place. In essence, the complex data mappings have to be performed to compose a valid mashup. Moreover, Yahoo pipes does not support UI integration, and support for service integration is still poor and is out of an end-user technical reach.  

Likewise, instead of domain-specific, the generic nature of the components that Yahoo Pipes offers are only understandable by programmers. However, we believe that a domain expert (i.e., a non-IT user) is still not able to get fruitful results from this tool. Because, one of the main reasons is that it restricts domain experts by not offering those domain constructs and terminologies they are familiar with. It is almost as generic as only understandable by IT experts as it exposes programming notations.

%\begin{figure}
%  \centering
%    \includegraphics[width=\columnwidth]{figs/Popfly_Mashup.pdf}
%  \caption{Microsoft Popfly: Composition Editor}
%  \label{fig:mspopfly}
%%\vspace{-3mm}
%\end{figure} 

\smallskip
\spara{Microsoft Popfly}\footnote{http://www.popfly.ms} Among the other popular mashup-based tools, Microsoft introduced Microsoft Popfly. This mashup tool targets universal integration (i.e., data, application and UI). Of the other tools (e.g., game creator, web creator) it offers mashup creator, a tool that offers pre-built components and let users to mash them together to make applications, but also in this case the end-users were generally not able to develop real-life applications. Popfly has been discontinued from August 2009 onwards. 

\smallskip
\spara{Intel Mash Maker} \citep{ennals2007intel} provides a different kind of mashup approach which mainly focuses on data (i.e., online content) of a user's interest. Intel mash maker is a Firefox extension that runs on the client-side browser and adds a toolbar to the browser with a set of buttons representing various functionalities. It basically monitors the user's behavior that checks what information a user visit or is interested in and automatically builds a mashup application that could be of interest to the user, even when the user was not aiming at building a mashup application. The tool mainly extracts relevant data from Web pages but does not provide any data integration functionality. Moreover, no UI or data presentation features are provided, likewise it does not allow service composition. The proper use of the tool, especially use of the advanced features, requires programming skills that non-technical end-users lack.

\smallskip
\spara{mashArt} Among the academic projects a noticeable example is mashArt \citep{daniel2009} project. The tool mainly aimed at a universal integration approach for UI components and end-user centric development. Aiming at these objectives, mashArt comes with models and languages able to accommodate all the three types of needed components (i.e., data, services and UIs) and with a simple web-based editor and an integrated lightweight runtime environment (allowing for instantaneous previewing) targeted at non IT-expert –skilled web users. Although mashArt achieved universal integration, yet it is not able to effectively target end-users. The tool does not solve the problem of complex data mappings, and also mainly the components that mashArt provides are of generic types.

\smallskip
\spara{JackBe Presto} 
JackBe Presto\footnote{http://www.jackbe.com} is one of the popular commercial products. The Presto suite is constituted by several distinct tools. One constitutes the composition development environment, Presto Wires, which adopts a Pipes-like approach for mashing up data from enterprise internal and external sources. It also allows a portal-like aggregation of UI widgets (so-called mashlets developed through the Presto Mashlet tool) visualizing the output of such mashups on a dashboard. Each mashlet is independent from the others, thus, and the synchronization at presentation level is limited. This enterprise solution focuses on integrating enterprise internal or external data and on visualizing them in the form of widgets. The portal-like approach, in general, provides a satisfying level of usability for end-users. However, universal integration is not actually achieved.

\smallskip
\spara{Taverna}
is a mashup like application, which allows the integration of the existing data sources (i.e, molecular biology sources) available on the Web \citep{hull2006taverna, sroka2010formal}. The tool allows users to design, execute and share workflows made-up using web services in the domain of molecular biology and bio-informatics. Components can be added and connected visually in a drag drop way and different kinds of services can be added to the service panel of the tool. Because mashups are intended to integrate data from one or more sources, the previous version of Taverna \citep{hull2006taverna} cannot be considered as a mashup tool, since it only focused on the integration of services, but the current version, named Taverna 2 \citep{sroka2010formal}, provides support for data streaming through pipelining and so data-driven workflow computation can be performed. Despite many claims, although the tool focuses on a particular domain, even then it is not suitable for the non-technical users because of its complexities related to the web service usage, and complex data mapping etc.

\smallskip
\smallskip
\noindent Similarly, the \emphbf{CRUISe project} \citep{pietschmann2009} specifically focuses on composability and context-aware presentation of UIs, but does not support the seamless integration of UI components with web services.The \emphbf{ServFace project} \footnote{\url{http://www.servface.eu}}, instead, aims to support normal web users in composing semantically annotated web services. The result is a simple, user-driven web service orchestration tool, but UI integration and process logic definitions are rather limited and again basic programming knowledge is still required.

Although a number of other mashup-related tools and platforms exist (e.g., Deri Pipes\footnote{\url{ http://pipes.deri.org/}}, , Dapper\footnote{\url{http://dapper.net/}}, to name a few), they all show similar limitations as of the others presented solutions (i.e., lack of universal integration support and/or simplicity of use for non-technical users, complex data mappings and so on). Our analysis on current the mashup initiatives highlights that none of the proposed solutions is able to successfully empower end-users to develop the applications actually supporting their daily activities. This is mainly due to the fact that, although through intuitive visual metaphors, most of them still expose programming concepts which, according to \citep{deangeli2010}, have semantics that end-users do not understand and do not want to learn. In the following section we summarize and present of analysis of all of these directions.

\section{Analysis and Discussion}

End user development comprises several alternative approaches, spanning from mashup development, to software configuration, to simple programming tasks. These approaches are often authentic, but sometimes they can be combined together to exploit the respective strength points \citep{bozzon2011development}. For instance, while users are getting more and more used to configure applications, also thanks to the pervasiveness of mobile and gaming software, mashup platforms for the development of simple Web applications are also gaining popularity. 

Yet, mashups were actually born as a hacking phenomenon, where very expert developers build applications by integrating reusable content and functionality sourced from the Web, for instance, see programmableweb\footnote{www.programmableweb.com}, and – despite the numerous attempts – mashup development is still for skilled programmers only. For instance, a very popular mashup tool  Yahoo Pipes! (as mentioned in the previous section) provides a mashup environment with a variety of components. These components wrapped very generic programming features thus providing a set of high-level functionalities such as loops, if-conditions, parameter passing, web-service binding etc. These high-level functionalities neither understandable by end-users nor used in their daily life application development purposes. 

Actually, mashup tools initially targeting end-users slowly moved towards the expert user, then to the developer, and finally to the expert developer. To this end, in fact, both model-driven web engineering \citep{ceri2003morgan} and mashup development \citep{daniel2009hosted} has shown that there are basically only two users classes in the real world. The first class represents \emphbf{developers}, who want to see the source code and to write imperative code by their own. These users do not trust model-driven approaches, because they feel this can reduce their freedom in application development. The second class represents \emphbf{non-developers}, who want to ignore all the technical issues and have simple, possibly visual or parameter-based configuration environments for setting up their applications.

A possible stratification of users into "developer" class could be \emph{expert users, entry-level developers, developer/designer} that can be theoretically defined does actually not exist. Recognizing the distinction of only two major user classes, empowering non-developers become more focused and challenging, yet non-trivial. As presented in this chapter that many approaches have been proposed to help these users develop their own applications, and we see largely they failed to do so. Among the reasons non-technical users found these enabling solutions difficult for their practical use is the language they speak, which is what constructs, concepts, modeling paradigm they use, is not understandable by the users. To provide \emph{non-developers}, which is our target user class, an end-user development platform whose main theme to speak the language of a user. That means, we present a domain-specific approach that leverage mashup strengths to offer an intuitive, easy-to-use yet flexible end-user development platform that would ultimately speak a user's language by incorporating domain concepts, terminologies, rules, and syntax a user is familiar with.

\newpage

%Example scenario
% !TeX root = ../phd-thesis-arxiv.tex

\chapter{Research Evaluation Example Scenarios and Requirements Understanding}
\label{sec:scenarios}

\section{Overview}
To obtain important conceptual as well as those low level details of a domain that can never be considered and incorporate without thorough analysis, we first present a few real evaluation procedures related to the domain of \emph{research evaluation}. For our selected domain, we asked and gather different evaluation procedures from different domain-experts working in different departments in our and other Universities. The domain-experts who perform or were involved in these evaluation tasks include professors, PostDoc, administrative personnel and also PhD students. They were involved in some kind of research evaluation tasks ranging from simple tasks to complex ones. The obtained procedures helped us to examine the domain thoroughly and so to extract domain as well as users' requirements. In the following sections we state these evaluation procedures , their relevant details to better understand the problems, requirements, and associated important concepts. In the end we also present a set of general requirements those extracted from the analysis of all the procedures.

\section{University of Trento Department Evaluation Procedure}
\label{sec:unitn-scenario}

As an example of a domain-specific application scenario, let us describe the evaluation procedure used by the central administration of the University of Trento (UniTN) for checking the productivity of each researcher who belongs to a particular department. The evaluation is used to allocate resources and research funds to the university departments. In essence, the algorithm compares the quality of the scientific production of each researcher in a given department of UniTN with respect to the average quality of researchers belonging to similar departments (i.e., departments in the same disciplinary sector) in all Italian universities. Impact measure of each researcher then collectively contributed to their particular department. The comparison uses the following procedure based on one simple bibliometric indicator, i.e., a weighted publication count metric. 

\begin{enumerate}

\item{A list of all researchers working in the selected department as well as in the Italian universities is retrieved from a national registry, and a reference sample of faculty members with similar statistical features (e.g., belonging to the same \emph{disciplinary sector}) of the evaluated department is compiled}. 

\item{Publications for each researcher of the selected department and for all Italian researchers in the selected sample are extracted from an agreed-on data source (e.g., Microsoft Academic, Scopus, DBLP, etc.)}.

\item {The publication list obtained in the previous step is then weighted using a venue classification. That is, the publications are classified by an internal committee in three categories, which represent quality of a particular venue, mainly based on ISI Journal Impact Factor: A/1.0 (top), B/0.6 (average), C/0.3 (low). For each researcher a single weighted publication count parameter is thus obtained with a weighted sum of his/her publications.}

\item {The statistical distribution -- more specifically, a negative binomial distribution -- of the weighted publication count metric is then computed out of the Italian researchers' reference sample}.

\item {Each researcher in the selected department is then ranked based on his/her weighted publication count by comparing this value with the statistical distribution. That is, for each researcher the respective percentile (e.g., top 10\%) in the distribution of the researchers in the same disciplinary sector is computed.}

\end{enumerate}

\begin{figure}[t]
 \centering
   \includegraphics[width=1\columnwidth]{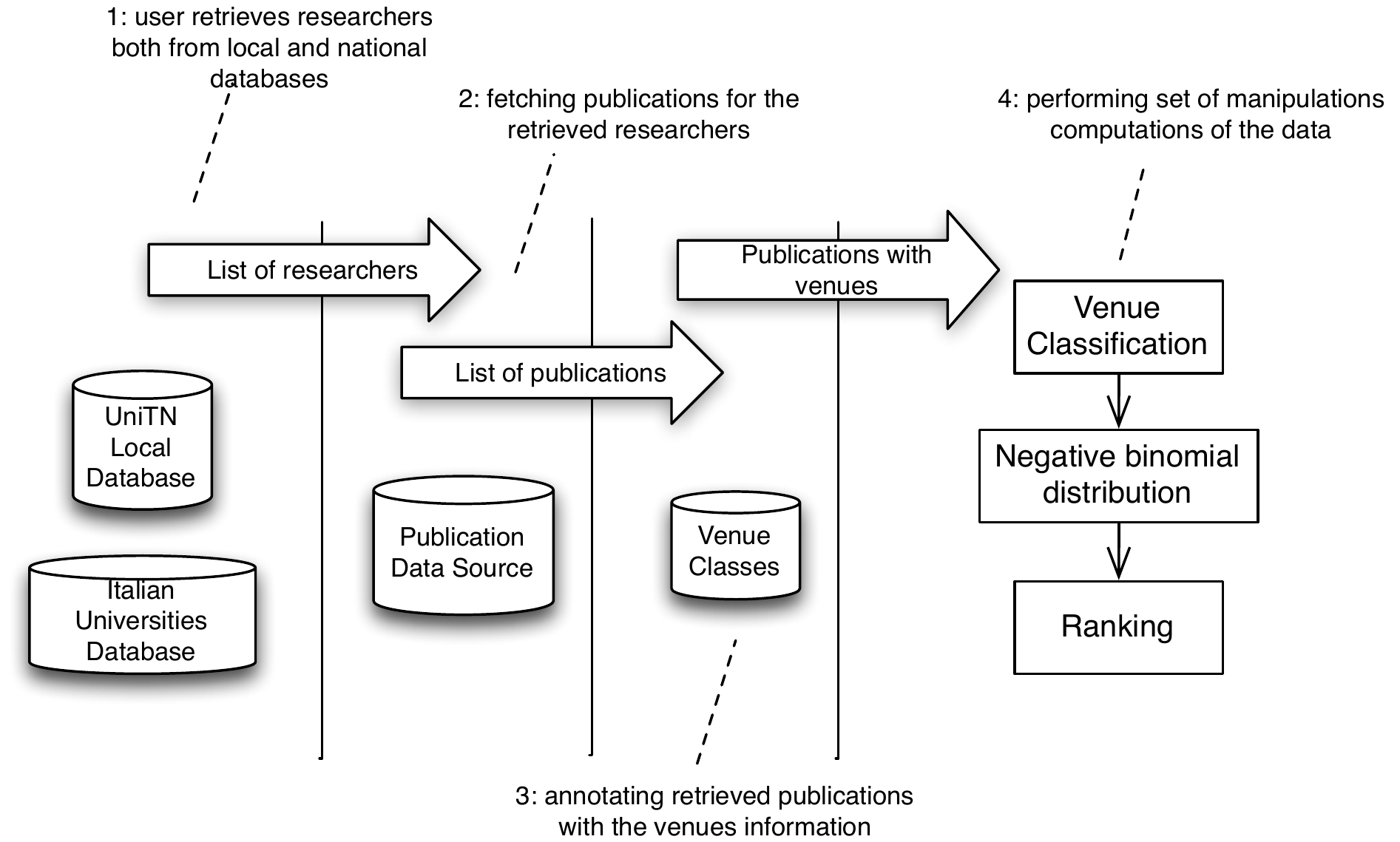}
 \caption{University of Trento department evaluation procedure, depicting steps a user performs manually}
 \label{fig:unitn-eval-flow}
\end{figure}

In Figure \ref{fig:unitn-eval-flow} we illustrate the steps a user has to perform to complete the described evaluation task. As it is shown in the Figure (step-1), the process starts from fetching researchers from UniTN local repository, and then also fetching list of all the researchers those belong to all Italian universities from a national repository (i.e., a web site, which provides data in excel format) those belong to the same discipline as of the UniTN discipline. In step 2, user has to retrieve publications from a data source for both UniTN and Italian researchers, a task that is beyond an effort a human can perform. Next, these publications must be annotated with the venue classification defined by the University management. That means, each publication is assigned a weight depending what venue it belongs to. These annotated publications are then used to compute the statistical distribution (i.e., negative binomial distribution) and then ranked based on the percentile accordingly. Finally, the results have to be presented in some visual format (e.g., charts, graphs etc.).

The percentile for each researcher in the selected department is considered as an estimation of the publishing profile of that researcher and is used for comparison with other researchers in the same department. As one can notice, plenty of effort is required to compute the performance of each researcher, which is currently mainly done manually. Fervid discussion on the suitability of the selected criteria often arises, as people would like to understand how the results would differ changing the publications ranking, the source of the bibliometric information, or the criteria of the reference sample. Indeed all these factors have a big impact on the final result and have been locally at the center of a heated debate. Many researchers would like to use different metrics, like citation-based metrics (e.g., h-index). Yet, computing different metrics and variations thereof is a complex task that costs considerable human resources and time and thus beyond human capacity.

\section{Italian Professorship Selection Scenario}
This section presents an evaluation procedure, which was adopted by the National Agency for the Evaluation of Universities and Research Institutes (ANVUR) in 2012 for hiring and promoting professors. The actual procedure is written in Italian language, however, in following we present the English translation.

According to the original evaluation procedure document, it states that based on the regulations for national scientific qualification establish that some of the indicators/indexes, when used for candidates for the national scientific qualification, should be normalized according to the academic age (i.e., the number of years starting from the first publication of a researcher) of the candidate. The normalization criteria varies based on a particular type of research output. In following we describe the normalization procedure in detail.

\begin{enumerate}
\item The number of articles in magazines/journals present in the major international databases and published in the consecutive 10 years previous to the date of publication of the decree (i.e., another regulation), normalization must be performed only if academic age is $<$ 10 years, and will be performed multiplying number of articles by 10, and dividing by academic age. 

\item The total number of received citations related to the whole scientific production, normalization should be performed dividing the number of citations by academic age. 

\item The number of books with ISBN published in the consecutive 10 years previous to the date of publication of the decree, the normalization must be done only if academic age is $<$10 years, and is performed multiplying the number of articles by 10, and dividing by the academic age. 

\item The articles in magazines/journal and chapters in books with ISBN published in the consecutive 10 years previous to the date of publication of the decree, only if academic age is $<$10 years, normalization is done multiplying number of articles in magazines and chapters of books by 10 and dividing by the academic age. 

\item The number of articles in magazines/journals that belong to "class A"  published in the consecutive 10 years previous to the date of publication of the decree, normalization must be done only if academic age is $<$10 years, and will be performed multiplying the number of articles by 10, and dividing by the academic age. 

\end{enumerate}

In addition to the normalizations that described above, the procedure also uses a customized version of the \emph{h-index}. The customized version is called \emph{contemporary h-index}. The \emph{ch-index} is different from the $h-index$, that is, it uses normalized citations of the normalized papers those selected for evaluation. The \emph{ch-index} is defined using following formula.

$ S (i, t) = \frac {4}{(t - t_1 + 1)} C (i, t) $ for  $  t \geq t_1 $

where the value of $C(i,t)$ is the number of citations observed in the database at time $t$ for i-th article. $ t_1$ is the year of publication of the article. Thus $S(i, t)$ is the value of citation indicator for the i-th article at time $t$.

The normalized results of all the indicators are then used to compare the threshold values that ANVUR has selected as the research quality threshold for a specific research area. The intention is that a candidate performing well above the defined thresholds will then be considered for hiring or promotion. 

\section{Analysis and Domain-Specific Requirements}
\label{sec:it-prof.scenario}
If we carefully look at the described scenarios, we see that these are all of \emph{domain-specific} type, i.e., these are entirely based on concepts that are typical of the research evaluation domain. For instance, the described evaluation procedures process domain objects (researchers, publications, metrics, and so on), use domain-specific computation logic, specific data sources (i.e., localized venue classification), customized evaluation metrics (i.e., \emph{ch-index}), and likewise these procedures use a set of domain-specific normalization rules. Despite many research evaluation approaches and tools those are made for evaluation purposes, as presented in the chapter \ref{sec-stateoftheart}, could hardly anticipate these domain-specific requirements thus resultantly failed to facilitate end-users. 

For example, the requirement we extract from these scenarios are that we need to empower people involved in the evaluation process (i.e., non-programmers, the average faculty member or the administrative persons in charge of it) so that they can be able to define and compose relatively complex evaluation processes, taking and processing data in various ways from different sources, and visually analyze the results. A tool having such provisions should allow to extract, combine, and process data and services from multiple sources, and to integrate these ingredients as user-defined way, finally representing the information in visual components. These are all the characteristics that a mashup can have, especially if the mashup logic comes from the users.

In order to enable the development of an application for the described evaluation procedures, there is no need for a composition or a mashup environment that supports as many composition technologies or options as possible. The intuition we elaborate is that, instead, a much more limited environment that supports exactly the basic tasks described in the scenarios (e.g., fetch the set of Italian researchers) and allows its users to mash them up in an as easy as possible way (e.g., without having to care about how to transfer data between components) will be more effective. However, to this end, the challenge lies in finding the right trade-off between flexibility and simplicity. The former, for example, pushes toward a large number of basic components, the latter towards a small number of components. As we will see, it is the nature of the specific domain that tells us where to stop.

To convey better understandings, in the following sections, we will therefore show how the development of a mashup tool that capable to run these example scenarios can be aided by being domain-specific. Moreover, based on the type of people involved who perform these evaluation tasks, we learned a number of requirements that we present in following. Turning the previous consideration into practice, the development of this tool will be driven by the following key principles:

\subsection{End-user centric requirements}
\label{sec:dsm-req}
\begin{enumerate}

\item \emphbf{Intuitive user interface} Enabling domain experts to develop their own research evaluation metrics, i.e., mashups, requires an intuitive and easy-to-use user interface (UI) both in terms of a tool's overall user experience as well as the modeling metaphors used for building mashups based compositions. For example, starting from the very first step, that is when users choose some components, those themselves be visually understandable for the users. %components should be visually      based on the concepts and terminology that the target domain expert is acquainted with. Research evaluation, for instance, speaks about metrics, researchers, publications, etc.

\item \emphbf{Intuitive modeling constructs} Next to the look and feel of the platform, it is important that the functionalities provided through the platform (i.e., the building blocks in the composition design environment) resemble the common practice of the domain. For instance, we need to be able to compute metrics, to group people and publications, and so on.

\item \emphbf{No data mapping} Our experience with prior mashup platforms, i.e., mashArt \citep{DanielER09} and MarcoFlow \citep{DanielBPM10}, has shown that data mappings are one of the least intuitive tasks in composition environments and that non-programmers are typically not able to correctly specify them. We therefore aim to develop a mashup platform that is able to work without the definition of data mappings.

\item \emphbf{Intuitive execution paradigm} When it comes to the question about how mashup tools, during run-time, exchange and flow data between components, end-users feel unattended about what is happening behind the scene. However, we aim to follow a data flow paradigm that end-users are familiar with in their daily life work. Moreover, we aim to reflect the execution states so that they become aware of what is being processed and how.

\end {enumerate}

The state of the art analysis about end-user development and mashup presented in the chapter \ref{ch:eud-soa} show that service composition, business process management (BPM), and mashup tools fail in providing end-users with intelligible concepts and constructs. That is because of various reasons like, complex user-experience, complex modeling constructs (i.e., components), complex data mappings and so on. Moreover, we will see that the naive approach of simply equipping a mashup tool with a set of domain-specific components is not enough, in order to obtain a tool that can be called \emph{domain-specific} and that can be amenable to end-users require a comprehensive analysis of domains to be considered along with a proper methodology that we present in the next chapter.
\newpage

% Chapter 4: Second contribution of the thesis.
% !TeX root = ../phd-thesis-arxiv.tex
%%% Chapter start. %%%
\chapter{End-User Oriented Mashup Platform Development Methodology}
\label{chp-method}
\section{Overview}
In the previous chapter, we presented a few real research evaluation scenarios, their analysis and the requirements that must be addressed for the practical success of a mashup tool. In chapter \ref{ch:eud-soa}, we presented in detail various aspects related to the mashups and end-user development.  We reported on the well-known approaches and also analyzed that these approaches, to a large extent, failed to facilitate end-users for their daily life development needs. We mainly identified that the generic nature of these approaches restricted end-users to comfortably adapt them. The reason behind is the interaction gap between the two sides (i.e., the end-user and technology). An end-user (i.e., a domain expert) lives and knows better within his domain of expertise, whereas, demands for more technical interaction kept increasing that certainly keeping apart both ends.

However, in this chapter we present our proposed \emphbf{methodology} for the development of mashup based tools that can lower the barriers for end-users by providing them a tool that speaks their language. For this reason, throughout this chapter we show how we have developed a mashup platform for our reference domain, in order to illustrate how its development can tackle the challenges systematically mentioned in the previous chapters. The development of the platform has allowed us to conceptualize the necessary tasks and ingredients and to structure them into a \emphbf{methodology} for the development of domain-specific mashup platforms. The methodology encodes a top-down approach, which starts from the analysis of the target domain and ends with the implementation of the specifically tailored mashup platform. In the next section, we first start from the essential concepts and definitions which are required to be defined before we proceed to the domain analysis step. %In following sections we state and define all the ingredients for developing a domain-specific mashup platform. %Specifically, developing a domain-specific mashup platform requires:

\section{Concepts \& Definitions}
\label{sec:concepts}

Before going into the details, we introduce the necessary concepts. First of all, leveraging from the interpretation of web mashups \citep{YuIC08}:

\begin{definition}\label{def:mashup}
A \emphbf{web mashup} (or \emph{mashup}) is a web application that integrates data, application logic, and/or user interfaces (UIs) sourced from the Web. Typically, a mashup integrates and orchestrates two or more elements.
\end{definition}

Most of the scenarios mentioned in chapter \ref{sec:scenarios} require all three ingredients listed in the definition: we need to fetch researchers and publication information from various Web-accessible sources (the data); we need to compute indicators and rankings (the application logic); and we need to render the output to the user for inspection (the UI). We generically refer to the services or applications implementing these features as \emph{components}. Components must be put into communication, in order to support the described evaluation algorithm. 

Simplifying this task by tailoring a mashup tool to the specific domain of research evaluation first of all requires understanding what a domain is. We define a domain and, then, a domain-specific mashup as follows:

\begin{definition}\label{def:domain}
A \emphbf{domain} is a delimited sphere of concepts and processes; \emph{domain concepts} consist of data and relationships; \emph{domain processes} operate on domain concepts and are either atomic (activities) or composite (processes integrating multiple activities), defined according to  \emph{domain rules}.
\end{definition}

\begin{definition}\label{def:dommashup}
A \emphbf{domain-specific mashup} is a mashup that describes a composite \emph{domain process} that manipulates \emph{domain concepts} via \emph{domain activities and processes} following \emph{domain rules}. It is specified in a domain-specific, graphical modeling notation.
\end{definition}

A \emphbf{domain-specific mashup} is therefore a \emph{web mashup} specified with a domain-specific model. 
The domain defines the "universe" in the context of which we can define domain-specific mashups. It defines the information that is processed by the mashup, both conceptually and in terms of concrete data types (e.g., XML schemas). It defines the classes of components that can be part of the process and how they can be combined, as well as a notation that carries meaning in the domain (such as specific graphical symbols for components of different classes). 

As we will see later in detail, every mashup can only use components that conform to the domain process model and that exchange data which belongs to the conceptual model. This means that each component can send or receive data based on the entities or relationships of the conceptual model. Finally, the domain defines rules that represent invariants to be met by each mashups. It has a \emph{static} part, which describes the concepts that are proper of the domain, and a \emph{dynamic} part, which describes the modifications the concepts may be subject to. For instance, in our reference scenario,  concepts include  \emph{publications}, \emph{researchers}, \emph{metrics}, etc. The process models define classes of components such as data extraction from digital libraries, metric computation, or filtering and aggregation components. A domain rule could, for instance, disallow the use of a specific information source for the computation of a given metric. These domain restrictions and the exposed domain concepts at the mashup modeling level is what enables simplification of the language and its usage. 

Generic mashup tools are neither aware of these concepts, nor of these operations. Given Definition \ref{def:domain} we can therefore say that our reference scenarios ask for a mashup that is specific to the domain of research evaluation, i.e., it asks for a domain-specific mashup. So following this we can define a domain-specific mashup tool as:

\begin{definition}\label{def:platform1}
A \emphbf{domain-specific mashup tool} (DMT) is a development and execution environment that enables \emph{domain experts}, i.e., the actors operating in the domain, to develop and execute \emph{domain-specific mashups} via a \emph{syntax} that exposes all features of the \emph{domain}. 
\end{definition}

A DMT is initially "empty". It then gets populated with specific components that provide functionality needed to implement mashup behaviors. For example, software developers (not end-users) will define libraries of components for research evaluation, such as components to extract data from Google Scholar, or to compute the h-index, or to group researchers based on their institution, or to visualize results in different ways. Because all components fit in the classes and interact based on a common data model, it becomes easier to combine them and to define mashups, as the DMT knows what can be combined and can guide the user in matching components. The domain model can be arbitrarily extended, though the caveat here is that a domain model that is too rich can become difficult for software developers to follow.

%================ Challenges and Problems ==================
\section{Challenges and problems}
\label{sec:challenges}

Given these definitions, the \emphbf{problem} we solve is that of providing the necessary concepts and a methodology for the development of domain-specific mashup models and DMTs. The problem is neither simple nor of immediate solution. While domain modeling is a common task in software engineering, its application to the development of mashup platforms is not trivial. For instance, we must precisely understand which domain properties are needed to exhaustively cover all those domain aspects that are necessary to tailor a mashup platform to a specific domain, which property comes into play in which step of the development of the platform, how domain aspects are materialized (e.g., visualized) in the mashup platform, and so on.

The DMT idea is heavily grounded on a rich corpus of research in \emph{Human-Computer Interaction} (HCI), demonstrating that consideration of user knowledge and prior experience are required to create truly usable and inclusive products, and are key considerations in the performance of usability evaluations \citep{Nielsen93}. The prior experience of products is important to their usability, and the transfer of previous experience depends upon the nature of prior and subsequent experience of similar tasks \citep{Thomas99}. Familiarity of the interface design, its interaction style, or the metaphor it conforms to if it possesses one, are key features for successful and intuitive interaction \citep{Okeye98}. 

More familiar interfaces, or interface features, allow for easier information processing in terms of user capability, and the subsequent human responses can be performed at an automatic and subconscious level. \citep{Karlsson06} identified that the use of semantics could be an effective tool for enhancing product design and use, particularly for novel users, as they can indicate how the product or interface will behave and how interaction is likely to occur. Similarly, \citep{Monk98} stressed that to be usable and accessible, interfaces need to be easily understood and learned, and in the process, must cause minimal cognitive load. Effective interaction consists of users understanding potential actions, the execution of specific action, and the perception of the effects of that action.

As we cannot exploit the users' technical expertise, we propose here to exploit their knowledge of the task domain. In other words, we intend to transform mashups from technical tools built around a computing metaphor to true cognitive artifacts \citep{Norman91}, capable to operate upon familiar information in order to "serve a representational function that affect human cognitive performance.''

%===================== APPROACH =================
\section{Methodology}
\label{sec:approach}

In order to develop a DMT, we have to look into the details of three incremental aspects, i.e., the domain concepts, the domain processes, and the implementation of the DMT. In following we state and define all the ingredients for developing a domain-specific mashup platform. Specifically, developing a domain-specific mashup platform requires:

\begin{enumerate}

\item Definition of a \emph{domain concept model} (DCM) to express domain data and relationships. The concepts are the core of each domain. They drive the implementation of the DMT and of its data types and components. It is therefore crucial to precisely delimit the concepts that characterize the domain, in order to instruct the tool how to use them and to develop components that understand them. The specification of domain concepts allows the mashup platform to understand what kind of \emph{data objects} it must support. This is different from generic mashup platforms, which provide support for generic data formats, not specific objects.

\item Identification of a generic \emph{mashup meta-model}\footnote{We use the term \emph{meta-model} to describe the constructs (and the relationships among them) that rule the design of mashup \emph{models}. With the term \emph{instance} we refer to the actual mashup application that can be operated by the user.} (MM) that suits the composition needs of the domain and the selected scenarios. A variety of different mashup approaches, i.e., meta-models, have emerged over the last years, e.g., ranging from data mashups, over user interface mashups to process mashups. Before thinking about domain-specific features, it is important to identify a meta-model that is able to accommodate the domain processes to be mashed up.

\item Definition of a \emph{domain-specific mashup meta-model}. Given a generic MM, the next step is understanding how to inject the domain into it so that all features of the domain can be communicated to the developer. We approach this by specifying and developing:

\begin{enumerate}

\item A \emph{domain process model} (PM) that expresses classes of domain activities and, possibly, ready processes. Domain activities and processes represent the dynamic aspect of the domain. They operate on and manipulate the domain concepts. Injecting the domain into the tool means introducing domain-specific extensions into the mashup meta-model, e.g., to take into account the nature of domain activities. The activities that can be composed in order to form new processes indicate which mashup components in terms of data, application logic, and UI components are needed to implement the domain-specific mashups. In the context of mashups, we can map activities and processes to reusable components of the platform. 

\item A \emph{Domain rule model} that may constrain the use of processes or activities, in order to guarantee the correct use of concepts and components in the tool. We specify domain rules in a \emph{domain rule model}.

\item A \emph{domain syntax} that provides each concept in the domain-specific mashup meta-model (the union of MM and PM) with its own symbol. The claim here is that just catering for domain-specific activities or processes is not enough, if these are not accompanied with visual metaphors that the domain expert is acquainted with and that visually convey the respective functionalities.

\item A set of \emph{instances of domain-specific components}. This is the step in which the reusable domain-knowledge is encoded, in order to enable domain experts to mash it up into new applications. 

\end{enumerate}

\item \emph{Implementation} of the DMT as a tool whose expressive power is that of the domain-specific mashup meta-model and that is able to host and integrate the domain-specific activities and processes. 
\begin{enumerate}
\item \emph{DMT.}  The DMT must support all features that are specified in both the domain-specific mashup meta-model and the domain concept model. Specifically, the extended mashup meta-model determines the \emph{expressive power} of the DMT.

\item \emph{Components.} The components instantiate the concepts in the domain-specific meta-model extension and implement the domain activities identified in step 3(d).
\end{enumerate}

\end{enumerate}

The above steps mostly focus on the \emph{design} of a domain-specific mashup platform. Since domains, however, typically \emph{evolve} over time, in a concrete deployment it might be necessary to periodically update domain models, components, and the platform implementation (that is, iterating over the above design steps), in order to take into account changing requirements or practices. The better the analysis and design of the domain in the first place, the less modifications will be required in the subsequent evolution steps, e.g., limiting evolution to the implementation of new components only.  

In the next subsections, we expand each of the above design steps starting from the domain concept model.
\section{The Domain Concept Model}
\label{sec:domain}

%In order to model a domain as a triple $D=\langle CM,PM,RM \rangle$, with $CM$ being the concept model, $PM$ being the process model, and $RM$ being the rule model, we propose the use of three simple modeling formalisms. We introduce each of them next. 

%\subsection{Domain knowledge}
It is important to precisely delimit the concepts that characterize the domain, in order to instruct the tool how to use them and to develop components that understand them. We specify domain knowledge in the form of a \emph{domain concept model}. The domain concept model is constructed by the IT experts via verbal interaction with the domain experts or via behavioral observation of the experts performing their daily activities and performing a suitable task-analysis. The heart of each domain is represented by the information items each expert of that domain knows and understands. 

The concept model represents the information experts know, understand, and use in their work. Modeling this kind of information requires understanding the fundamental information items and how they relate to each other, eventually producing a model that represents the knowledge base that is shared among the experts of the domain. In domain-specific mashups, the concept model has three kinds of \emphbf{stakeholders} (and usages), and understanding this helps us to define how the domain should be represented.
\begin{itemize}

\item The first stakeholders are the mashup modelers (domain experts), i.e., the end-users that will develop different mashups from existing components. For them it is important that the concept model is easy to understand, and an entity-relationship diagram (possibly with a description) is a commonly adopted technique to communicate conceptual models.

\item The second kind of stakeholders are the developers of components, which are programmers. They need to be aware of the data format in which entities and relationships can be represented, e.g., in terms of XML schemas, in order to implement components that can interoperate with other components of the domain. 

\item The third stakeholder is the DMT itself, which enforces compliance of data exchanges with the concept model.

\end{itemize}
Therefore:

\begin{definition}\label{def:conceptmodel}
The \emphbf{domain concept model (DCM)} describes the \emph{conceptual entities} and the \emph{relationships} among them, which, together, constitute the domain knowledge. 
\end{definition}

A DCM is an example of data that is used to be as input to or output from a mashup component. Modeling DCM is also an attempt to separate out what doesn't vary much from what does in a particular domain. These first-class concept types are constrained by the domain rules.

\begin{figure}[t]
 \centering
   \includegraphics[width=1\columnwidth]{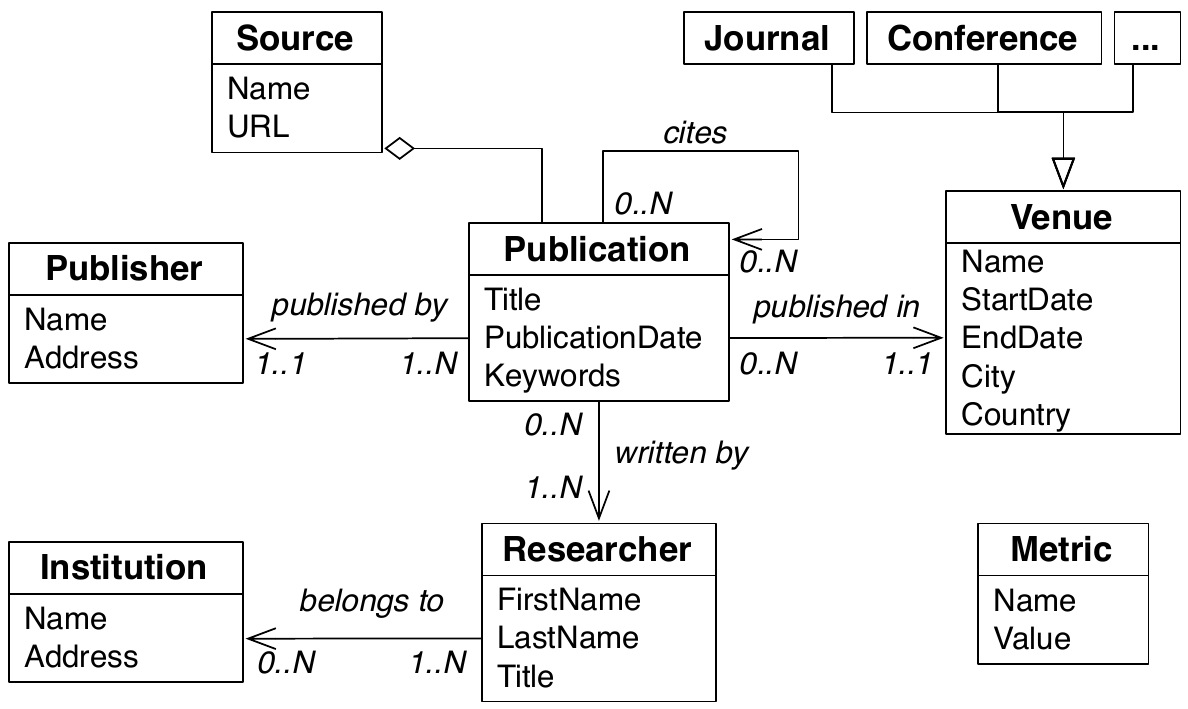}
 \caption{Domain concept model, covering main concepts required for the referenced research evaluation scenarios}
 \label{fig:concept-model}
\end{figure}  

We express the domain-model as a conventional entity-relationship diagram. It also includes a representation of the entities as XML schemas. For instance, in Figure \ref{fig:concept-model} we put only main concepts we could identify in our reference scenarios into a DCM, detailing entities, attributes, and relationships. The core element in the evaluation of scientific production and quality is the \emph{publication}, which is typically published in the context of a specific \emph{venue}, e.g., a conference or journal, by a \emph{publisher}. It is written by one or more  \emph{researchers} belonging to an \emph{institution}. Increasingly -- with the growing importance of the Internet as an information source for research evaluation -- also the \emph{source} (e.g., Scopus, the ACM digital library or Microsoft Academic) from which publications are accessed is gaining importance, as each of them typically provides only a partial view on the scientific production of a researcher and, hence, the choice of the source will affect the evaluation result.
%(indeed the source selection is one of the critical and most debated topics when defining metrics).
The actual evaluation is represented in the model by the \emph{metric} entity, which can be computed over any of the other entities.

In order to develop a DMT, the ER (Entity-Relationship) model has to be generated through several interactions between the domain expert and the IT expert, who has knowledge of conceptual modeling. The IT expert also generates the XML schemas corresponding to the ER model, which are the actual artifacts processed by the DMT.

In fact, although the ER model is part of the concept model, it is never processed itself by the DMT. It rather serves as a reference for any user of the platform to inform them on the concepts supported by it. In principle, other formalisms can be adopted (such as UML Class diagrams). We notice that each concept model implicitly includes the concept of \emph{grouping} the entities in arbitrary ways, so groups are also an implicitly defined entity.

\section{The Generic Mashup Meta-Model}
\label{sec:mm}

When discussing the domain concept model we made the implicit choice to start from \emph{generic} (i.e., domain-independent) models like Entity-Relationship diagrams and XML, as these are well established data modeling and type specification languages amenable to humans and machines.
For end-user development of mashups, the choice is less obvious since it is not easy to identify a modeling formalism that is amenable to defining end-user mashups (which is why we endeavor to define a domain-specific mashup approach). If we take existing mashup models and simply inject specific data types in the system, we are not likely to be successful in reducing the complexity level. However, the availability of the DCM makes it possible to derive a different kind of mashup modeling formalism, as discussed next.

To define the \emphbf{type of mashups} and, hence, the modeling formalism that is required, it is necessary to model which features (in terms of software capabilities) the mashups should be able to support. Mashups are particular types of web applications. They are component-based, may integrate a variety of services, data sources, and UIs. They may need an own layout for placing components,  require control flows or data flows, ask for the synchronization of UIs and the orchestration of services, allow concurrent access or not, and so on. Which exact features a mashup type supports are described by its \emph{mashup meta-model}.

Besides specifying a type or class of mashups, the mashup meta-model (MM) specifies how to draw the actual mashup (process) models. In the following, we first define a generic mashup meta-model, which may fit a variety of different domains, then we show how to define the domain-specific mashup meta-model, which will allow us to draw domain-specific mashup models.

%For each given domain, a domain-specific mashup (process) model will then be derived from the (generic) mashup meta-model (MM).

\begin{definition}\label{def:metamodel}
The generic \emphbf{mashup meta-model (MM)} specifies a \emph{class} of mashups and, thereby, the \emph{expressive power}, i.e., the concepts and composition paradigms, the mashup platform must know in order to support the development of that class of mashups.
\end{definition}

%\begin{figure}[t]
%  \centering
%    \includegraphics[width=0.7\textwidth]{figs/MM2.pdf}
%  \caption{Mashup meta-model supporting universal integration and rule enforcement}
%  \label{fig:meta-model}
%\end{figure} 

\begin{figure}[t]
 \centering
   \includegraphics[width=1.05\columnwidth]{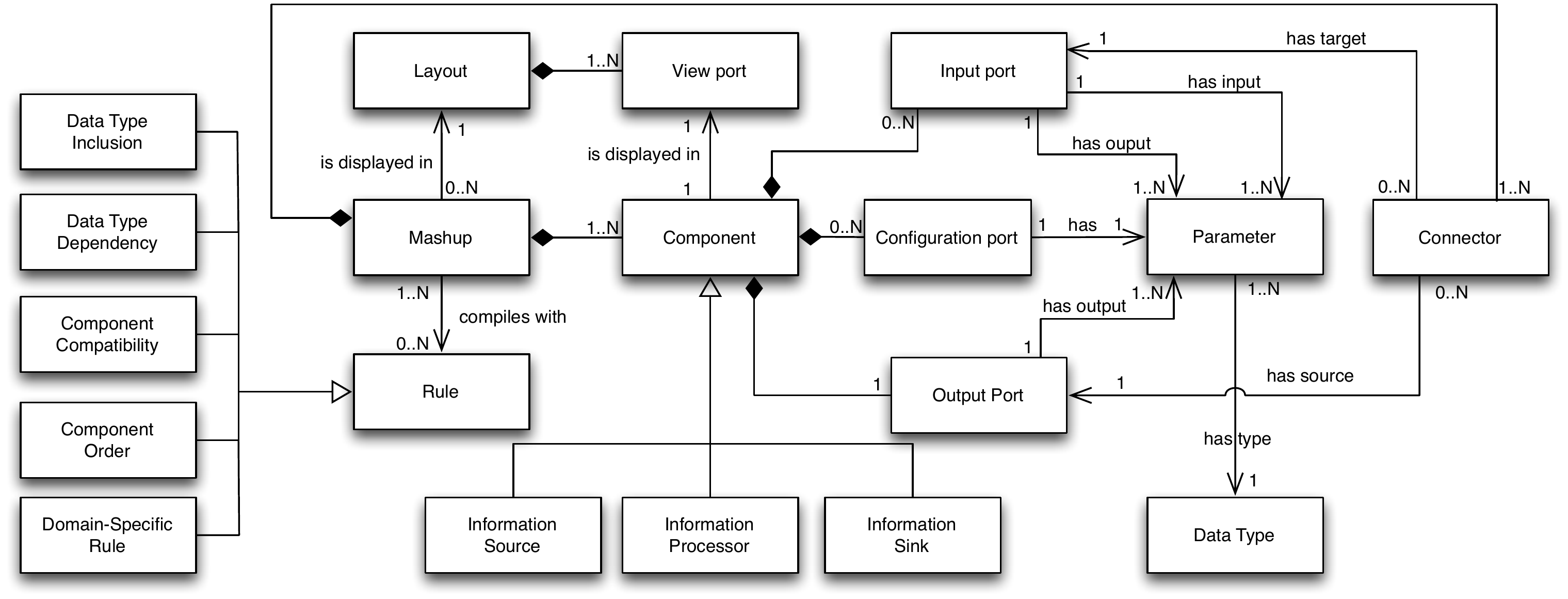}
 \caption{Mashup Meta-model supporting, domain-specific concepts, processes, rules, and universal integration}
 \label{fig:mmm}
\end{figure}

The MM therefore implicitly specifies the expressive power of the mashup platform. Identifying the right features of the mashups that fit a given domain is therefore crucial. For instance, our research evaluation scenario asks for the capability to integrate data sources (to access publications and researchers via the Web), web services (to compute metrics and perform transformations), and UIs (to render the output of the assessment). We call this capability \emph{universal integration}. Next, the scenario asks for data processing capabilities that are similar to what we know from Yahoo! Pipes, i.e., data flows. It requires dedicated software components that implement the basic activities in the scenario, e.g., compute the impact of a researcher (the sum of his/her publications weighted by the venue ranking), compute the percentile of the researcher inside the national sample (producing outputs like "top 10\%"), or plot the department ranking in a bar chart. Figure \ref{fig:mmm} depicts our mashup meta-model that supports the universal integration and also enforce various rules that of a domain-specific type or a of generic nature. In following we describe the details of the proposed mashup meta-model.

\subsection{The mashup meta-model} % I think write The Mashup meta-model

We start from a very simple MM, both in terms of notation and execution semantics, which enables end-users to model own mashups. %Indeed, it can be fully specified in one page:

\begin{enumerate}
\item As shown in the Figure \ref{fig:mmm}, a \emphbf{mashup} $m=\langle C,P, R, VP,L \rangle$, defined according to the meta-model MM, consists of a set of \emph{components} $C$, a set of \emph{connectors} (i.e., data pipes) $P$, a set of rules $R$, a set of view ports $VP$ that can host and render components with own UI, and a \emph{layout} $L$ that specifies the graphical arrangement of components. 

A mashup compiles with a set of rules. These rules can be of various types, for example, \emph{data type inclusion}: validates when a new data type introduced to the mashup; \emph{data type dependency} confirms inheritance of those data types that are already exist;\emph{component compatibility} validates components compatibility upon connecting two components. Sometimes two components seem compatible to each other even then their ordering (i.e., the position of a component in a mashup) could make problems. So \emph{component order} checks for right ordering. Finally, there could be many domain-specific rules that a mashup must consider. The detail of domain-specific rules are given later in this chapter. 

\item A \emphbf{component} $c =\langle IPT, OPT, CPT, type, desc \rangle$, where $c\in C$, is like a task that performs some data, application, or UI action. 

Components have \emph{ports} through which pipes are connected. Ports can be divided in \emph{input} ($IPT$) and \emph{output} ports ($OPT$), where input ports carry data into the component, while output ports carry data generated (or handed over) by the component. Each component must have at least either an input or an output port. Both IPTs and OPTs can have \emph{parameters} of specific \emph{data types}. The data types include both primitive and domain-specific types once the MM gets extended for a domain.

\emph{Configuration} ports ($CPT$) are used to configure the components. They are typically used to configure filters (defining the filter conditions) or to define the nature of a query on a data source. The configuration data can be a constant (e.g., a parameter defined by the end-user) or can arrive in a pipe from another component. Conceptually, constant configurations are as if they come from a component feeding a constant value.

A component can be of type \emph{information source}, \emph{information processor}, or \emph{information sink}. Components with no input ports are called \emph{information sources} and work as data source by supplying data to other components. Components with no output ports are called \emph{information sinks}. All UI components are always of information sink type. They do not perform business logic on the consumed data, but to visualize it to users. Components with both input and output ports are called \emph{information processors}. These components take data, process it and produce results.

The type ($type$) of the components denotes whether they are \emph{UI components}, which display data and can be rendered in the mashup's layout, or \emph{application components}, which either fetch or process information or a data source components. Mainly the type information is used by the internal's logic but it could also be used to arrange components for better presentation for end-users. 

Components can also have a description \emph{desc} at an arbitrary level of formalization, whose purpose is to inform the user about the data the components handle and produce.

\item A \emphbf{pipe} (i.e., connector) $p \in P$ carries data (e.g., XML/JSON documents) between the ports of two components, implementing a data flow logic. So, $p \in  IPT \times (OPT \cup CPT) $.

\item A \emphbf{view port} $vp \in VP$ identifies a place holder, e.g., a DIV element or an IFRAME, inside the HTML template that gives the component its graphical identity. Typically, a template has multiple placeholders.

\item Finally, the \emphbf{layout} $L$ defines which component with own UI is to be rendered in which view port of the template. Therefore $l \in C \times VP$.

\end{enumerate}

Each mashup following this MM must have at least a source and a sink, and all ports of all components must be attached to a pipe or manually filled with data (the configuration port).

This is all we need to define a mashup and as we will see, this is an executable specification. There is nothing else besides this picture. This is not that far from the complexity of specifying a flowchart, for example. It is very distant from what can be an (executable) BPMN specification or a BPEL process in terms of complexity.

In the model above there are \emph{no variables} and \emph{no data mappings}. This is at the heart of enabling end-user development as this is where much of the complexity resides. It is unrealistic to ask end-users to perform data mapping operations.
Because there is a DCM, each component is required to be able to process any document that conforms to the model. This does not mean that a component must process every single XML element. For example, a component that computes the h-index will likely do so for researchers, not for publications, and probably not for publishers (though it is conceivable to have an h-index computed for publishers as well). So the component will "attach" a metric only to the researcher information that flows in. Anything else that flows in is just passed through without alterations. The component description will help users to understand what the component operates on or generates, and this is why an informal description suffices. 
What this means is that each component in a domain-specific mashup must be able to implement this \emph{pass-through} semantics and it must operate on or generate one or more (but not all) elements as specified in the DCM. Therefore, our MM assumes that all components comply to understand the DCM.
% and there cannot be ways in which components are combined and give errors. 

Furthermore, in the model there are also \emph{no gateways} as in BPMN, although it is possible to have dedicated components that, for example, implement an if-then semantics and have two output ports for this purpose. In this case, one of the output ports will be populated with an empty feed. 
Complex routing semantics are  virtually impossible for non-experts to understand (and in many cases for experts as well) and for this reason if they are needed we delegate them to the components which are done by programmers and are understood by end-users in the context of a domain. 

\subsection{Operational semantics}
\label{sec:gen-exe-semantics}
The behavior of the components and the semantics of the MM are as follows:

\begin{enumerate}
\item Executions of the mashups are \emph{initiated} by the user. A user have to explicitly start the execution using some user interface means (e.g., a button click).

\item Components that are \emph{ready} for execution are identified. A component is ready when all the input and configuration ports are filled with data, that is, they have all necessary data to start processing.

\item  All ready components are then \emph{executed}. They process the data in input ports, consuming the respective data items from the input feed, and generate output on their output ports. The generated output fills the inputs of other components, turning them executable. 

\item The execution proceeds by identifying ready components and executing them (i.e., \emph{reiterating} steps 2 and 3), until there are no components to be executed left. However, during the execution if in case some component requires user interaction (e.g., an input) before it proceed, then the execution stops and starts again after user acts as needed. This means it is possible to interact with the mashup execution during runtime. At this point, all components have been executed, and all the sinks have received and rendered information.
\end{enumerate}
%Figure \ref{fig:scenario} shown earlier shows an example of a mashup that follows the model. The figure shows the component with their input and output ports (in white, and typically on the left and right of the component, respectively) and the configuration ports (in grey). Its semantic is the one described in the scenario text. 

\subsection{Generic mashup syntax}

\begin{figure}[t]
  \centering
    \includegraphics[width=1\columnwidth]{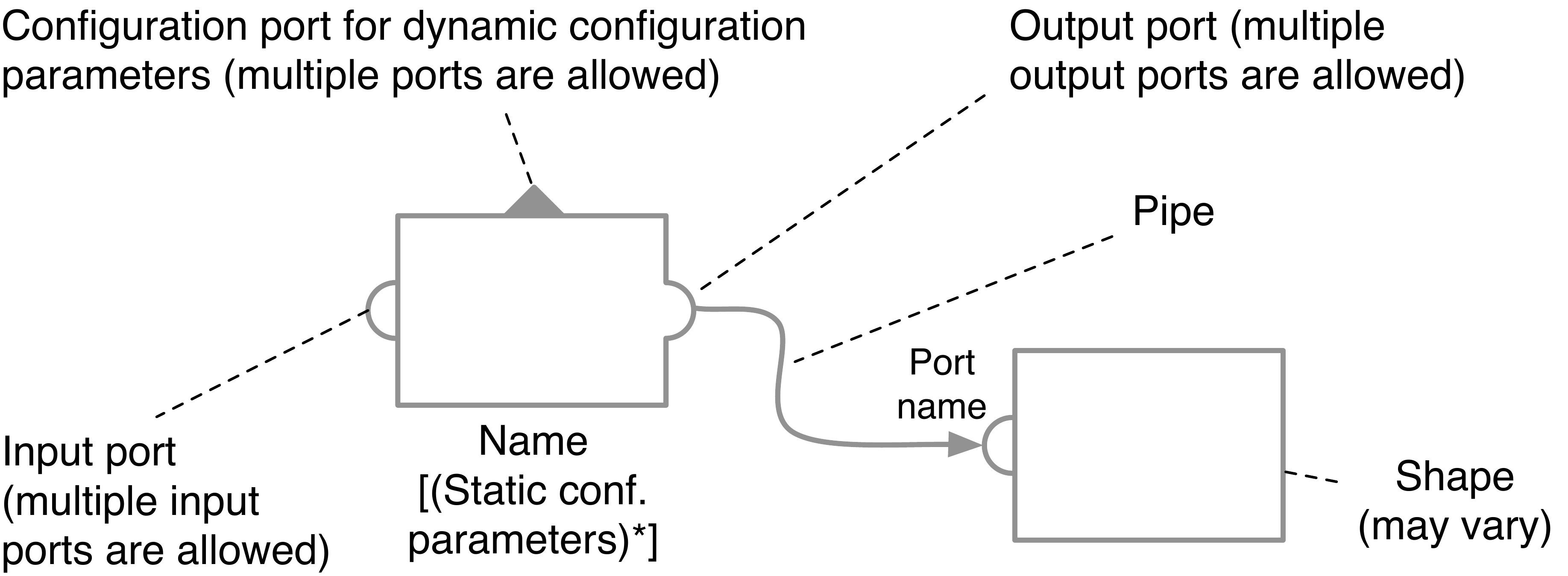}
  \caption{Basic syntax for the concepts in the mashup meta-model.}
  \label{fig:basic}
\end{figure}

Developing mashups based on this meta-model, i.e., graphically composing a mashup in a mashup tool, requires defining a \emphbf{syntax} for the concepts in the MM. In Figure \ref{fig:basic} we map the above MM to a basic set of generic graphical symbols and composition rules. In the next section, we show where to configure domain-specific symbols.

\section{The Domain-Specific Mashup Meta-Model}
\label{sec:mashup}

%\subsection{Domain processes}

The mashup meta-model (MM) described in the previous section allows the definition of a class of mashups that can fit in different domains. Thus, it is not yet tailored to a specific domain, e.g., research evaluation. Now we want to push the domain into the mashup meta-model constraining the class of the mashups that can be produced to that of our specific domain. Despite the relative simplicity, providing users with a DCM-restricted mashup meta-model is still not likely to be sufficient in terms of ease of use.  The user will still be faced with a large number of possible components to be placed on a canvas.

The next step is therefore understanding the dynamics of the concepts in the model, that is, the typical classes of processes and activities that are performed by domain experts in the domain, in order to transform or evolve concrete instances of the concepts in the DCM and to arrive at a structuring of components as well as to an intuitive graphical notation. What we obtain from this is a \emph{domain-specific mashup meta-model}. Each domain-specific meta-model is a specialization of the mashup meta-model along four dimensions: 

\begin{enumerate}
\item Domain-specific activities and processes
\item Domain-specific rules
\item Domain-specific syntax
\item Domain instances
\end{enumerate}

The domain-specific meta-model extension extends the MM with domain-specific \emph{sub-types} of the component entity in the MM. Sub-types allow the injection of classes of domain processes or activities into the MM and, hence, the introduction of domain-specific \emph{terminology} and \emph{syntax}. In figure \ref{fig:mm-extension} we show domain-specific meta-model extension, and describe its details as follows.
%Composing, i.e., mashing up, processes or activities operating on instances of domain concepts into composite processes enables the construction of complex, possibly value-adding process logics, which is exactly the goal of mashups in general. The composability of processes and activities is guaranteed by the CM, which is shared by all processes and activities and limits the types of outputs and inputs to the set of its concepts.

\subsection{Domain process model}
\label{sec:dpm}
%We define the domain process model as follows:
\begin{definition}\label{def:activitymodel}
The \emphbf{domain process model (PM)} describes the classes of \emph{processes} or \emph{activities} that the domain expert may want to mash up to implement composite, domain-specific processes.
%, as well as a respective graphical notation. It may also include a set of built-in components.
%We express domain processes as a set of tuples $\langle Name,Parameters,In\-puts,Out\-puts \rangle$.
\end{definition}

%\begin{figure}[t]
%  \centering
%    \includegraphics[width=1\columnwidth]{meta-model-extension.pdf}
%  \caption{Mashup meta-model extension to its various extensible ports extending it to domain-specific mashup meta-model }
%  \label{fig:mm-extension}
%\end{figure}  

Operatively, the process model is again derived by specializing the generic meta-model based on interactions with domain experts, just like for the domain concept model. This time the topic of the interaction is aimed at defining classes of components, their interactions and notations. 
%Figure \ref{fig:mm-extension} depicts an extension of the mashup meta-model to our reference domain (i.e., research evaluation). 
In the case of research evaluation, this led to the identification of the following classes of activities, i.e., classes of components:

For simplicity, we discuss only the processes that are necessary to implement the reference scenarios. 

\begin{enumerate}
  \item \emphbf{Source extraction} activities. They are like queries over digital libraries such as DBLP or Google Scholar. They may have no input port, and have one output port (the extracted data). These components may have one or more configuration ports that specify in essence the "query". For example a source component may take in input a set of researchers and extract publications and citations for every researcher from Google Scholar. 
  
\item \emphbf{Metric computation} activities, which can take in input institutions, venues, researchers, or publications and attach a metric to them. The corresponding components have at least one input and one output port. For example, a component determines the h-index for researchers, or determines the percentile of a metric  based on a distribution.

\item \emphbf{Aggregation} activities, which define groups of items based on some parameter (e.g., affiliation).  

\item \emphbf{Filtering} activities, which receive an input pipe and return in output a filtering of the input, based on a criterion that arrives in a configuration port. For example we can filter researchers based on the nationality or affiliation or based on the value of a metric. 

\item \emphbf{UI widgets}, corresponding to  information sink components that plot or map information on researchers, venues, publications, and related metrics.
\end{enumerate}

%    $\langle \textbf{Aggregate},AggregationFunction,[Publication|Researcher|Institution|$
%
%$Venue]+,[Publication|Researcher|Institution|Venue]+ \rangle$
%
%$\langle \textbf{Filter},FilterCondition+,[Publication|Researcher|Institution|Venue]+,$
%
%$[Publication|Researcher| Institution|Venue]+ \rangle$
%
%$\langle \textbf{PlotChart},ChartType,Metric+,- \rangle$
%    
%   come in two flavors: those with no configuration ports, in which case each 
%  
%  
%  $\langle \textbf{LoadStaticSource},\langle URL,Selector? \rangle,-,[Publication|Researcher|$
%$Institution|Venue|Metric]+ \rangle$
%
%   $\langle \textbf{LoadParametricSource},\langle URL,Selector? \rangle,Query?,[Publication|$
%
%$Researcher| Institution|Venue|Metric]+ \rangle$
% $\langle \textbf{ComputeMetric},MetricType,[Publication|Researcher|Institution|$
%
%$Venue]+,Metric \rangle$
%
%  
%We identify the following processes in our reference scenario (using basic BNF quantifiers to express cardinalities in the PM):
%
%
%
%$\langle \textbf{PlotMap},MapType,[Institution|Venue]+,- \rangle$
%
%\medskip
%
%For example, the the $LoadStaticSource$ process loads a set of $publications$, $researchers$, $institutions$, $venues$, or $metrics$, depending on the $URL$ from which to load the data; specifying the optional $selector$ expression allows one to limit the load to a subset of the available data (for instance, only the researchers of UniTN). 

%
%\begin{figure}
%  \centering
%    \includegraphics[scale=0.6]{figs/extension.pdf}
%  \caption{Domain-specific meta-model extension for research evaluation}
%  \label{fig:extension}
%\end{figure}  

\subsection{Domain rules}
As a domain comprises of domain concepts, activities/processes and rules (i.e., constraints, restrictions on concepts and activities) to prescribe and/or restrict the way in which domain experts use domain activities and processes to achieve their goals. These domain rules can be defined in a way like integrity constraints e.g., from the cardinalities between concepts in a domain concept model. However domain rules not only cover integrity constraints but usually also allow or restrict domain behaviors (i.e., domain activities/processes). For example, a rule could be that DBLP cannot be used for computing H-index metrics and thus can be instantiated as a \emph{Component compatibility (mashup)} rule disallowing the usage of these two components in the same mashup composition). The rule enforcement in the DMT provides assistance and guidance to the domain-experts improving usability, composition correctness and development errors reduction.

\begin{figure}[t]
  \centering
    \includegraphics[width=.70\columnwidth]{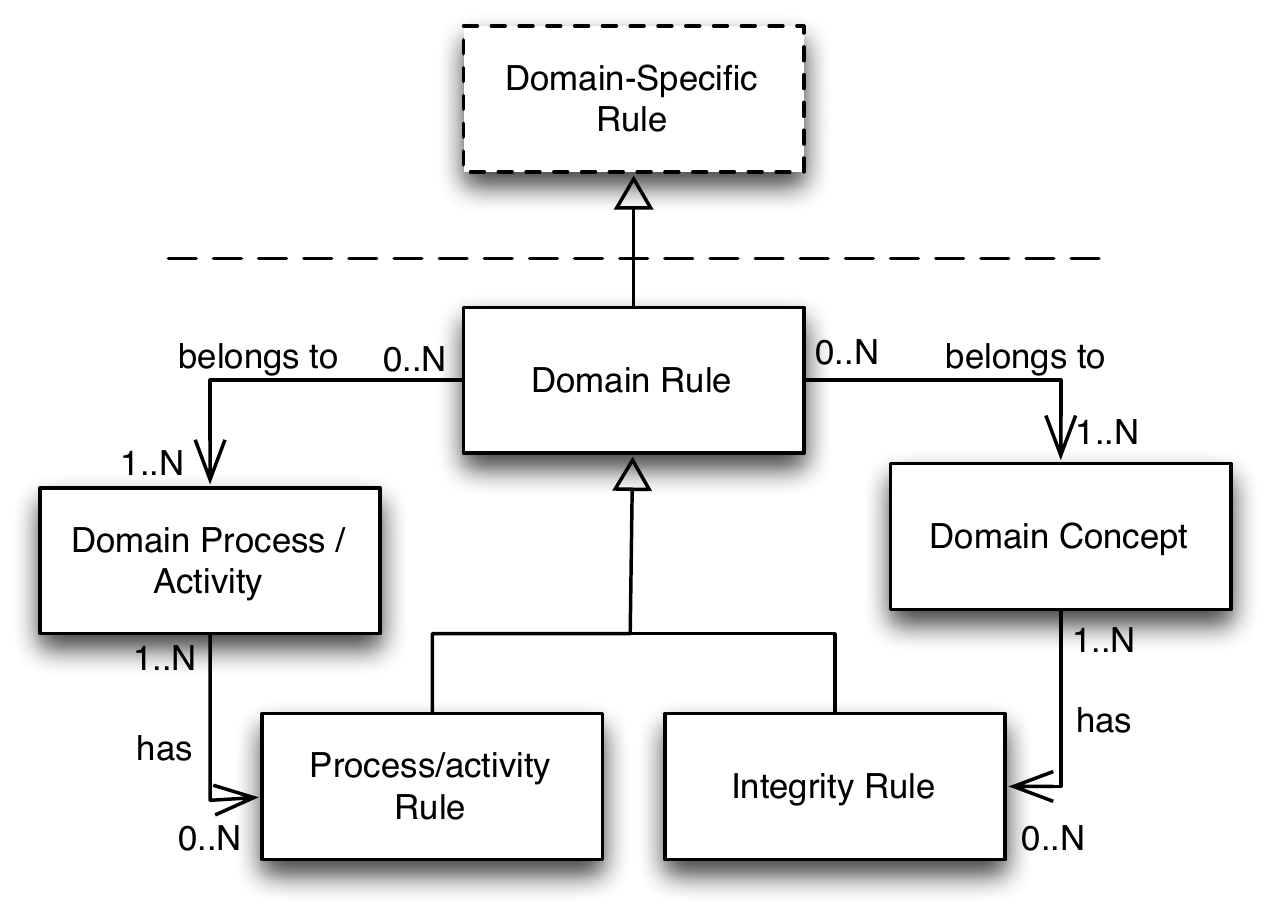}
  \caption{Extension to the domain-specific rules.}
  \label{fig:dsr-mm-extension}
\end{figure}  

A well know way to define these rules is through ECA structure (event, condition, activity) which means: if the \emph{event} occurs and \emph{conditions} are met, then execute the \emph{activity} \cite{herbst1996business}. Figure \ref{fig:dsr-mm-extension} depicts the extension of MM along with rule model. Domain rules can be classified in many different ways. When analyzing rules in the context of domain processes, following are the two rule types we identify: \emph{Activity rules:} domain rules related to the a particular activity or subset of an activity. \emph{Integrity rules:} are related to the domain objects and their relationships, for example, the value of the H-index metric cannot be negative.

\subsection{Domain syntax}
\label{sec:dsm-domain-syntax}
\begin{figure}[t]
  \centering
    \includegraphics[width=1\columnwidth]{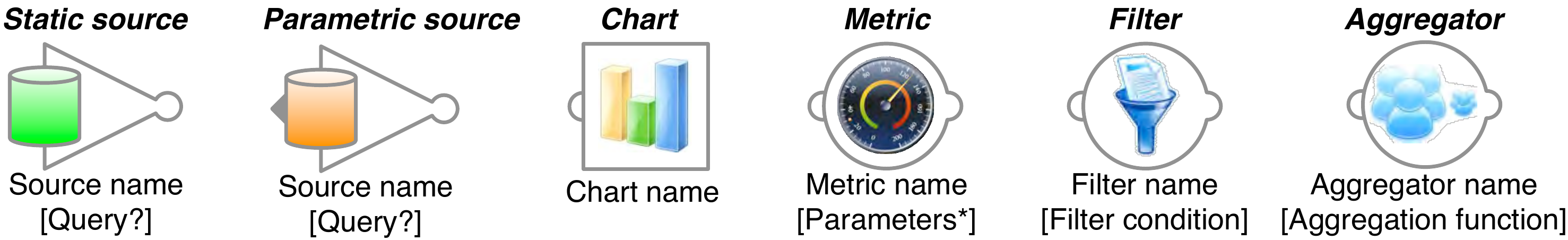}
  \caption{Domain-specific syntax for the concepts in the domain-specific meta-model extension}
  \label{fig:syntax}
\end{figure}

A possible \emphbf{domain-specific syntax} for the classes in the PM (derived from the generic syntax presented in Figure \ref{fig:basic}) is shown in Figure \ref{fig:syntax}.%, which is used for our reference scenario in Figure \ref{fig:scenario} shown earlier. 
Its semantics is the one described by the MM in Section \ref{sec:mm}. In practice, defining a PM that fully represents a domain requires considering multiple scenarios for a given domain, aiming at covering all possible classes of processes in the domain. 

\subsection{Domain instances}
\begin{figure}[t]
  \centering
    \includegraphics[width=1\columnwidth]{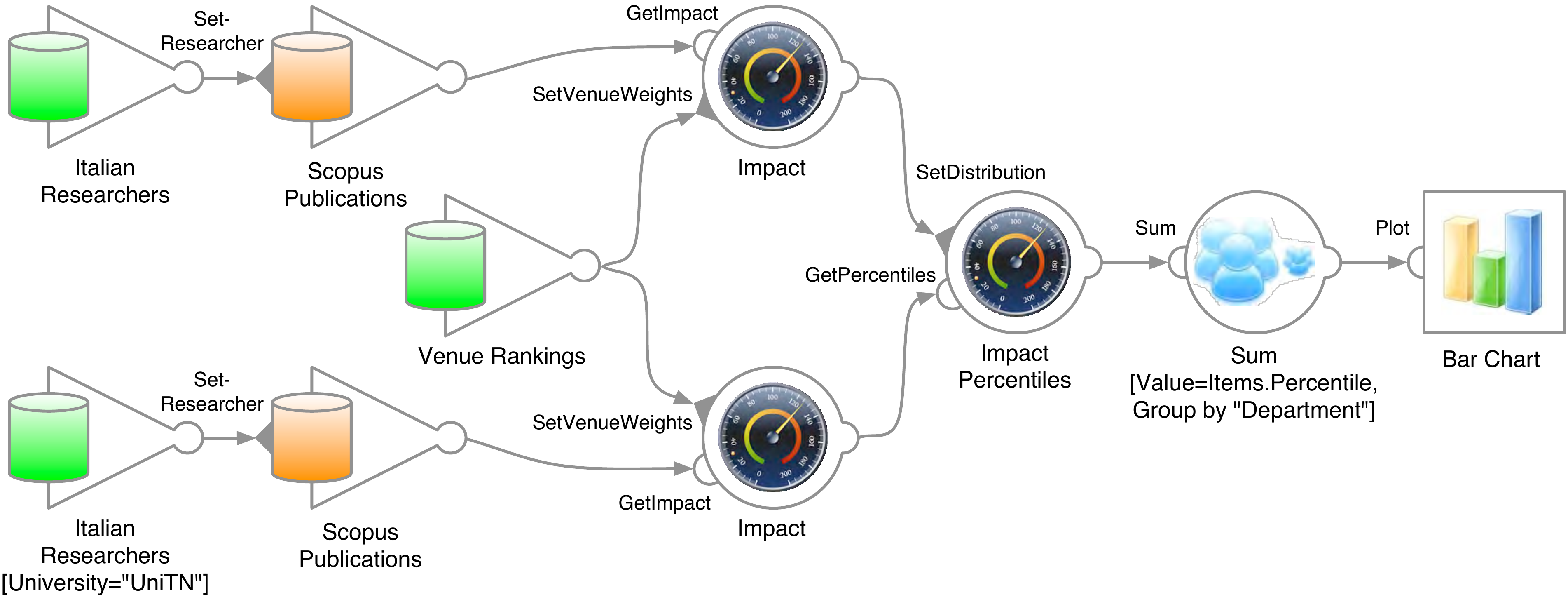}
  \caption{An example of the use of instances of domain-specific components}
  \label{fig:scenario}
\end{figure}

Domain instances are fully functional domain-specific components that are ready to be used in mashup compositions. These domain-specific components with domain syntax (i.e., domain symbols) implements domain activities and processes, consuming and producing domain-specific concepts at input and output ports. Figure \ref{fig:scenario} actually exemplifies the use of \emph{instances} of domain-specific components. For example, the \emph{Microsoft Academic Publications} component is an instance of  \emph{source extraction} activity with a configuration port (\emph{SetResearchers}) that allows the setup of the researchers for which publications are to be loaded from Microsoft Academic. The component's symbol is an instantiation of the parametric source component type in Figure \ref{fig:syntax} without static query. Similarly, the \emph{Italian Researchers} (source extraction activity), the \emph{Venue Ranking} (source extraction activity), the \emph{Impact} (metric computation activity), the \emph{Impact Percentiles} (metric computation activity), and the \emph{Bar Chart} (UI widget) components.

In summary, what we do is limiting the flexibility of a generic mashup tool to a specific class of mashups, gaining however in intuitiveness, due to the strong focus on the specific needs and issues of the target domain. Given the models introduced so far, we can therefore refine our definition of DMT given earlier as follows:

\begin{definition}\label{def:platform2}
A \emphbf{domain-specific mashup tool (DMT)} is a development and execution environment that (i) implements a domain-specific mashup meta-model, (ii)  exposes a domain-specific modeling syntax, and (iii) includes an extensible set of domain-specific component instances. 
\end{definition}

Once the domain models are ready, the IT expert can then customize a mashup platform that meets the requirements that emerge from the domain model. The DMT will therefore expose not only a concept model, but also a process model that specializes MM and that presents to the user a set of components grouped in a domain-meaningful way and with a graphical appearance that makes sense for the domain. Doing so implies, first, understanding which type of mashups the platform should support and, then, tailoring the mashup platform to the specific domain. To this end, the next chapter presents the implementation related details that implements a generic mashup platform following the mashup-meta model.

\newpage

% Chapter 5: Third contribution of the thesis.
% !TeX root = ../phd-thesis-arxiv.tex

%%% Chapter start. %%%
\chapter{Domain-Specific Mashup Platform Development}
\label{sec-contrib03}

\section{Overview}
In the previous chapter we have presented the methodology for the development of a domain-specific mashup platform. The methodology clearly separate domain-independent concerns (i.e., in the form of \emph{mashup meta-model}) from what of domain-specific ones (i.e., \emph{domain-specific mashup meta-model} and its extensions). A mashup platform whose development follows the defined methodology initially stays empty in terms of domain-specific knowledge (i.e., terminologies, concepts, rules, activities etc.) that is then injected tailoring it to a domain-specific mashup tool. That is how we enable generic platform to be tailored for a specific domain. In this chapter we present how we developed the domain-specific platform that will be then tailored for our reference domain (that is presented in the next chapter). However, in this chapter we specifically focus on the technical concerns and technological design decisions that we have taken. 

The mashup meta-model proposed in the previous chapter explains well the capabilities a mashup platform can offer whose implementation follows the specified model. For example, the provision of \emph{universal integration} is achieved through the support of components those can be of type service, UI or data, that is, the platform from its architectural design supports this capability. The \emph{easy-of-use} feature, that is to effectively enable non-technical users in development, is achieved through via no-complex mapping concept. \emph{Intuitiveness} is achieved via introducing domain-specific syntax for composition constructs. These are all the fundamental characteristics that help our platform to provide an effective end-user development environment. In this chapter we not only present steps in the development of a mashup platform that should on one side reduce the complexity of creating mashups for non-technical users but on the other side support developers in the process of developing new components. Primarily focusing on domain-specific mashups greatly help us achieve these objectives. 

To this end, we first present our baseline mashup engine that is comprised of various modules, which are explained later in this chapter. We aim for a very lightweight yet powerful mashup engine that can easily run in web browsers, that is, at the client-side with no need to download any extra software. For this purpose, the engine is implemented using JavaScript language and runs at client-side in a web browser. The choice of using JavaScript language over other languages is highly motivated by the fact that most Web 2.0 Ajax based web applications whose major goals are to offer fast interactive yet attractive designs and user interfaces, use client-side languages like JavaScript. 

\section{Components \& Compositions Execution Insights}

Before describing the technical details of the mashup platform, we first present a few design aspects that must be considered in order to get maximum benefit of our mashup meta-model. Just to clarify a few terminologies, we refer a "composition" to a set of components connected together to make a mashup. Components are fundamental units contain presentation or application logic and perform certain operations. They usually take some input and generate some output (i.e., the result of their operation). Several components can be connected, so that the output of one component serves as input for another component, which forms a mashup composition (i.e., also referred as a mashup or simply a composition). In the next sections, both "mashup" and "composition" terms interchangeably used, though representing the same meaning. 

\subsection{Orchestration style} Of the many important aspects, orchestration is a key aspect to be considered prior to the development of a mashup platform. Generally, the orchestration, where multiple complex computing units involved, manages their coordination while their execution. Similarly, in our case, it specifies how to synchronize the execution of components in a composition making better coordination among them. Mostly, there are three prominent approaches that have been adopted \citep{yu2008understanding}: 

\emphbf{Flow-based} approach maintains orchestration as sequencing of components that is also a kind of flowchart based approach, where multiple units (e.g., components) connect together whose execution happens according to a defined sequence. 

\emphbf{Event-based}  approach offers a publish-subscribe way, where pub/sub models maintain synchronous behavior among components. When a component behaves like publisher sends messages to a queue which then consumed by all those components (subscribers) who are interested.

\emphbf{Layout-based} approach place components in a composition into a common layout that then each component's behavior is specified individually by accounting for the other components' reactions to user interactions. 

We use a combination of flow-based and event-based approaches. That is, generally components execution takes place following flow-based style, however, components' operations can subscribe to the various data buses that then received required data when an event triggered. 

\subsection{Data-passing style}
The choice of data-passing approach is another pivotal aspect, which describes the behavior through which data flows among various components. This important property alone can be used to effectively distinguish among various mashup tools, especially when the target end-user belongs to a non-technical user class. Mainly two approaches have been followed in the past, which are \emph{data-flow} and \emph{blackboard-based} \citep{yu2008understanding}. 

According to the \emph{data-flow} approach, actual data flow from a component to another component. A component starts its execution upon receiving data its waiting for, and once the execution completes it sends the data to the next component in the flow. The data-flow approach considered more intuitive for non-technical users as it follows the philosophy of a natural workflow in daily life work. On the other hand, according to the \emph{blackboard} based approach data is written to variables, which serve as the source and target of an operation invocation on components, much like in programming languages. 

In our case, we follow the data-flow based approach. For example, a data source component produces data (after fetching from a database/web service etc.) and hand over it to the next connected component that then consumes it for further processing. To convey the execution status of a component that would also reflect a composition execution status, we aim to present to end-users the execution status of individual components. Another aspect related to the data-flow approach, which we describe later in this chapter, is that sometimes on the background instead of passing the actual data we pass control data. This scenario gets activated for components whose implementation is of a web-service type. So far we have presented the different types of components, while we will present how these components can be implemented (i.e., as a web service, or as a client side implementation) later.

\subsection{Compositions execution}
A mashup composition, which comprises of several connected components, executes to achieve its goals. The execution of a composition means, running its components in an order that is defined by the end-user. The general execution semantics of a composition/mashup is described in section \ref{sec:gen-exe-semantics}. However, in this section we look at whether the execution follows an instance-based or a continuous approach \citep{yu2008understanding}. An instance-based model is the traditional service composition model, in which a certain kind of message's arrival activates a new instance of the composition, and the system executes the instance within the same main thread and context (much like a program run). On the other hand, the continuous model has one instance per component in a composition model. Each component works as a thread, processing the input data feed and transforming or filtering it to generate the output. The strategy we follow is continuous model based, that is, to allow various components to execute using their own threads and also communicate between them when required.

%\subsection{Mashup Engine Characteristics}
%Our reference scenario is data driven, which means that our mashup platform mainly has to support the retrieval, processing and presentation of data. We can therefore identify the following characteristics for our engine:
%
%\begin{itemize}
%\item A component can either be of type data (DA), application logic (AL), or user interface (UI), depending on whether it acts as a pure data source, a component containing or providing access to application logic, or a component that also provides a GUI to users. The engine and the underlying conceptional model have to take into account all three component types.
%
%\item Compositions are made by expert users for themselves or other users familiar with the domain. Therefore, we require the composition output to be a user interface which presents the results to the user in an suitable and understandable way. This will be realized through UI components.
%
%\item As already mentioned, we mostly have to do with data integration, conversion and presentation. For that reason our approach is data-flow oriented. Each component accepts some data, performs its operations on it and passes the resulting data to the next component or displays the result. Concerning domain modeling, the components implement the atomic domain processes and compositions the composite processes.
%
%\end{itemize}

\section{Components Definitions}
\begin{figure}[t]
 \centering
   \includegraphics[width=1\columnwidth]{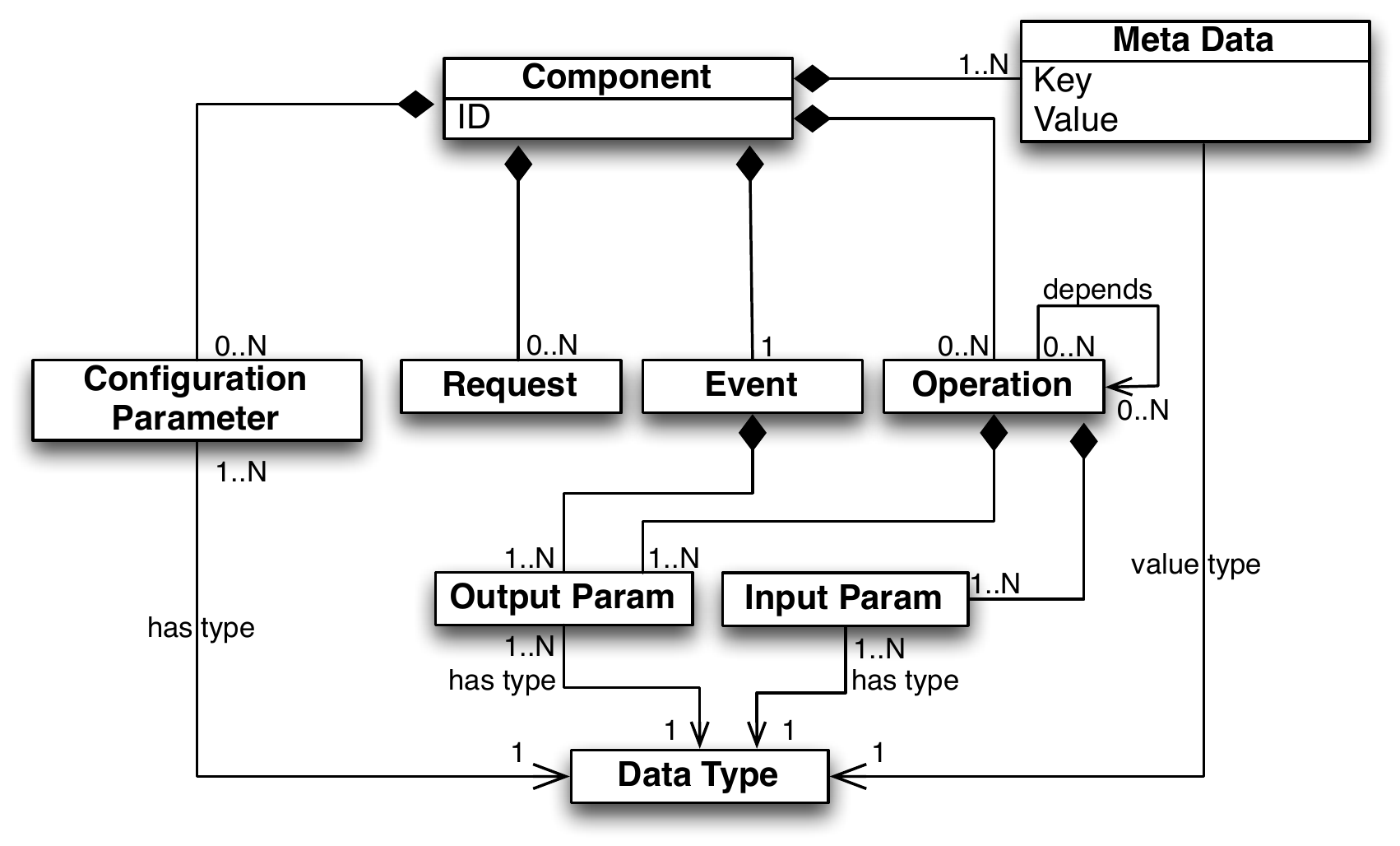}
 \caption{Model Representing Component Definition Language (CDL)}
 \label{fig:cdl}
\end{figure}

Given the above insights, now we first detail on \emph{Component Definition Language (CDL)}, which is build based on components capabilities described in the mashup meta-model. CDL represents just the technical version of what a component is defined by the mashup meta-model. As components are the main building blocks of a mashup, they consume data, perform certain actions/manipulation and produce results.  From a software development point of view, a component can be seen as a function or method. It can take one or more input and produce one or more outputs. The input might come from another component or from an external service or from component's own UI. Furthermore, a component might require direct user interaction and provide a corresponding UI (i.e., configuration UI). Considering this idea, we show in Figure \ref{fig:cdl}, the component definition language model and in Figure \ref{fig:comp-comm} depicts the component communication mechanism. In following we elaborate the details of both aspects.  %define a model with a minimal set of characteristics (see also Figure 5) required for a component to be operable in our environment.

\subsection{Component Definition Language (CDL)}

From a technical point of view CDL is comprised of the following elements to build a component:

\begin{itemize}
\item \emphbf{Operations}: A component exposes a set of actions that it can perform by means of operations. Operation can be seen as the input configuration ports (i.e., IPT's) as defined in the mashup meta-model. Operations are invoked through events (described below) and can accept one or more parameters as input and can produce one or more output parameters. Each parameter can have a certain data type. The generic model does not constrain the types, although the mashup tool can restrict possible types that is from a domain concept model (DCM) in the form of an XSD. To complete its computation, a component might need to have operations to be called in a certain order. Therefore, operations can be dependent on each other. Ideally, an operation should expect only one input, to make its purpose more intelligible to the composition designer. This does not necessarily restrain the capabilities of components: A multi-input operation can be split up into several dependent operations.

\item \emphbf{Events}: Events are the way to propagate results of a component's action to other components. Events implement the output ports (i.e., OPT) defined in the mashup meta-model. They are either generated programmatically, for example after an operation is completed, or through the user interacting with a component's UI. Like the input for operations, the output contained in the event data should conform to one or more data types. Creating a composition mainly consists of connecting events with operations which accept the same data types (i.e., domain concepts).

These two concepts, operations and events, are general enough to cover any kind of interaction between components as they are essentially a mixture of the Observer pattern \citep{vlissides1995design} and the more general Publish/Subscribe pattern \citep{eugster2003many}. These are common patterns used in Model View Controller (MVC) architectures, providing a way to decouple different parts of an application and make them easily exchangeable. From this point of view, applying such a concept seems to be a logical step: the components are the different parts of a composition (the application) and they need to be highly exchangeable due to the dynamic nature of mashups.

\begin{figure}[t]
 \centering
   \includegraphics[width=1\columnwidth]{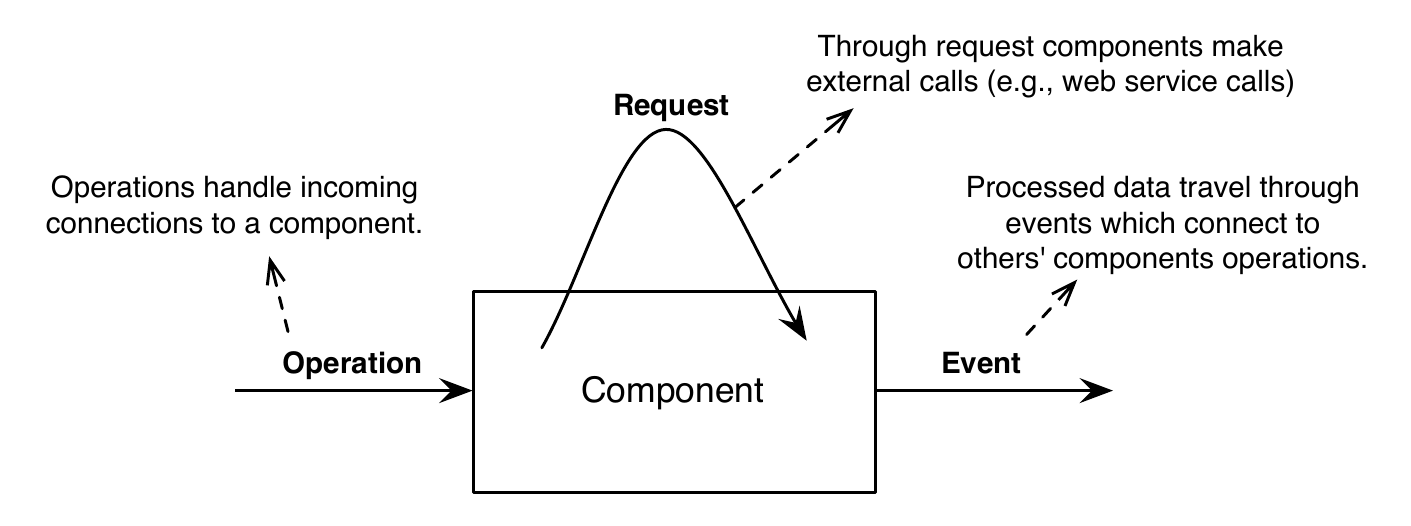}
 \caption{Component Communication}
 \label{fig:comp-comm}
\end{figure} 

\item \emphbf{Requests}: In a composition components send data to each other through operations events connections, but this is not the only type of communication that happens in a composition, instead, often times components communicate with external services or API to fetch or to process data. In the previous chapter we described that components can be of an information source type that act like data sources or can be of type information processor that implement business logic. For these kinds of components it is common that they call external services to accomplish their task. External interactions can be triggered from an operation or from a component's UI (i.e., against a user UI interaction). Although these interactions can be seen as internal calls without the relevance to the environment, we think that a formal specification of this interaction can be useful for the platform provider and lead to a more comprehensive specification of the component interface. We call this characteristic a \emph{request}. The model is deliberately kept universal in this regard and does not require a detailed specification of the possible type of a service or the request and response formats, though the current implementation expects requests to be executable by means of Ajax. Interpreting the response is in the responsibility of the component implementation.  

\item \emphbf{Configurations}: Each component can have a set of configuration parameters (Configuration ports (i.e., CPTs)) according to the mashup meta-model. Users use these configuration parameters to configure components through the component's UI. These parameters must have a data type that can be a primitive or a platform specified type (e.g., domain type) and each parameter value (i.e., user supplied value) of a component belongs to that specific instance of the component and the composition, hence not shared among other compositions.

\item \emphbf{Meta-data}: Finally, a component can have an arbitrary set of meta-data associated with it, typically in the form of key-value pairs. This can be leveraged by the mashup platform to store necessary, platform specific information. For example, the description and type attribute of a component can be defined using the meta-data feature. 
\end{itemize}

\subsection{Component Definition Language in Action}
Listing \ref{list:cdl} shows a simplified version of the model definition of a component, which is responsible for retrieving a list of researchers based on its configuration. The component has a fully qualified name as ID, a descriptive name (line 1) and a more detailed description (line 3). Furthermore, it has four configuration parameters (line 5 to 25) , {\tt sectionId, uniId, departmentId} and {\tt facultyId} defined using {\tt config} tag. These configuration parameters are the examples of filters, which restrict the result-set using various filtering criteria. Each {\tt config} parameter definition contains further information for displaying the configuration fields, like the {\tt label} \& {\tt Sector} under {\tt option} tag. There are various types of options that can be set, for example, as in {\tt sectionId} configuration parameter (line 6 to 12). These include:

 - {\tt label} defines a label using its {\tt value} attribute.

 - {\tt renderer} renders a UI field based on selected renderer. In this case a rendered of type {\tt jsm.ui.input.Autocomplete} is used.

- {\tt url} specifies a url of an external service if the data has to be fetched from it.

- {\tt search parameter} is like the query string value in web service calls.

- {\tt value} specifies what field used as value-field in the UI.

- {\tt display} specifies the display field used to populate a UI field (e.g., a text field)

We explain how all these parameters work collectively in section \ref{sec:config:interface}. Moreover, this CDL example only expends first configuration parameter details for the purpose of conveying the understanding.

The event (line 17 to 19) returns a collection of researchers, denoted by its data type. It gets triggered by the request (line 21 to 29), which connects to some service with the given url, sending the configuration parameters, denoted by the {\tt name} syntax. Presuming the web service correctly returns a list of researchers, no further implementation has to be provided by the developer. With this definition, the engine generates a generic configuration interface and can manage the request to the web service automatically. Apart from simplifying the development process for data source components, it also completely hides the technical details of service calls from the composition designer. 

This example also shows the usage of meta-data in the definition. The conceptual model does not require a name or description, but it can still be provided through meta-data. That means, the implementation of the component does not process this data but other routines can access it. For example, the user interface of the mashup tool can use this data to present more information about a component for a better user experience. In this example, we do not show the details about how a service call can be configured. However, section \ref{sec:dataprocessor} provides in-depth details of this aspect.

\begin{lstlisting} [caption={A component definition following CDL}, label={list:cdl}]
<component id="org.reseval.ItalianResearchers" name="Italian Researchers">
    
    <description>Gets a list of Italian researchers, optionally filtered</description>
    
    <config ref="sectionId">
        <option name="label" value="Sector"/>
        <option name="renderer">
            <option name="type" value="jsm.ui.input.Autocomplete"/>
            <option name="url" value="http://example.com/italianSource/sector/name/autocomplete"/>
            <option name="search_parameter" value="input"/>
            <option name="value" value="{id}"/>
            <option name="display" value="{name}"/>
        </option>
    </config>
    <config ref="uniId">
        <option name="label" value="University"/>
	...
    </config>
    <config ref="departmentId" dependsOn="uniId">
        <option name="label" value="Department"/>
	...
    </config>
    <config ref="facultyId" dependsOn="uniId">
	...
    </config>
   ...
    <event name="Researchers loaded" ref="researchers_loaded">
        <output name="researchers" type="Researcher[id][name][masID][dblpID]" collection="true"/>
    </event>
    
    <request name="Get Researchers" ref="get_researchers" triggers="researchers_loaded">
        <url>http://example.com/italianSource/getResearchers</url>
        <parameters>
            <parameter name="uniID" value="{uniId}"/>
            <parameter name="facID" value="{facultyId}"/>
            <parameter name="depID" value="{departmentId}"/>
            <parameter name="secID" value="{sectionId}"/>
        </parameters>
    </request>
...
</component>
\end{lstlisting}

\section{Mashup Compositions Definitions}
Given the component definition language, we now present \emph{Mashup Definition Language} (MDL), a technical version of what a mashup is defined by the mashup meta-model. The mashup definition language provides a way in which a mashup composition can be defined in terms of components, connections among them through pipes, their states (i.e., parametric values they hold), their instances information and the layout information. We pursue the same goal for the mashup model as for the component model: A minimal set of characteristics that is necessary to represent a functional composition. 

In essence, a mashup composition, which is formed using multiple components, is defined connecting events (i.e., output ports) with the operation (i.e., input ports). An event emits data which passes through a pipe and finally consumed by an operation. Components in a composition may hold a user-defined configuration parameters. A composition contains its layout information, which is then used to render components to their proper layout and position. So basically an MDL that is capable to accommodate the above mentioned snippets of information is suitable for our purpose from a technical point of view. 

\subsection{Mashup Definition Language (MDL)}
Technically, based on the MDL a mashup is comprised of three basic things, as described below:

\begin{itemize}

\item \emphbf{Components} Many components form compositions hence play major role in building mashups. A composition can have multiple components connected together. The MDL defines connections among components in terms of source and target components. A source component is the one whose event is connected to another component's operation that is called a target component in this case. Each component assigned an instance id in a composition along with its full qualified name. Other information like a component's configuration details also preserved in the MDL. 

\item \emphbf{Connections} When two components connect, resultantly form a connection. A connection information in the form of source and target components is maintained by the MDL. That is how the association between events and operations of two components is maintained. That is then this information used by the mashup engine to work as a publish-subscriber approach to hand over data to the target operation emitted by the source event.

\item \emphbf{Meta-Data} Mashup definition language also permits to define arbitrary parameters in the form of meta-data (i.e, specifically key-value pair). These parameters can be used to define some special cases such as mashup composition state, permissions. This also provides a way of extending the definition language with details which are not anticipated yet.

\end{itemize}

\subsection{Mashup Definition Language in Action}
Although, an MDL is an internal document of the mashup platform, even then describing its details would further help in case of an extension to the platform if needed. Listing \ref{list:mdl} shows a short sample definition of a mashup composition which consists of only two components and connection between them, just for the sake of understanding. 

\begin{lstlisting} [caption={Example of a composition using composition definition language}, label={list:mdl}]
{
...
  "components": [
    {
      "instance_id": "2",
      "component_id": "org.reseval. ItalianResearchers",
      "config": {
        "facultyId": {
          "value": "",
          "display": "All"},
        "departmentId": {
          "value": "82",
          "display": "INGEGNERIA E SCIENZA DELL INFORMAZIONE- DISI"},
        "uniId": {
          "value": 83,
          "display": "TRENTO"},
        "sectionId": {
          "value": "",
          "display": ""}},
      "data": {
        "name": "DISI Researchers",
        "minimized": false,
        "position": [
          21,
          56]}},
    {
      "instance_id": "4",
      "component_id": "org.reseval.MAS",
      "config": {
        "endYear": {
          "value": "2010",
          "display": ""
        },
        "startYear": {
          "value": "2008",
          "display": ""}},
      "data": {
        "name": "MAS",
        "minimized": false,
        "position": [
          218,
          55]}}],
  "connections": [
    {
      "source": "2",
      "event": "researchers_loaded",
      "target": "4",
      "operation": "set_researchers"
    }
...
 "data": {
	"public":true,
	"name":"DISI-ItaliaEvaluation-MASBased",
	...
	}
\end{lstlisting}

As one can notice that the presented MDL lists down the details of the components used in the composition. Just like a database table, each component with its various attribute represents a tuple. These attributes, to name a few include a component's configuration parameters and their values, component's position in the overall mashup layout, its UI status like minimized or not etc. 

Listing \ref{list:mdl} shows a composition of two components in JSON\footnote{http://json.org/} format. The Italian Researchers component (line 5 to 25) we showed in the previous example, and the Microsoft Academic component (line 26 to 42), which accepts a list of researchers and adds a list of publications to each researcher. Of the other parts of this mashup definition, the components, the connections and further meta-data are the important ones. For each component, the MDL stores {\tt instance-id} (line 4), {\tt component id} (line 6), {\tt config} (i.e., the configuration parameters and their instance values) (line 7 to 19), {\tt name} (i.e., component's name) (line 21), and layout {\tt position} (line 23 to 25). The instance IDs remain unique within a composition. Whereas a component ID is the one given by component developer and it remains unique among all other components in the platform. 

The important information about connections described for this composition on line 43 to 49. For each individual connection the MDL preserves source component's instance ID (line 45), its event name (line 46) and the target component's instance ID (line 47) and its operation name (line 48). Other information about a composition's name and its visibility status is defined using {\tt data} tag (line 51 to 53).

%\begin{figure}[t]
% \centering
%   \includegraphics[width=1\columnwidth]{generic-architecture.pdf}
% \caption{Overall architecture of the platform}
% \label{fig:achitecture}
%\end{figure} 

\section{The Mashup Engine}
\label{sec:mashup-engine}

Given the CDL that defines components and the MDL that defines mashup compositions (or mashups), we now pursue for a mashup engine that allows the development of components following CDL, composing mashup compositions and finally running those compositions following MDL. The mashup engine must be able to incorporate the above described aspects like orchestration and data-passing style. We aim for data-flow paradigm as a general approach, which must also be conveyed and understandable by the end-users, and sometimes we use control-signals to decrease the data passing overhead hence to increase the overall performance. However, mashup engine's decision on when specifically data or control-signal flows, is described in the next chapter there we first introduce necessary concepts for its understanding.

The \emph{Mashup Engine} is a core part of the platform, which manages various modules and all communications that take place among these modules. One of the main objectives, which drive along the development of the mashup engine, was to keep separate platform or environment specific requirements to that of a mashup tool's specific ones (i.e., domain-specific) and in parallel to provide a consolidated platform that can easily be tailored to a specific domain. That's the reason, throughout the elaboration of various steps of the mashup engine, which we described in the next sub-sections, we mainly focus on those set of generic aspects whose design and implementation is not dependent but of course inspired of a domain. This allows us to use the engine for other domains with similar characteristics as of our reference domain (i.e., research evaluation). 

\begin{figure}[t]
 \centering
   \includegraphics[width=1\columnwidth]{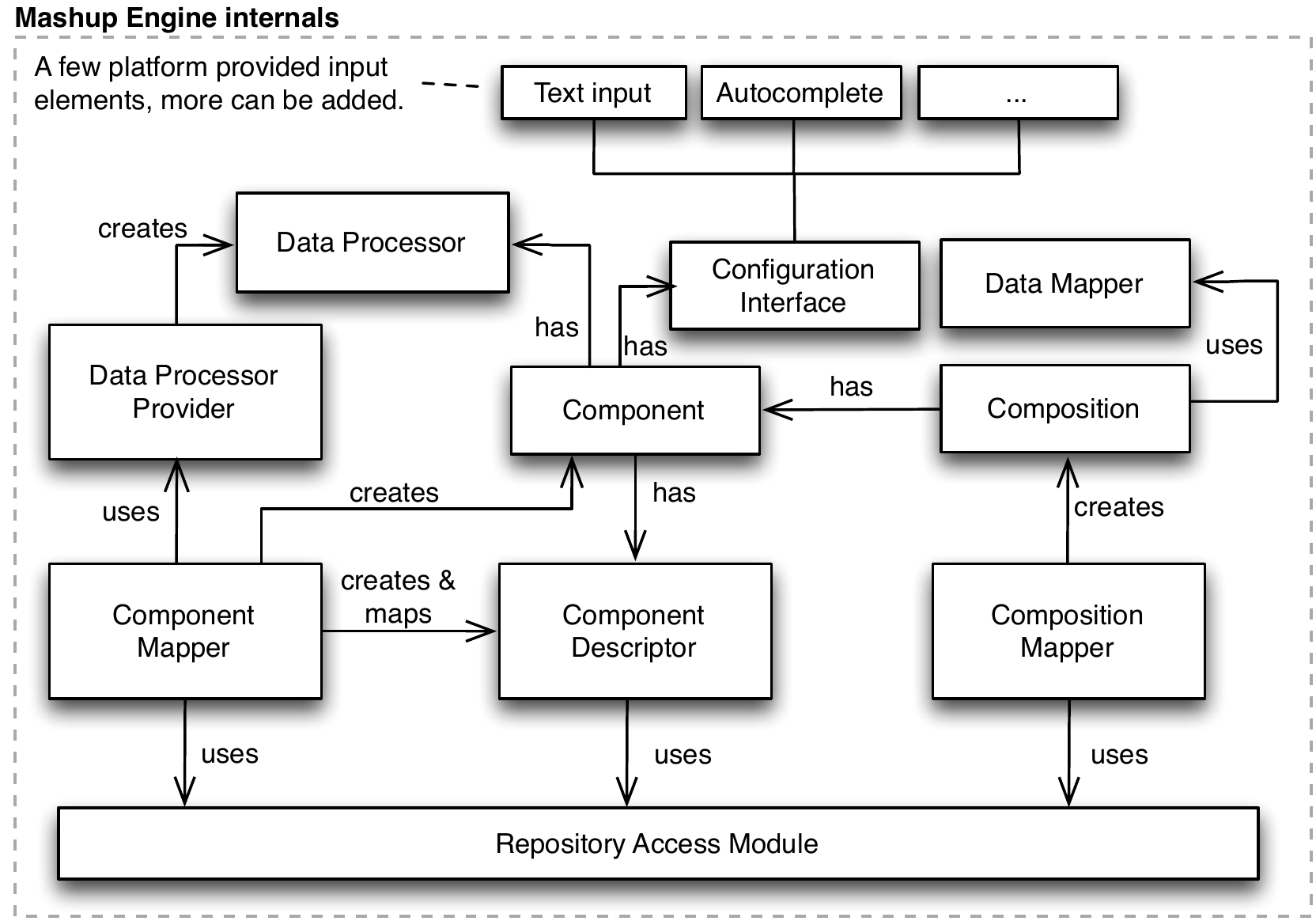}
 \caption{Mashup Engine Internals: various modules inside mashup engine and their interactions}
 \label{fig:engine}
\end{figure} 

\subsection{Mashup Engine Architecture}
The mashup engine is designed and developed for the client side technologies (e.g., web browser). JavaScript and the Goolge Closure Library\footnote{https://code.google.com/closure/library/} were the languages used for the development. In following, we describe the main modules, their roles and relations to each other. Figure \ref{fig:engine} depicts an overall architecture of the mashup engine. The modules presented in the architecture are the main building blocks of the engine. Later, we will present how UI of a mashup tool can interact with them to give information and control to users. Some modules only describe an abstract interface, for which the platform provider has to provide a concrete implementation adapted to the environment of mashup tool.

\subsection{The Repository Module}
The repository module is the one responsible for performing typical CRUD operations (i.e., create, read, update and delete) for managing components and compositions and other external calls that the engine needs to perform. For example, this is the place which is used to access web services that a developer defines inside a component definition. Moreover, the repository module does not fix the way components and compositions are stored to some persistent storage. Instead, the repository interface is designed for the usage of synchronous and asynchronous storage facilities.  For example one can use the HTML 5 local storage API,  or a server side storage accessed with Ajax or any other means. As an overall this gives provision to the platform provider to decide on how and where the storage will be done based on his requirements and possibilities.

In addition to that, the repository module also independent of what representation is used to describe components or composition (e.g., as in our case either XML or JSON). Hence, this task is delegated to component and composition mappers (described below) to transfer whatever representation of components and composition to \emph{component Descriptor} or \emph{Composition} which are then understandable by the components and composition classes.

\subsection{Component- and Composition Mapper}
The mashup engine does not restrict on a specific component, and compositions representational format. This characteristic introduced because there are many other formats that can be used like to name a few famous ones include W3C Widgets\footnote{http://www.w3.org/TR/widgets/} and OpenAjax Widgets\footnote{http://www.openajax.org/member/wiki/OpenAjax Metadata 1.0
Specification Widget Overview} and there might be others to be developed in the future. For this reason we introduced the concept of mapper in the platform that actually maps a particular format to the system's internal one. In the previous sections we showed the components and compositions representations that the mashup engine followed in which the default mappers work.

So, to not restrict the engine to a specific representation, the mapper's perform the conversion to a specific representation into our internal model and vice versa. It might not be possible to map any component description to our component model, but the idea leads to a certain degree of independence and leaves room for extensions. This again leaves the choice which representations to use and support on platform provider.

\subsection{Component Descriptor and Component} 
As described earlier that a  component, which is a basic building block, can be used in several compositions and can also occur more than once in a composition. Therefore, it is necessary to distinguish between the properties shared by all the instances of the component, like the operations and events and those of instance-specific properties, like a composition-specific name, it's id, configuration settings etc.

The \emph{component descriptor} and \emph{component class} are used to keep track of these aspects of a component. Basically, the component descriptor is a software artifact that represents a component's model definition, which is stored in its CDL document. The information that a component descriptor gets populated after parsing component's CDL include operations, events, requests and configuration parameter details of a component. An instance of the component class is generated to represent this and component's instance-specific information like the values of configuration parameters, instance id, and it also includes the basic execution logic needed to run a component's action. 

% A component descriptor is a container for the information extracted from the component model, in particular it holds the information about which configuration parameters, operations, events and requests a component has. A component instance is derived from a specific component descriptor and contains the instance specific information, the configuration interfaces if one is defined and the basic execution logic which is needed to perform the component's actions, that is, executing operations and emitting events. It also contains the specific implementation logic a developer may define. This may also include a UI interface if the developer specified one. 

The user interface of the mashup tool directly interacts with component descriptor and component instances to provide information about them and let the user manipulate them through the configuration interface and the UI of a component.

\subsection{Composition} 
Compositions represent user defined workflows that accomplish some tasks. As compositions comprised of multiple components connected together in an order to interact with each other by means of connections made between events and operations. The composition module task is to register those connections and performs the communication between components. Basically, the composition listens to each component's event and notifies the connected components. It's also responsible for passing the event's emitted data to the operation. Finally, in order to inform the users about a possible state of a composition while running, the composition module is used by the UI of a mashup tool to convey such indicators. 

%It is the composition's responsibility to keep track of these connections and manage the communication between the components when the computation started. The composition is listening to each component's events and notifying the connected components, passing along which operation has to be executed and the arguments from the event. The UI of the mashup tool can interact with Composition objects in various ways and provide means to modify a composition or track its progress while its execution.

%\subsection{Extensions} 
%There are a couple of modules which are optional, that is, they are not necessary for the basic functionality of the engine. However, with them, the platform provider can extend the basic functionality with his own application logic. This is how we inject our domain-specific requirements and handling of data into the engine, which is explained in the next section. These modules are listed under extensions, but this should not be mistaken for a lack of importance. In fact they are as essential for our purpose as the core modules.

\subsection{Data Mapper}
During a composition's run-time, all components that belong to a composition and are connected with each other communicate by means of sending and receiving data among them. Basically, the actual communication takes place between an event and an operation. Event emits data and operation consumes it. The data mappers are responsible for the conversion of the data received from the event so that the data is understandable by the operation. Data mappers would not worthwhile for components those understand and built based on and for a specific domain, for example, in our case many components understand our reference domain DCM (e.g., XSD), but components such as bar charts, pie charts, and other visualization components understand a predefined data-format, which depends on what visualization API is used (as we use Google Charts API). Without the conversion functionality a new instance of a same visualization component would be needed that specifically implements a component's specific data visualization requirements that ends in developing too many new UI components.

This is the reason, we use data mappers to encapsulate the data conversion logic so that the data is converted in a format that is understandable by the target component. The current implementation of the data mappers follows our current domain model and hence the conversion is automatically performed for such components. Simply, a new data-mapper will be needed in case of a different domain.

\subsection{Data Processor}
\label{sec:dataprocessor}
\begin{figure}[t]
 \centering
   \includegraphics[width=1\columnwidth]{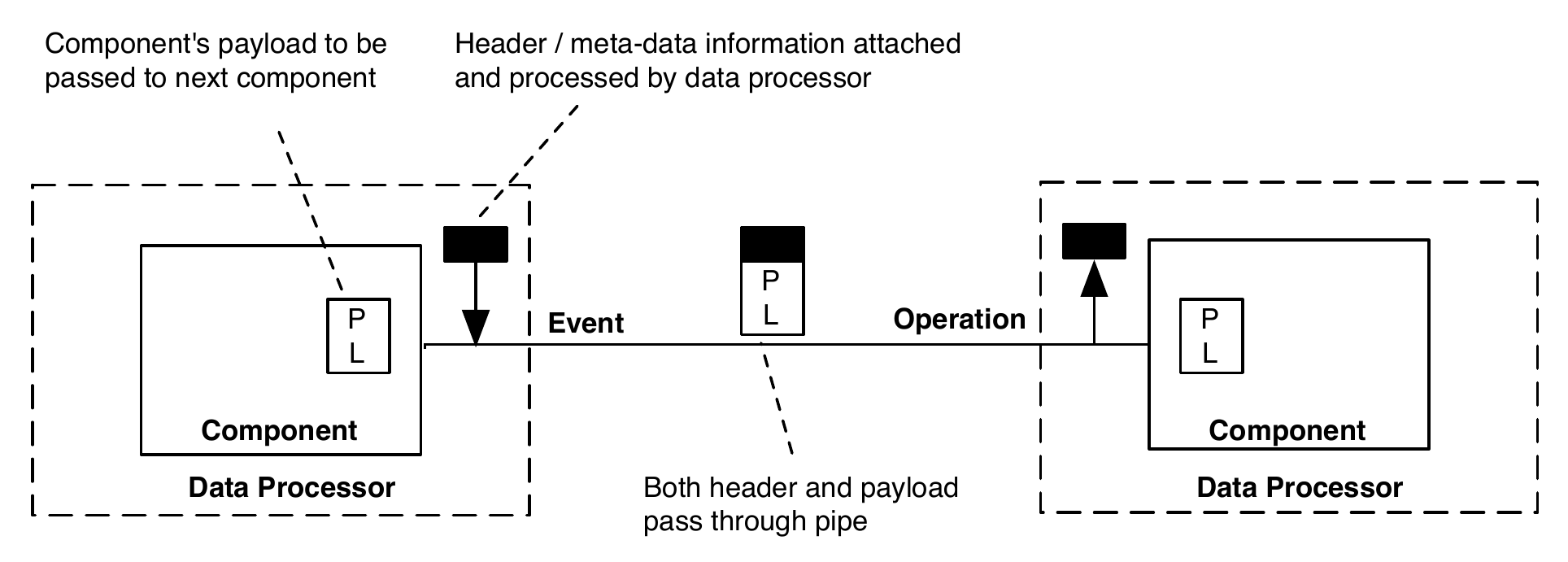}
 \caption{Message passing between component with payload and header information}
 \label{fig:msg-passing}
\end{figure}

To perform tasks a component is responsible for, component may require to call external services to fetch data or to perform computations. Also, at run-time mashup engine needs to inspect the data that is being transferred between components for various reasons, like, to read/write meta-data, calling external services, conversion of data etc.  For this reason, data processors provide a way to the platform to intercept any communication that a component initiates to either interact with other components or to call services. The data processors allow for pre and post processing of the data that components use. Figure \ref{fig:msg-passing} depicts how a data processor can be seen as a wrapper around a component. Another advantage of the data processors is for the component developers. A component developer can develop and focus on components related implementation concerns without having to worry about platform-specific requirements or conditions. That means, it also increases the reusability of components those belong to the same domain and share similar characteristics.  

To further understand the potential of data processors, in following we describe the details of how components communicate with each other having the data processor in place to intercept their calls. As explained already that a composition keeps the information about connected components in terms of events and operations references. These events and operations exchange messages. A message, which is passed from an event to an operation, consists of two parts, i.e., a {\tt body} and a {\tt header}, exactly as HTTP header and body means to HTTP protocol. The header contains meta-data information, whereas the body contains the actual payload. In our case, the header contains arbitrary meta-data that is not of concern to a component's implementation. Whereas, the data in a message's body populated by the data an event emits. A component implementation can only access to the body of a message not to the header as it's only understandable by the platform.

It is then a data processor's responsibility that it intercept the communication and process header section accordingly. Data processors make different components to synchronize themselves if they understand the header information. Figure \ref{fig:msg-passing} depicts the communication between two components along with different stages a message passes through. The figure shows, a white box inside written PL (i.e., payload, the actual data), a black box represents header. This is also the point where the mashup engine decides about switching data-flow approach (the default approach) to control-flow approach. We explain the switching mechanism in the next chapter in section \ref{sec:df-cf-flow} after conveying the necessary concepts that are required for its understanding.  

%As data processors are able to intercept calls to operations and events of a component, they can inject and extract data into and from the header of a message. This allows data processors of different components to synchronize themselves if they understand the header data. An illustration of this communication is shown in Figure \ref{fig:msg-passing}, where the black box represents header data and the white box the data processed by the components (i.e., payload).

%This gives the platform provider an unobtrusive way to tailor the functionality of components to his needs. In the next chapter, in which we describe the ResEval Mash mashup tool, we will explain how we facilitate data processors to transform a data-flow oriented mashup into a control-flow oriented one.

\subsection{Configuration Interface}
\label{sec:config:interface}
A mashup platform whose target users are non-technical, must offer a user-interface that is intuitive, easy to use and consistent. Less-skilled users tend to prefer high intuitiveness, with relatively less but consistent UI elements to play with configuration settings. We noticed that some mashup tools come with too complex user interface options (e.g., Yahoo Pipes) and others with too limited options (e.g., mashArt). However, a right balance is preferable between the two extremes. 

To this end, one possibility to let components developers decide entirely on what type of configuration interface a component should provide. The problem with this approach is, it could lead a component's configuration interface to arbitrarily complex level, and it might result in major inconsistencies with other components. In order to solve all these issues, the choice we made is to provide a basic set of UI configuration elements along with the parameters the input needs. Component developers can choose among the provided set of input elements, or in case if a new element is needed, then he can add it too.   

In listing \ref{list:config}, we present an example of two configuration input elements. The first element (line 2 to 11) is a simple text-field, which is connected with a web service to provide auto-complete. This text field offers auto-completion of the text when a user starts typing. The second input element (line 13 to 25) generates a drop-down field and also connected with a service, that is where it gets data from. One can notice that the second input element is dependent on the first element, which is set using {\tt dependsOn} attribute (line 13). That means, the {\tt departmentId} element gets updated whenever the {\tt uniId} element will be changed. The auto-updation is achieved through the {\tt value} and the {\tt display} tags in {\tt departmentId} element, both with values {\tt \{id\}} and {\tt \{name\}} respectively (line 18 \& 19). The use of curly braces makes a variable that is then accessible throughout a model. As both {\tt id} and {\tt name} are already defined by the {\tt uniId} element so it is possible to use them anywhere in the component's definition. 

Most of the remaining options have been already explained in earlier sections. The engine includes a few basic input fields and it is possible to add new elements as required. If a certain type is not available or none is specified, a simple text field is used instead. 
%The input type is specified using the {\tt type} option. The {\tt name} is a fully qualified name, similar to the name of the component itself. This particular input field, jsm.ui.input.Autocomplete, renders a text field and opens a list of suggestions when the user starts typing (see Figure \ref{fig:uni-autocomplete}). The suggestions are loaded from the provided URL. The other options define how to map the incoming data from the web service to the data the field needs. 
%\begin{figure}[t]
% \centering
%   \includegraphics[width=.50\columnwidth]{uni-autocomplete}
% \caption{An autocomplete field created with the configuration using Listing \ref{list:config}}
% \label{fig:uni-autocomplete}
%\end{figure}

\begin{lstlisting} [caption={Configuration interface definition}, label={list:config}]

<config ref ="uniId ">
	<option name="label "value="University"/>
	<option name="renderer ">
		<option name="type " value="jsm.ui.input.Autocomplete " />
		<option name="url " value="http:// ... /university/autocomplete" />
		<option name="search_parameter" value="input" />
		 <option name="value" value="{id}"/>
            <option name="display" value="{name}"/>
	</option>
</ config>
	...
<config ref="departmentId" dependsOn="uniId">
        <option name="label" value="Department"/>
        <option name="renderer">
            <option name="type" value="jsm.ui.input.Dropdown"/>
            <option name="url" value="http://.../{uniId}/departments"/>
            <option name="value" value="{id}"/>
            <option name="display" value="{name}"/>
            <option name="default">
                <option name="value" value=""/>
                <option name="display" value="All"/>
            </option>
        </option>
   </config>

\end{lstlisting}

%For this to work, all objects representing an input field have to implement a certain interface defined by the mashup engine. Still, the actual representation of the input field and its complexity is implementation detail. The mashup engine also handles specified dependencies between configuration fields, that is, fields can be informed when the value of another field changes and update their state accordingly.

\newpage

% Chapter 6: Implementation of a framework, tool, etc.
% !Tex root = ../phd-thesis-arxiv.tex
%%% Chapter start. %%%
\chapter{ResEval Mash: A Domain-Specific Mashup Tool}
\label{chp-reseval}

\section{Overview}

The domain-specific mashup platform described in the previous chapter provides a consolidated ground for the development of a domain-specific mashup tool. In this chapter, we present how we have developed such a mashup tool for our reference domain. \emphbf{ResEval Mash}\footnote{http://open.reseval.org} \citep{imran2012systematic} \citep{imran2012resevalmash} is a mashup tool tailored to the research evaluation field, i.e., for the assessment of the productivity or quality of researchers, teams, institutions, journals, and the like. The tool is specifically tailored to the need of sourcing data about scientific publications and researchers from the Web, aggregating them, computing metrics (also complex and adhoc ones), and visualizing them. ResEval Mash is a hosted mashup platform \citep{daniel2012developing} with a client-side editor and runtime engine, both running in a common web browser. It supports the processing of also large amounts of data, a feature that is achieved via the sensible distribution of the respective computation steps over client and server. 

In the following, we first present the important design principles which have been learned from our past works, and also in result of those interactions that we did with domain experts. Moreover, we show how ResEval Mash has been implemented, starting from the domain models introduced throughout the previous sections.

\section{Design Principles}
\label{sec:requirements}
Starting from the considerations that we presented in the section \ref{sec:concepts}, the implementation of ResEval Mash is based on a set of design principles (described below), which we think are crucial for the success of a mashup platform like ResEval Mash. These design principles stem both from the earlier work on this direction \citep{NamounECOWS2010} \citep{casati2012developing} and also from the requirements that we have presented in the section \ref{sec:dsm-req} those gathered from both domain and end-users. Moreover, these are also based on our past experience with the similar problems in the context of the LiquidPub European project\footnote{\url{http://liquidpub.org/}}.

% The first and last principles stem from our prior work and user studies \citep{NamounECOWS2010} in the context of web service composition and end-user development, while the second and third principles respond to domain-specific requirements identified during the  analysis of the domain and are based on our past experience with similar problems in the context of the LiquidPub European project\footnote{\url{http://liquidpub.org/}}.

\subsection{Intuitive graphical user interface}
The user interfaces of development tools may not be a complex theoretical issue, but acceptance of programming paradigms can be highly influenced by this aspect too. The user interface comprises, for instance, the selection of the right graphical or textual development metaphor so as to provide users with intelligible constructs and instruments. It is worth investigating and abstracting the different kinds of actions and interactions the user can have with a development environment (e.g., selecting a component, writing an instruction, connecting two components), to then identify the best mix of interactions that should be provided to the developer. To this end, we built as very simple yet powerful interface of the tool, that implements domain-syntax model to its various visual parts \citep{muhammad2012resevalchi}. That is, the tool visualizes intuitive graphical symbols those of domain-specific nature and easily understandable by domain-experts. 

\subsection{Hidden data mappings} 
In order to prevent the users from defining data mappings, the mashup component used in the platform are all able to understand and manipulate the domain concepts expressed in the DCM, which defines the domain entities and their relations. That is, they accept as input and produce as output only domain entities (e.g., researchers, publications, metric values). Since all the components, hence, speak the same language, composition can do without explicit data mappings and it is enough to model which component feeds input to which other component. 

\subsection{Data-intensive processes}
 Although apparently simple, the chosen domain is peculiar in that it may require the processing of large amounts of data. For instance, we may need to extract and process all the publications of the Italian researchers, i.e., on average several dozens of publications by about sixty-one thousand researchers (as the scenario presented in the section \ref{sec:unitn-scenario} demands). Loading these large amounts of data from remote services and processing them in the browser on the client side is unfeasible due to bandwidth, resource, and time restrictions. Data processing should therefore be kept, especially for this kind of scenarios, on the server side (we achieve this via dedicated RESTful web services running on the server).

\subsection{Platform-specific services}
As opposed to common web services, which are typically designed to be independent of the external world, the previous two principles instead demand for services that are specifically designed and implemented to efficiently run in our domain-specific architecture. That is, they must be aware of the platform they run on. As we will see, this allows the services to access shared resources (e.g., the data passed between components) in a protected and speedy fashion. 

\subsection{Runtime transparency} 
Finally, research evaluation processes like our reference scenarios focus on the processing of data, which -- from a mashup paradigm point of view -- demands for a data flow mashup paradigm. Although data flows are relatively intuitive at design time, they typically are not very intuitive at runtime, especially when processing a data flow logic takes several seconds (as could happen in our case). In order to convey to the user what is going on during execution, we therefore want to provide transparency in the state of a running mashup. 

We identify two key points where transparency is important in the mashup model: component state and processing state. At each instant of time during the execution of a mashup, the runtime environment should allow the user to inspect the data processed and produced by each component, and the environment should graphically communicate the processing progress by animating a graphical representation of the mashup model with suitable indications (i.e., in our case we use different colors to represent different states).

These principles require ResEval Mash to specifically take into account the characteristics of the research evaluation domain. Doing so produces a platform that is fundamentally different from generic mashup platforms, such as Yahoo! Pipes\footnote{\url{http://pipes.yahoo.com/pipes/}}.

\section{ResEval Mash Architecture}
%\subsubsection{Overall Architecture}
\label{sec:architecture}

\subsection{Overview}
\begin{figure}
  \centering
    \includegraphics[width=0.82\columnwidth]{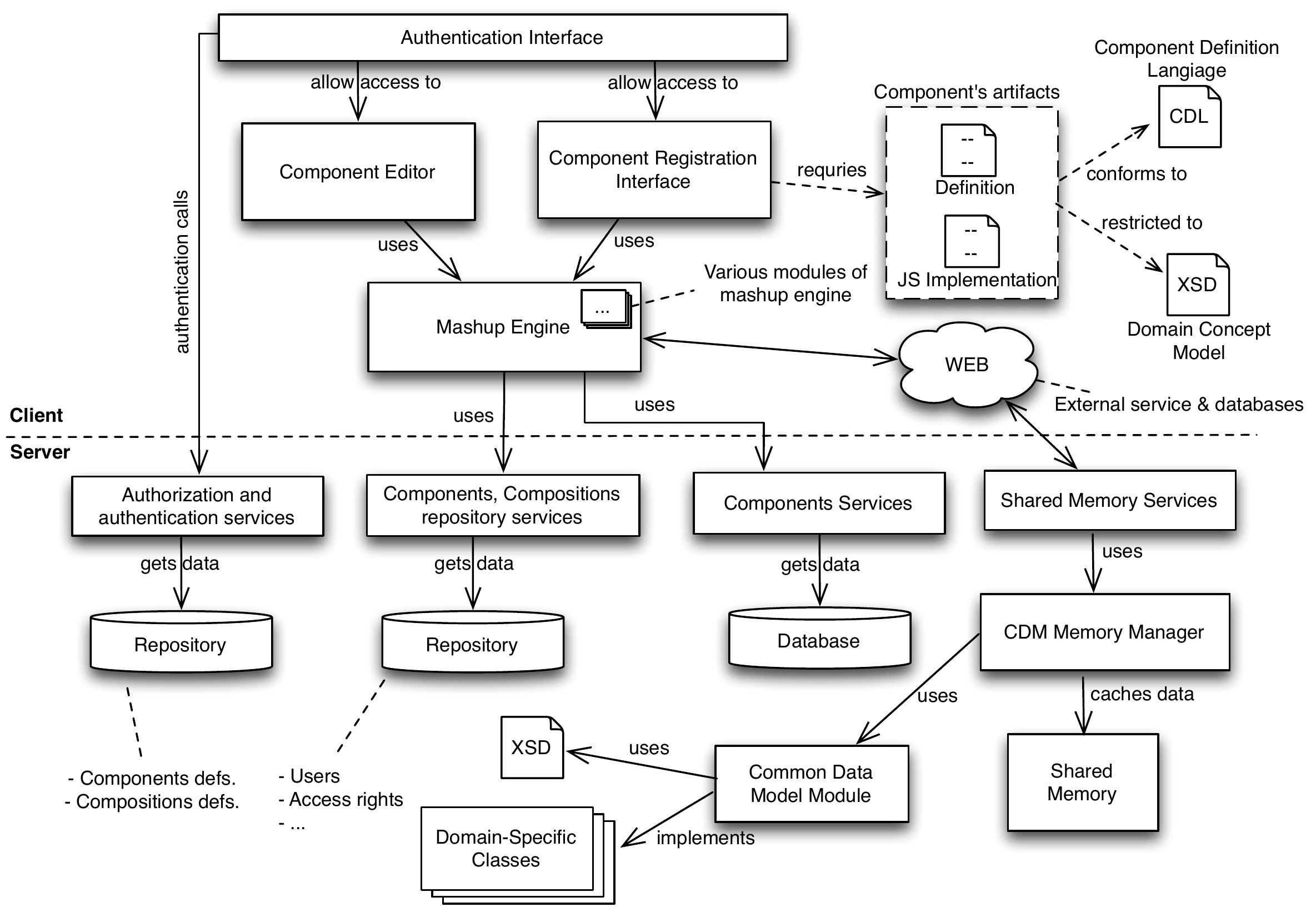}
  \caption{ResEval Mash Architecture presenting its core module both on client and server sides}
  \label{fig:architecture}
\vspace{-4mm}
\end{figure}

Figure \ref{fig:architecture} illustrates the internal architecture that takes into account the above principles and the domain-specific requirements introduced throughout the previous sections: \emph{Hidden data mappings} are achieved by implementing mashup components that all comply with the \emph{domain conceptual model} described in Figure \ref{fig:concept-model}. If all instances of domain activities understand this domain concept model and produce and consume data according to it, we can omit data mappings from the composition environment in that the respective components simply know how to interpret inputs. The \emph{processing of large amounts of data} is achieved at the server side by implementing platform-specific services that all operate on a \emph{shared memory}, which allows the components to read and write back data and prevents them from having to pass data directly from one service to another. To provide users with a mashup environment that has an \emph{intuitive graphical UI} we design first a domain syntax as explained in section \ref{sec:dsm-domain-syntax}, which provides each object in the composition environment with a visual metaphor that the domain expert is acquainted with and that visually convey the respective functionalities. For instance, ResEval Mash uses a gauge for metrics and the icons that resemble the chart types of graphical output components. 

The core of the platform is the functionalities exposed to the domain expert in the form of modeling constructs. These must address the specific domain needs and cover as many as possible mashup scenarios inside the chosen domain. To design these constructs, a thorough analysis of the domain is needed, so as to produce a domain process model as described in section \ref{sec:dpm}, which specifies the classes of domain activities and, possibly, ready processes that are needed (e.g., data sources and metrics).  The components and services implement the \emph{domain process model} i.e., all the typical domain activities that characterize the research evaluation domain. \emph{Runtime transparency} is achieved by controlling data processing from the client and animating accordingly the mashup model in the Composition Editor. Doing so requires that each design-time modeling construct has an equivalent runtime component that is able to render its runtime state to the user. The modeling constructs are the ones of the \emph{domain-specific syntax} illustrated in Figure \ref{fig:syntax}, which can be used to compose mashups like the one in our reference scenario (see Figure \ref{fig:scenario}). Given such a model, the Mashup Engine is able to run the mashup according to the \emph{meta-model} introduced in Section \ref{sec:mm}. The role of the individual module in Figure \ref{fig:architecture} is described as follows:

%
%Next, a set of instances of domain activities (e.g., an h-index algorithm) must be implemented, which can be turned into concrete mashup components. Finally, in order to relieve users from the definition of data mappings, ResEval Mash is based on an explicit domain concept model as explained in section \ref{sec:method} and shown in Figure \ref{fig:concept-model}, which expresses all domain concepts and their relationships. If all instances of domain activities understand this domain concept model and produce and consume data according to it, we can omit data mappings from the composition environment in that the respective components simply know how to interpret inputs.

\subsection{Mashup Engine} 
The most important part of the platform is the \emph{Mashup Engine}, which is developed for the client-side processing, that is we control data processing on the server from the client. We have already presented and described the details of the mashup engine's internal behavior in the previous chapter. However, here we just try to highlight its interaction with those modules that we introduced in the ResEval Mash's architecture. The engine is primarily responsible for running a mashup composition, triggering the component's actions and managing the communication between client and server. As a component either binds with one or more services or with a JavaScript implementation, the engine is responsible for checking the respective binding and for executing the corresponding action. The engine is also responsible for the management of complex interactions among components. A detailed view of these possible interaction scenarios is given later in this chapter.

\subsection{Composition editor} 
\begin{figure}[t]
  \centering
  \includegraphics[width=1.03\textwidth]{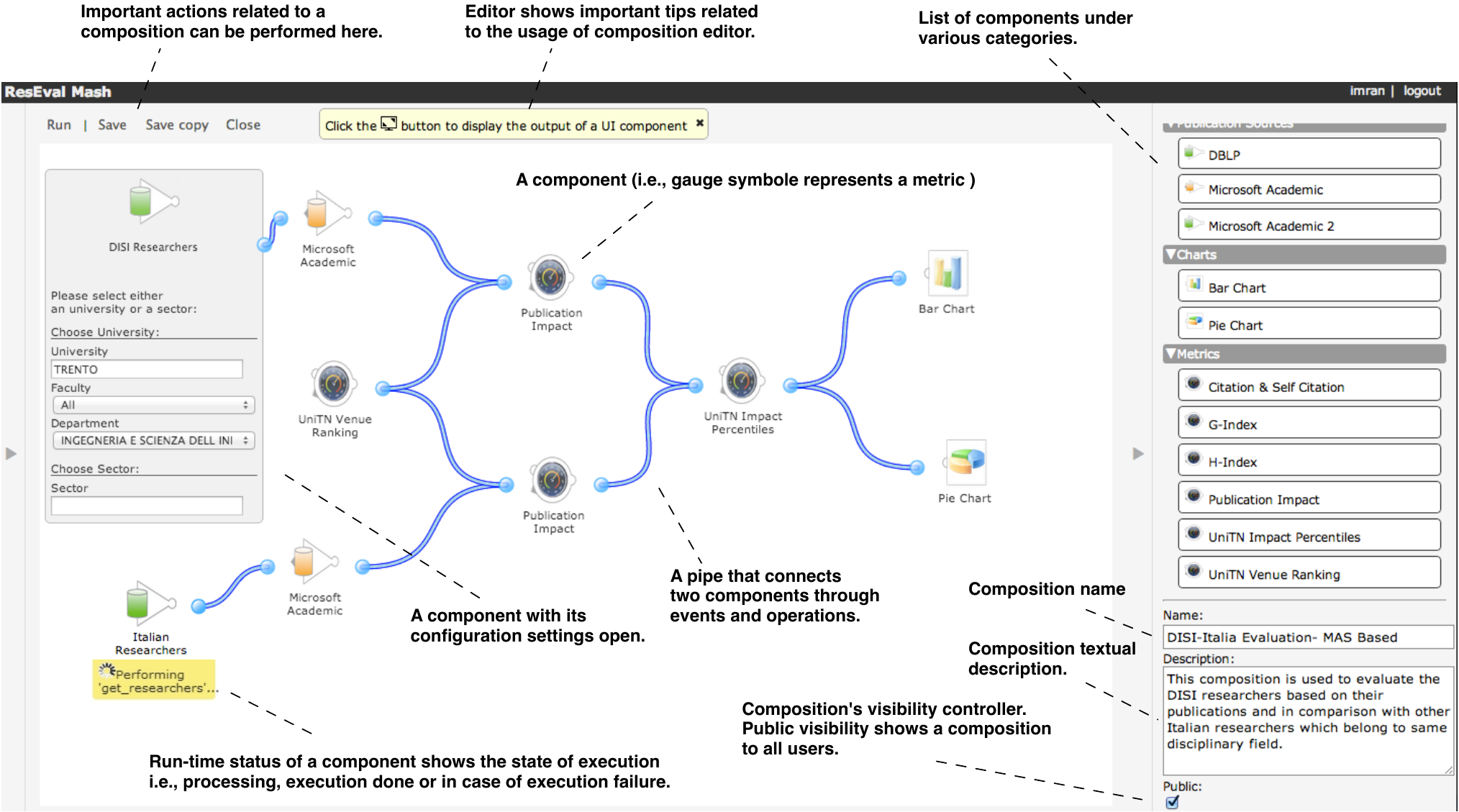}
  \caption{ResEval Mash's composition editor and its various parts}
  \label{fig:composition-ed}
\end{figure}

Figure \ref{fig:composition-ed} shows ResEval Mash's \emph{composition editor}. The composition editor provides the mashup canvas to the users. It shows a components list from which users can drag and drop components onto the canvas and connect them. The composition editor implements the \emph{domain-specific mashup meta-model} and exposes it through the \emph{domain syntax}. From the editor it is also possible to launch the execution of a composition through a run button and hand the mashup over to the \emph{mashup engine} for execution. 

\begin{figure}[t]
  \centering
  \includegraphics[width=0.45\textwidth]{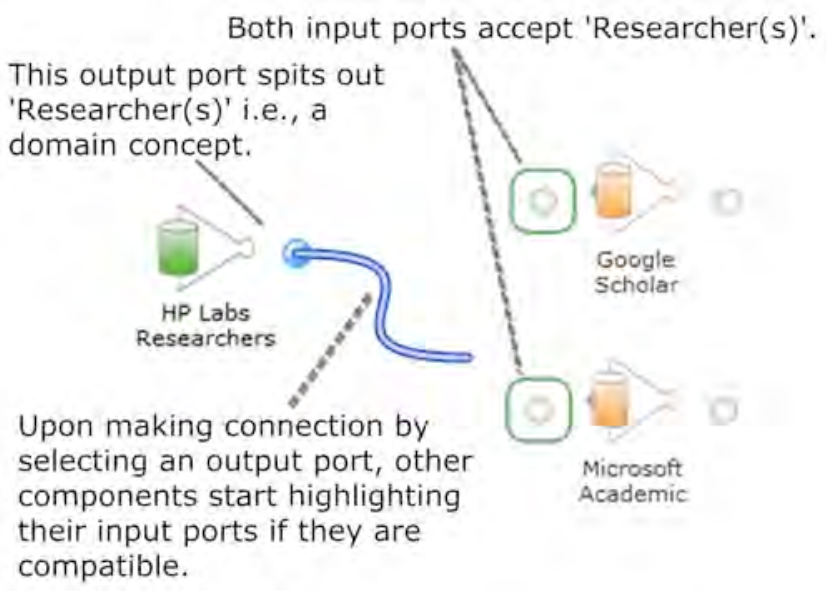}
  \caption{ResEval Mash's composition editor highlighting compatible ports upon making connections among components}
  \label{fig:-components-conn}
\end{figure}

Each container with a symbol on the canvas represents one component with its name, a symbol, configuration interface, and in case of a UI component it's output interface. On the left side of a component shows its input ports (i.e., where it accepts connections) and on the right side it shows output ports (i.e., where it emits data to the next connected components). Components can be expanded to view their configuration interface and various ports. To connect two components, a user has to click and drag from the output port of a source component and drop and release click on input port of a target component. Figure \ref{fig:-components-conn} depicts how connections can be performed. Composition editor highlights all compatible ports (i.e., checking domain concepts compatibility) of all components present on the canvas upon a mouse click on a component's event (i.e., output port), that is how editor knows that user intends to make a connection. 

To provide run-time transparency, which convey to users the state of each component's execution, composition editor shows various visual states of the components. A component can have and change among three visual states. These visual states correspond to a component's execution status, like a component which is not in the running state shows its label and boarder in black color. Whilst, a component which is in the running state shows an extra label in yellow color right below the component, which shows the operation name which is being executed. The third state represents the successful execution of a component and shows both component name and its boarder in green color. The fourth and the final state, which shows component's boarder and a notice in a red color that represents the component execution failed due to some reason.

\subsection{Component Registration Interface} 
\begin{figure}[t]
  \centering
  \includegraphics[width=1.03\textwidth]{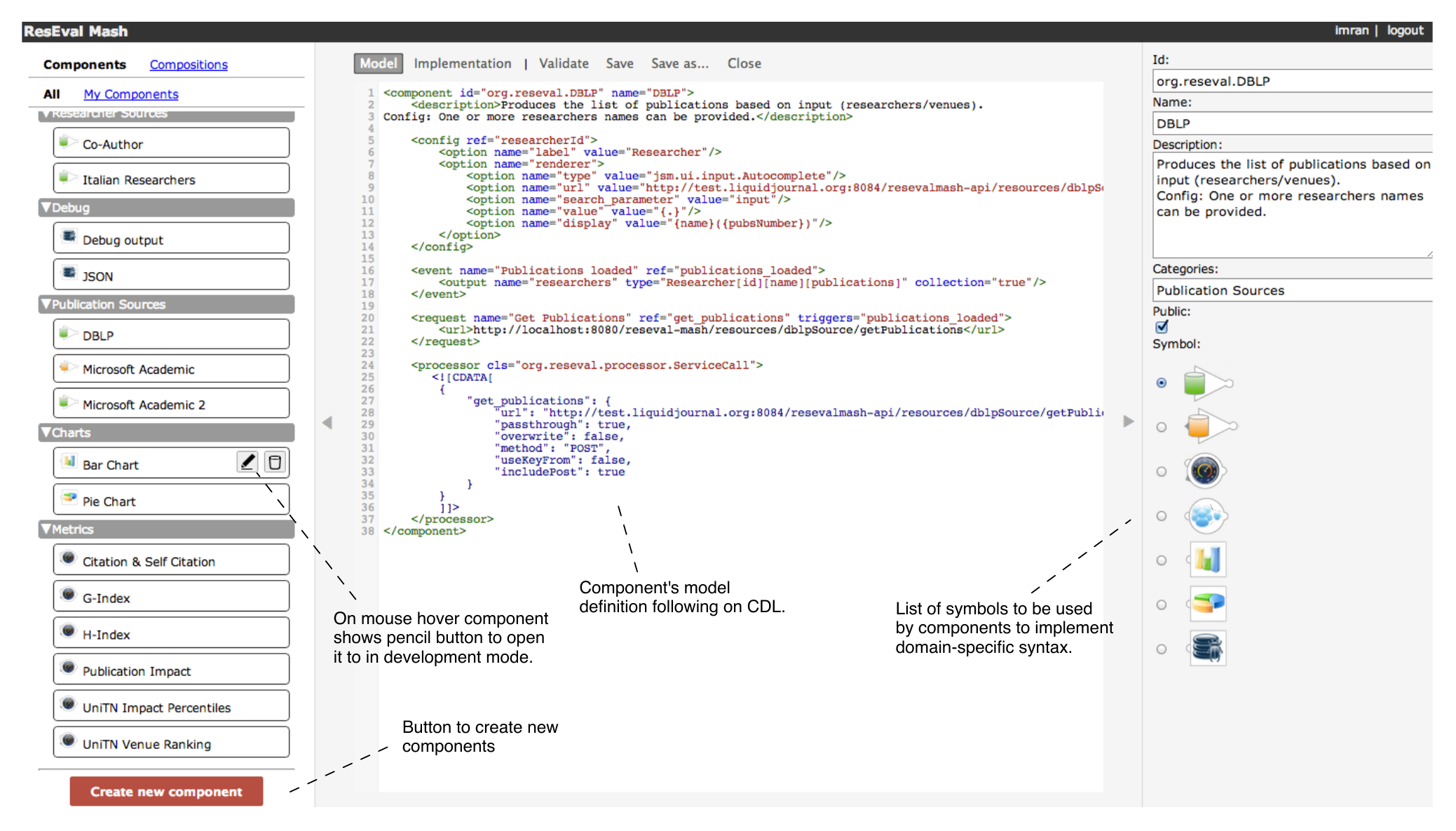}
  \caption{ResEval Mash's component registration interface showing a component's definition}
  \label{fig:-components-model-def}
\end{figure}

The tool also comes with a \emph{component registration interface} for developers, which aids them in the setup and the addition of new components to the platform. The interface allows the developer to define components starting from ready templates. In order to develop a component, the developer has to provide two artifacts: (i) a component definition (Figure \ref{fig:-components-model-def}) and (ii) a component implementation. The implementation consists either of JavaScript code for client-side components or it can be linked to a web service, which is achieved providing a binding to a web service for server-side components. 

To provide ease to component developers , especially with dynamic, untyped languages such as JavaScript, testing and debugging can take much time since errors are often only discovered when the code is executed, the editor provides a supportive interface which allows easy adjustments to the code directly in the browser. Developing inside the browser is not very popular yet, but is possible, especially for languages native to the browser environment, such as JavaScript. Developers do not need to upload code changes of a component repeatedly after some changes are made as the editor automatically identifies that new changes are available and hence it deploys new version.

The component editor consists of two separate editors, one for the component model definition and one for the implementation. Both editors provide syntax highlighting and rudimentary code completion support. For example, the component model editor offers code completion for the elements of our XML representation. This is realized with the help of the CodeMirror\footnote{http://codemirror.net/} library.

%\begin{figure}[t]
%  \centering
%  \includegraphics[width=1.03\textwidth]{figs/component-impl-def.pdf}
%  \caption{ResEval Mash's component registration interface showing a component's implementation}
%  \label{fig:-components-impl-def}
%\end{figure}

\subsection{Server-Side Services} 
On the server side, we have a set of RESTful web services, i.e., the \emph{repository services}, \emph{authentication services}, and \emph{components services}. Repository services enable CRUD operations for components and compositions, that is, the mashup engine interacts with these services to perform CRUD operations. Authentication services are used for user authentication and authorization. Components services manage and allow the invocation of those components whose business logic is implemented as a server-side web service. These web services, together with the client-side components, implement the \emph{domain process model}. The idea behind these services was to move the computation from the client side to the server side to improve performance by utilizing server's computational power and big memory hence reducing client side burden. The interaction details between client side components and their services is explained in section \ref{sec:component-service-interaction}. And, a detail explanation of how to develop a service for a component is given in section \ref{sec:dsm-ecosystem}.

\subsection{CDM Memory Manager, CDM Module \& Shared Memory} 
The \emph{common data model (CDM) memory manager} enforces and supports the checking of data types in the system. To use the domain-specific data-types (i.e., DCM concepts) for various modules on the server side, it is necessary to have an interface that can read a domain-specific model (e.g., an XML schema definition (XSD) in our case), and can parse it, generate implementation classes. These classes then expose their definitions to the other modules. In order to configure the CDM, the \emph{CDM memory manger} generates corresponding Java classes (e.g., in our case these classes are POJO, annotated with JAXB annotations) from an XSD that encodes the domain concept model. Having the required data-types context in place, the CDM memory manager is then responsible for the insertion and retrieval of data to and from the shared memory. 

On the server side, a \emph{shared memory} is maintained that \emph{CDM memory manager} uses it to read and write data. The shared memory can store multiple states of data, and also multiple instances of the same data. However, all the data must comply with the data-types provided by the common data model module. The insertion and retrieval to and from the shared memory follows key-value pair mechanism, where the value represents actual data and key works as an identifier. Client side mashup engine initially generates a key, which passes along the data that is then used by the CDM memory manger to store in the shared memory. Hence, all data processing services read and write to this shared memory through the CDM memory manager. That means, the CDM interacts with the \emph{shared memory} to provide a space for each mashup execution instance if required. In our first prototype we use the server's working memory (RAM) as \emph{shared memory}, which allows for high performance. Clearly, this solution fits the purpose of our prototype but it may not scale to in-production installations, which may need to deal with large numbers of users and large amounts of data that only hardly can be kept in RAM if it offers small memory. However, in our future work, we aim to develop a persistent database-based \emph{shared memory}. 

%\subsection{Server-engine} 
%All services are managed by the \emph{server-side engine}, which is responsible for managing all the modules that are on the server side, e.g., the CDM memory manager, the repository, and so on. The server-side engine is the place where requests coming from the client side are dispatched to the respective service implementing the required operations.

\subsection{Local Database and the Web} 
Both the \emph{database} and the \emph{Web} represent the data which is required and used by the component services. We as a platform provider provide a database\footnote{The database holds data that we have crawled, downloaded from various sources for performance purposes.} and a basic set of services on top of it. A third-party service can be deployed and thus it can use an external database anywhere on the Web. However, the development of a third-party service must comply with the specification presented later in the section \ref{sec:dsm-ecosystem}.

\section{Intelligent Switching between Data-flow and Control-flow}
\label{sec:df-cf-flow}
As explained earlier that the use of server-side web services is one way to implement business logic of a component, which can also be implemented through JavaScript that is we call these components as client-side components. For components whose business logic is implemented by server side services require to send data from client side to the server side for various processing. Moreover, if a composition comprised of more than one such component then for each such component data must be sent to the server side from the client and vice-versa. This scenario will be even worse, especially if the data (i.e., data which is being used in communications) are bigger; as in our case too, then following this strategy poses serious challenges in terms of speed that decreases an overall performance of the platform. To deal with data-intensive compositions, which deals with huge amounts of data, the traditional mechanism (i.e., data-flow back and forth between client and server) is not an appropriate choice. For this reason, the mashup engine adds a \emph{Control-flow layer}, which provides a substantial increase of performance during such situations. 

\begin{figure}[t]
  \centering
  \includegraphics[width=.60\textwidth]{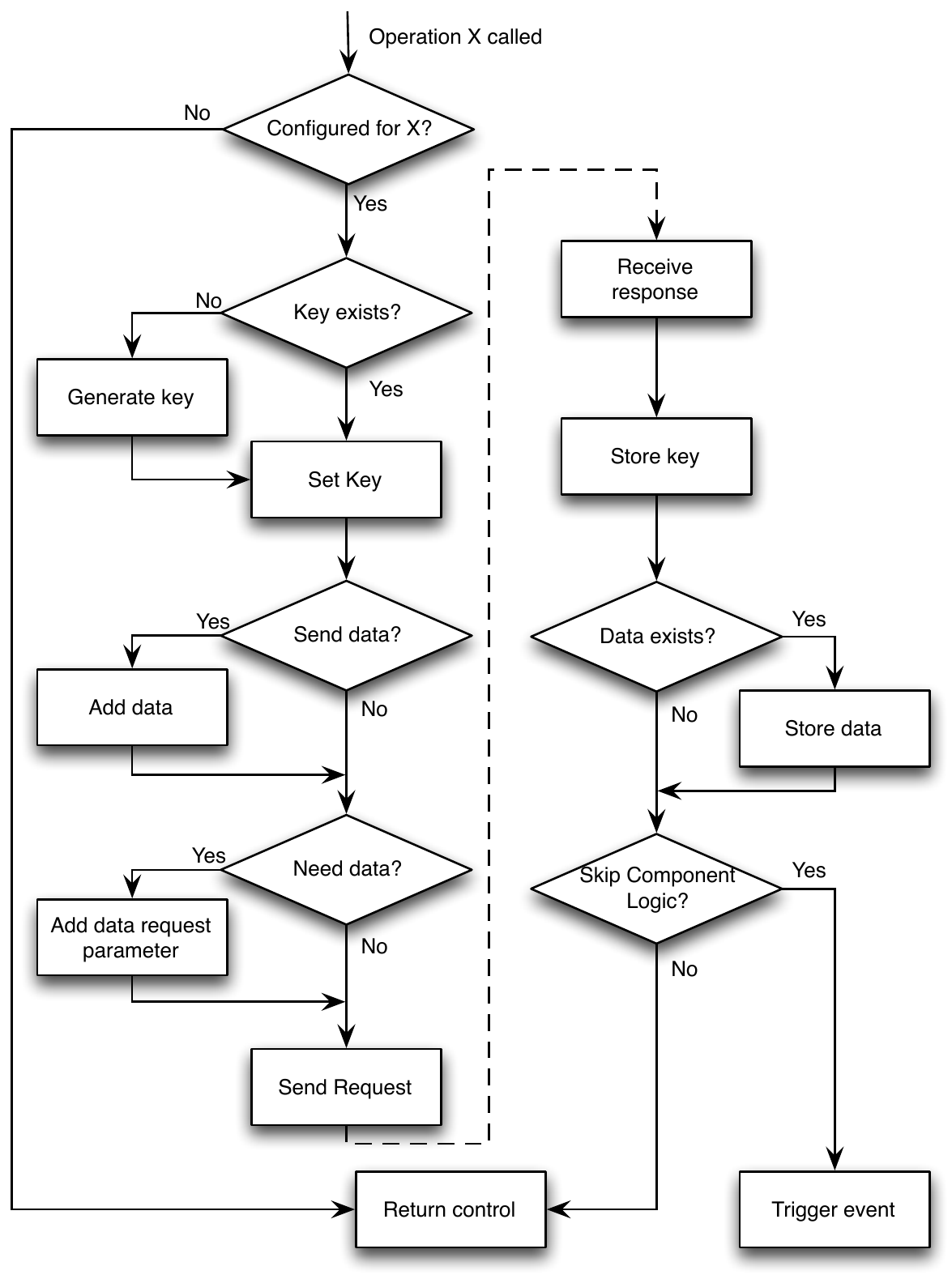}
  \caption{Service Call Data Processor Flow Chart}
  \label{fig:service-call-flow-chart}
\end{figure}

The platform achieves the functionality of intelligently switching between data to control flow and back to data-flow with the help of data processors presented in \ref{sec:dataprocessor}. The data processors are designed to intercept any operation call, request or event. To call a corresponding web service of a component, it is required to intercept whenever an operation of the component is called, and sending a request to the service. This is achieved by configuring a \emph{ServiceCall} data process, a specific type of the data processor as listed in the Listing \ref{list:dataprocessor-config}. The operation "get researchers" (line 5) is configured, having a service URL (line 6), with passthrough parameter as \emph{false} and overwrite parameter as \emph{true} and also mentioning what type of service it is (line 9). The \emph{passthrough} parameters represents if the data (i.e., the service response) will be needed by the component implementation (i.e., client-side implementation) for performing some actions before it will be handed over to the next component. The \emph{overwrite} parameter represents if the original data will be overwritten or not. These lines of configuration make sure that whenever an operation of a component is called, the data processor checks for its ServiceCall configuration.

\begin{lstlisting}[caption={Data processor configuration}, label={list:dataprocessor-config}]
...
 <processor cls="org.reseval.processor.ServiceCall">
       <![CDATA[
        {
            "get_researchers": {
                "url": "http://example.com/getResearchers",
                "passthrough": false,
                "overwrite": true,
                "method": "POST"
               }
            }
        }        
        ]]>
    </processor>
...
\end{lstlisting}

The overall procedure, which is to detect and to decide when and where to send data, is represented in the flow chart in Figure \ref{fig:service-call-flow-chart} and explained as follows. In case, if a servicecall is configured, the data processor inspects the header of the message for a \emph{key} (i.e., a numeric identifier comprises of composition id and time stamp information) which should be dispatched to the service to be called. The key is used on the server side and used as an identifier for the data in the shared memory. A new key is generated by the data processor if one is not found in the message, which also means the particular data does not exist on the server side. 

The data processor further checks if some data already exist in the message body that will be dispatched to the server along with the \emph{Data Request} parameter. The data request can be set either as  "yes" or "no", which describes whether the response of the service has to be sent back to the client side or not. The decision about setting data request as "yes" or "no" depends on the next component in the composition connected to the component that is being inspected. A component whose business logic is implemented as a JavaScript file requires the data on the client side for processing, on the other hand a component whose business logic is implemented by a service on the server side does not require the data to be present on the client side. That is how the mashup engine using data processors detects and makes decisions accordingly. The details of each particular interaction scenario between client and server side components as described in the next section \ref{sec:component-service-interaction}.

\begin{figure}[t]
  \centering
  \includegraphics[width=.35\textwidth]{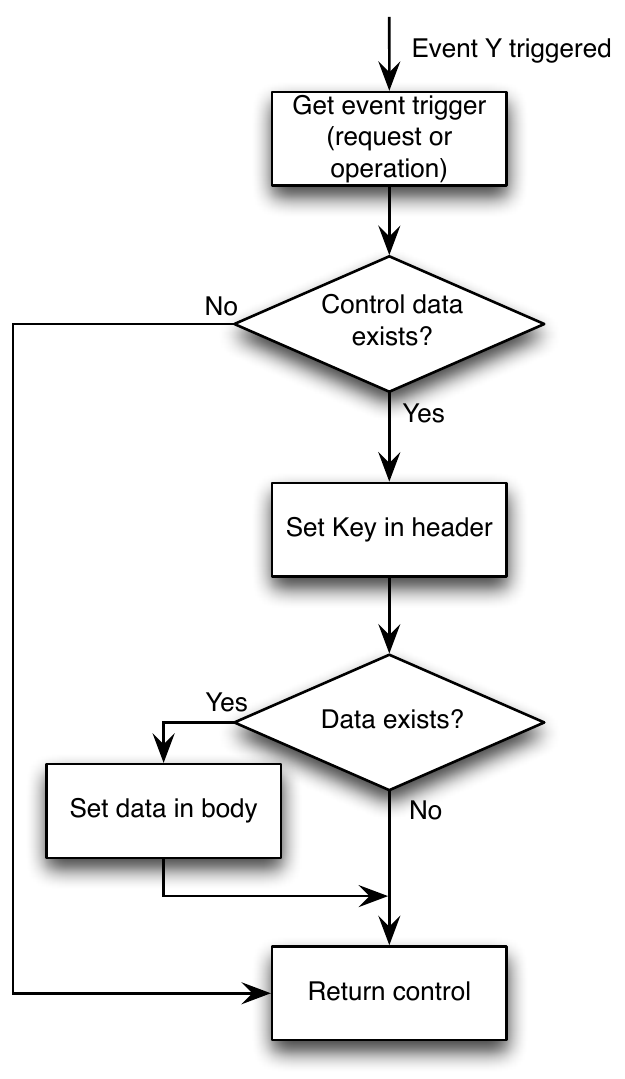}
  \caption{Service Call Data Processor Flow Chart: Event}
  \label{fig:service-call-flow-chart-event}
\end{figure}

\begin{figure}[t]
  \centering
  \includegraphics[width=.55\textwidth]{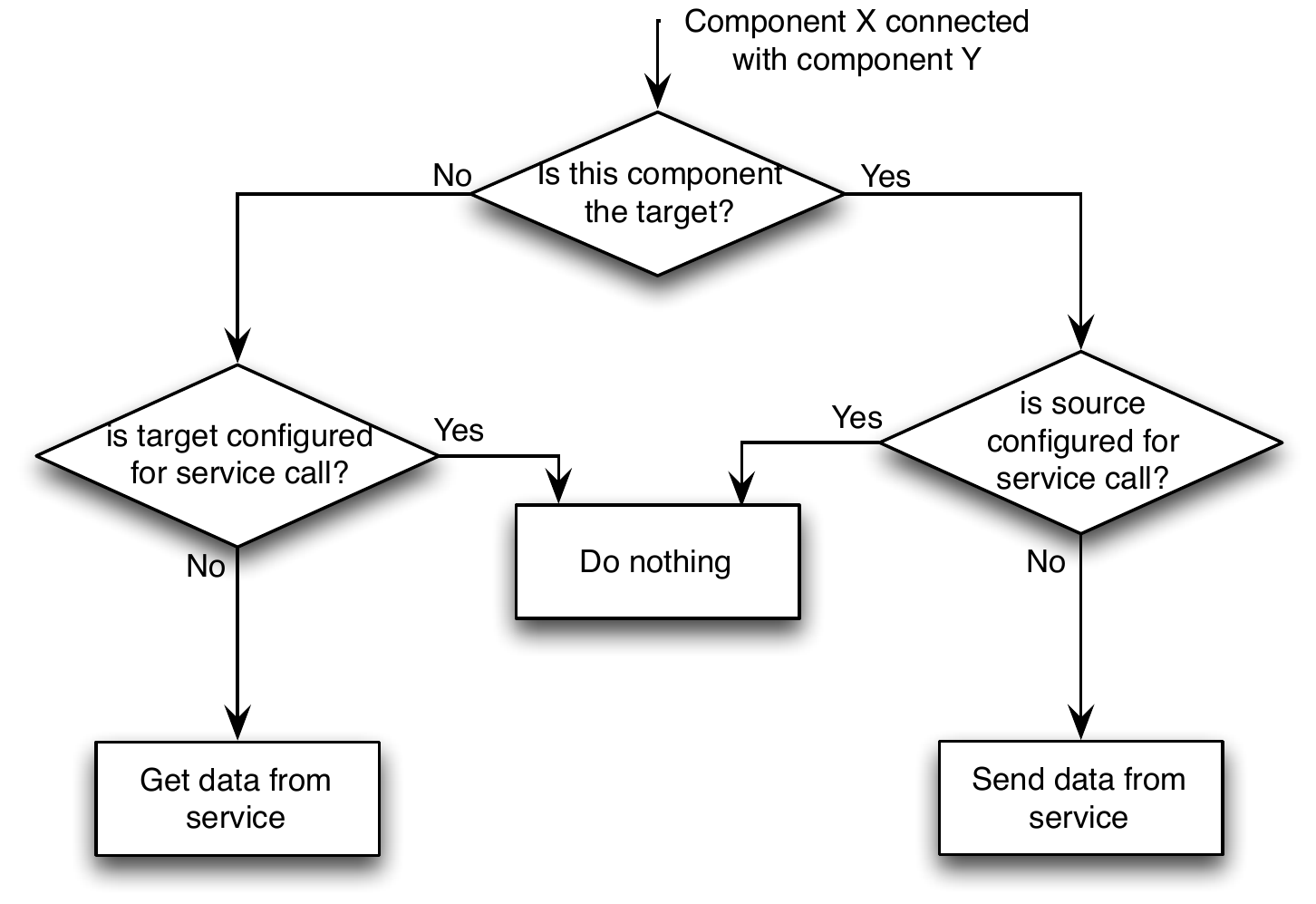}
  \caption{Detecting client-side and server-side components}
  \label{fig:detecting-cs-comp}
\end{figure}

Figure \ref{fig:service-call-flow-chart-event} depicts how the data processor intercept call upon an event trigger. First it checks whether the triggered event is the result of an operation or a request. In both cases, the control data presence is checked. If control data is available, the key is set in the header, the response data is set in the body of the message. Figure \ref{fig:detecting-cs-comp} depicts the flow chart of the mechanism used to detect whether the target component is a client-side or a server-side component. The result is then used to configure dataRequest parameter values that decides whether to fetch data or not on the client-side.

\section{Components Models and Data Passing Logic}
\label{sec:component-service-interaction}

There are two component models in ResEval Mash, depending on whether the respective business logic resides either on the client or the server side: \emph{server components} (SC) are implemented as RESTful web services that run at the server side; \emph{client components} (CC) are implemented as JavaScript file and run on the client side. Independently of the component model, each component has a client-side component front-end, which allows (i) the Mashup Engine to enact component operations and (ii) the user to inspect the state of the mashup during runtime. All communications between components are mediated by the Mashup Engine, internally implementing a dedicated event bus for shipping data via events. Server components require interactions with their server-side business logic and the shared memory; this interaction needs to be mediated by the Mashup Engine. Client components directly interact with their client-side business logic; this interaction does not require the intervention of the Mashup Engine.

The components consume or produce different \emph{types of data}: actual \emph{data} (D), \emph{configuration parameters} (CP), and control data like \emph{request status} (RS), a flag that conveys whether actual \emph{data is required} in output (DR), and a \emph{key} (K) identifying data items in the shared memory. All components can consume and produce actual data, yet, as we will see, not always producing actual data in output is necessary. The configuration parameters enable the setup of the components. The request status enables rendering the processing status at runtime. The key is crucial to identify data items produced by one component and to be "passed" as input to another component. As explained earlier, instead of directly passing data from one service to another, for performance reasons we use a shared memory that all services can access and only pass a key, i.e., a reference to the actual data from component to component.

\begin{figure}[t]
  \centering
  \includegraphics[width=0.95\textwidth]{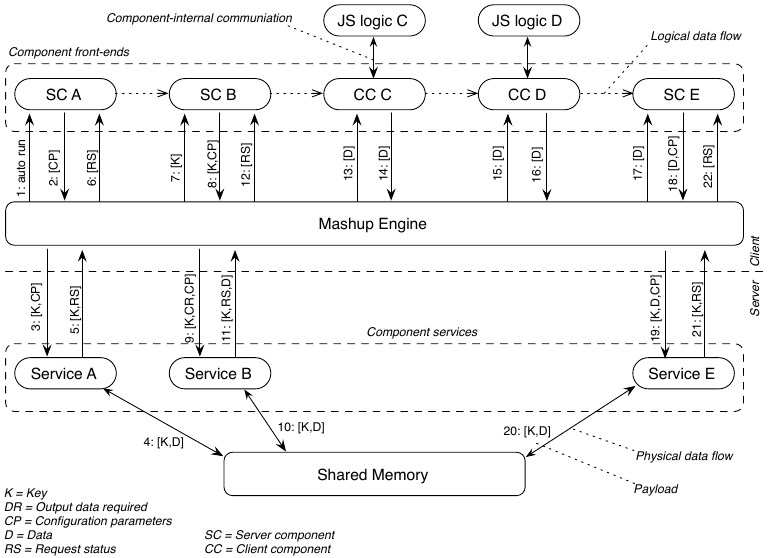}
  \caption{ResEval Mash's internal data passing logic.}
  \label{fig:datapassing}
\end{figure}

Based on the flow of  components in the mashup model, we can have different data passing patterns. Given the two different types of components, we can recognize four possible interaction patterns. The four patterns are illustrated in Figure \ref{fig:datapassing}. All of these interactions are mediated by the mashup engine, hence neither a composition composer nor a component developer needs to think about these complexities while component or composition development. In particular, we may have two types of interaction, that is, (i) the interaction among components that are connected in the designed composition and (ii) the interaction among a component and its server-side implementation (only in the case of components of type SC). Both these types of interaction are managed by the Mashup Engine. In the first type, the Mashup Engine manages the event bus used to publish the components' events (carrying associated data) and trigger the subscribed components' operations. In the second type, the Mashup Engine acts as a proxy for the web service operation invocation with the help of data processors. In both cases, the Mashup Engine has the role of managing and including  the correct control data in  all the events and service invocations, that is crucial for letting the platform work properly. In following we elaborate  individual interaction pattern separately. 

%Basically when an output port of source component is connected with an input port of a target component, an event is generated at that time and the engine listens to that event hence it knows that which is the next component to pass data to. 

%\begin{figure}[t]
%  \centering
%  %\includegraphics[width=\textwidth]{figs/user-studies-steps-college.pdf}
%  \includegraphics[width=\textwidth]{figs/Components-Services-Interactions.pdf}
%\vspace{-3mm}
%  \caption{Components and services interactions}
%\vspace{-3mm}
%  \label{fig:components-services-interaction}
%\end{figure}

%During the execution of a composition, a component performs two types of interactions: first with its corresponding business logic implementation (either a web service operation or JavaScript function) and, second, it interacts with other components to pass the data. 

%All the modules in the overall system interact with eachother as shown in Figure \ref{fig:architecture} but the most important interactions are component to component and component to their services. There are four types of scenarios that can happen when two component are connected. These four scenarios are described in detail later in this section. 
\begin{enumerate}
\item \emphbf{SC-SC interaction:}
As shown in the Figure \ref{fig:datapassing}, both the components (SC A \& SC B) are of type SC. Component A is connected with component B. Since component A is the first component in the composition and it does not require any input, it can start the execution immediately. It is the responsibility of the Mashup Engine to trigger the operation of the component A (step 1). At this point, component A calls its back-end web service through the Mashup Engine, passing only the configuration parameters (CP) to it (2). The Mashup Engine, analyzing the composition model, knows that the next component in the flow is also a server component (component B), so it extends component A's request adding a \emph{key}  control information to the original request, which can be used by component A's service to mark the data it produces in the shared memory. Hence, the Mashup Engine invokes service A (3). Service A receives the control data, executes its own logic, and stores its output into the Shared Memory (4). Once the execution ends, Service A sends back the control data (i.e., key and request status) to the Mashup Engine (5), which forwards the request status to component A (6); the engine keeps track of the key. With this, component A has completed and the engine can enable the next component (7). 
In the SC-SC interaction, we do not need to ship any data from the server to the client.

%In this scenario two components are connected and both are service components. In Figure \ref{fig:components-servies-interaction} SC component A is connected with SC component B. Assuming that component A is the first component in the composition and it does not required any input from other components so that it can start the execution immediately. So the mashup engine first triggers component A as shown in Figure \ref{fig:components-servies-interaction} call sequence number 1. As both components have their corresponding services so first of all component A calls its service through mashup engine by only passing configuration paramerters. The engine detects that the next connected component is also a service component (i.e., SC component B) so it wrapps the component A's request with adding a control data (i.e., key) and engine calls the target service (i.e., service A).  The service receives the control data and performs its execution (service execution details are described in section \ref{sec:components_implementation}). Service sends back control data (i.e., key, request status) to the clinet side mashup engine and engine sends it to the component A as shown in call sequency number 4 \& 5. At the end component A passes control data received from the enigne to the next component as shown in call sequency number 6.  

\item \emphbf{SC-CC interaction:}
Once activated, component B enacts its server-side logic (8, 9, 10). The Mashup Engine detects that the next component in the flow is a client component, so it adds the DR control data parameter in addition to the key and the configuration parameters, in order to instruct the web service B to send actual output data back to the client side after it has been stored in the Shared Memory. In this way, when service B finishes its execution, it returns the control data and the actual output data of the service (i.e., key, request status and output data) to the Mashup Engine (11), which then passes the request status to component B (12) and the actual data to the next component in the mashup, i.e., component C (13).

%This scenario shows the interaction between SC component B and CC component C as shown in Figure \ref{fig:components-servies-interaction}. Component B is a service component whereas component C is a client component (i.e., it needs data at client side for the processing). As component B has already received a key from component A, so it calls its corresponding service though the enigne as shown in call sequence number 7 \& 8. The engine intrupts the call and detects that the next connected component with component B is a client component. So the engine adds a control data param (i.e., data request (DR)) so to ask server side to send actual data back to the client side. After finishing the execution of service B, it returns the control data and the actual output data of the service (i.e., key, request status and data) to the client side engine which is then passed to the component B by the engine as shown in call sequence number 9 \& 10. Finally the component B passes the actual data to the component C as shown in call sequence number 11.
 
\item \emphbf{CC-CC interaction:}
Client component to client component interactions do not require to interact with the server-side services. Once the component C's operation is triggered in response to the termination of component B, it is ready to start its execution and to pass component B's output data to the JavaScript function implementing its business logic. Once component C finishes its execution, it sends its output data back to the engine (14), which is then able to start component D (15) by passing C's output data. 

%Client to client component interaction is a bit different than the previous two cases, as these components do not required to interact with the server side services. In our example which is shown in Figure \ref{fig:components-servies-interaction} the Component C has already received the actual data from the component B, so it is ready to start its execution. Once component C finished its execution it is ready to send data to the next connected component (i.e., component D). As all these communinication is managed by the engine so it is engine duty to send actual output data of component C to the component D. This interaction is shown in call sequence number 12. 

\item \emphbf{CC-SC interaction:}
After the completion of component D (16), the Mashup Engine passes the respective data to component E as input (17). At this point, component E calls its corresponding service E, passing to it the actual data and possible configuration parameters (18), along with the key appended by the Mashup Engine (19). Possibly, also the Output Data Request flag could be included in the control data but, as explained, this depends on the next component in the flow, which for presentation purpose is not further defined in Figure \ref{fig:datapassing}. Eventually,  service E returns its response (i.e., key and request status -- plus possible output data if the DR flag is present) to the Mashup Engine (21), which is then delivered to component E (22).

%This particular scenario is also quite interesting as source is a client component and target is a service component. In our example, once the component D is finished its execution so its data is ready to pass on to the next component. The engine passes this data to the component E as shown in call sequence number 13. Now component E can call its corresponding service through the engine. So it sends the actual data and configuration param (if required) to the service. The engine intrupts this call and adds a control param (i.e., key) and calls the service as shown in call sequence number 14 \& 15. What is added by the engine to this intrupted call depends on the next connected component, so here we assume a service component is connected next to the component E. Finally the service returns the response (i.e., key and request status) to the client side which is then delivered to the component E as shown in call sequence number 16 \& 17.

\end{enumerate}

While ResEval Mash fully supports these four data passing patterns and is able to understand whether data are to be processed at the client or the server side, it has to be noted that the actual decision of where data are to be processed is up to the developer of the respective mashup component. Client components by definition require data at the client side; server components on the server side. Therefore, if large amounts of data are to be processed, a sensible design of the respective components is paramount. As a rule of thumb, we can say that data should be processed on the server side whenever possible, and component developers should use client components only when really necessary. For instance, visualization components of course require client-side data processing. Yet, if they are used as sinks in the mashup model (which is usually the case), they will have to process only the final output of the actual data processing logic, which is typically of smaller size compared to the actual data sourced from the initial data sources (e.g., a table of h-indexes vs the lists of publications by the set of the respective researchers).

\section{The Domain-Specific Service Ecosystem}
\label{sec:dsm-ecosystem}
%Here we write about implementation concerns such as: components definitions, their client side java script implementation, server side services implementation. How a domain-specific service looks like, give an example of code or sequence diagram. Also describe about interaction between services and the server side engine in order to get and store data into the shared memory....

An innovative aspect of our mashup platform is its approach based on the concept of \emph{domain-specific components}. In Section  \ref{sec:architecture} we described the role of the Components services in the architecture of the system. These are not simply generic web services, but web services that constitute a \emph{domain-specific service ecosystem}, i.e., a set of services respecting shared models and conventions and that are designed to work collaboratively where each of them provides a brick to solve more complex problems proper of the specific domain. Having such an ecosystem of compatible and compliant services, introduces several advantages that make our tool actually usable and able to respond to the specific requirements of the domain we are dealing with.

Given the important role domain-specific components and services play in our platform, next we describe how they are designed and illustrate some details of their implementation and their interactions with the other parts of the system.

%\subsubsection{Implementing Domain-Specific Components for ResEval Mash} 
%\label{sec:components_implementation} 

%In this section we will show how we inject the domain into the mashup components. In particular, we will describe how a component for our domain-specific ResEval Mash platform must be designed and implemented. For the sake of clarity, we will do it with help of a concrete example component present in the tool, i.e., the Italian Researchers component, which is designed to return a list of Italian researchers filtered by university, faculty, department or disciplinatory sector.

\medskip

A ResEval Mash component requires the definition of two main artifacts: the component descriptor (i.e., following the component definition language specifications) and the component implementation.

The \emphbf{component descriptor} describes, to briefly mention, the main properties of a component, which are: 
\begin{enumerate}
	\item \emph{Operations}. Functions that are triggered as a consequence of an external event that take some input data and perform a given business logic.
	\item \emph{Events}. Messages produced by the component to inform the external world of its state changes, e.g., due to interactions with the user or an operation completion. Events may carry output data.
	\item \emph{Implementation binding}. A binding defining how to reach the component implementation.
	\item \emph{Configuration parameters}. Parameters that, as opposed to input data, are set up at composition design time by the designer to configure the component's behavior.
	\item \emph{Meta-data}. The component's information, such as name and natural language description of the component itself. 
\end{enumerate}

In our platform the component descriptors are implemented as XML file, which must comply with an XML Schema Definition (XSD). The XSD defines both the schema for the component descriptors and the admitted data types. Validating the descriptor against the data types definition we can actually enforce the adoption of the common domain concept model (DCM), which enable smooth composability and no need for data mapping in the Composition Editor, as discussed in Section \ref{sec:requirements}.

\begin{figure*}
  \centering
   \includegraphics[width=0.80\columnwidth]{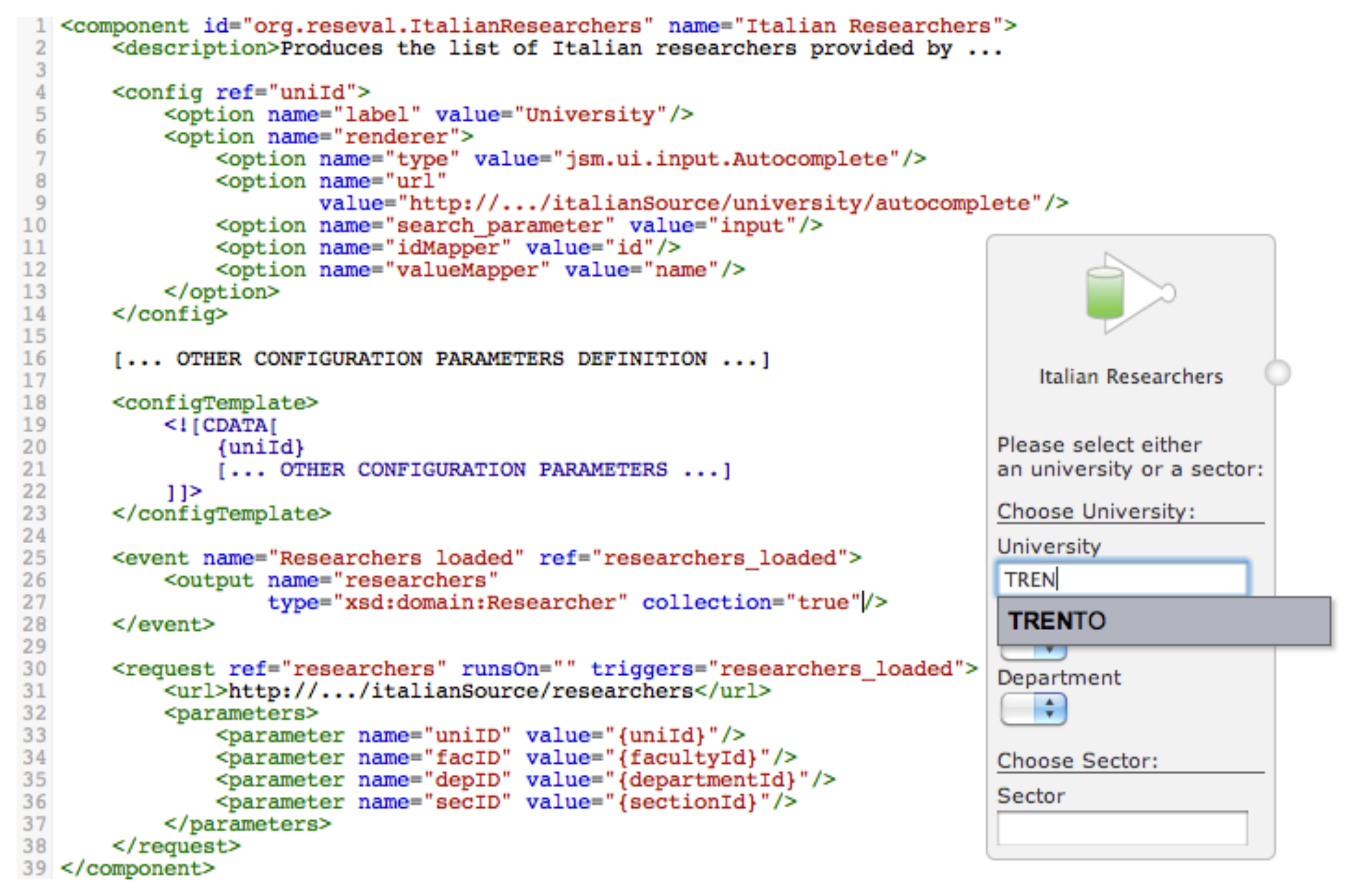}
  \caption{The descriptor of the Italian Researchers component along with its representation in the Composition Editor}
  \label{fig:component-descriptor}
\end{figure*}

For example, an excerpt of the Italian Researchers component descriptor along with its representation in the Composition Editor is shown in Figure \ref{fig:component-descriptor}.
The component is implemented through a server-side web service. Its descriptor does not present any operation and it has an event called {\tt Researchers Loaded}, which is used to emit the list of researchers that are retrieved by the associated back-end service. The binding between the service and its client-side counterpart is set up in the descriptor through the {\tt <request>} tag. As shown, this tag includes the information needed to invoke the service, i.e., its end-point URL and the configuration parameters that must be sent along with the request. In addition, the attribute {\tt triggers} specifies the event to be raised upon service completion. The attribute {\tt runsOn}, instead, specifies the component's operation that must be invoked to start the service call. In this particular case, since the component has no operations and no inputs to wait for, when the mashup is started the Mashup Engine automatically invokes the back-end service associated with the component, causing the process execution to start. If we were dealing with a component implemented via client-side JavaScript, we would not need the {\tt <request>} tag, and the implementation binding would be represented by the {\tt ref} attribute of the component operation or event, whose value would be the name of the JavaScript function implementing the related business logic. 

The component in Figure \ref{fig:component-descriptor} has different configuration parameters, which are used to define the search criteria to be applied to retrieve the researchers. We can see the {\tt uniId} parameter. Beside the name of the related label, we must specify the {\tt renderer} to be used, that is, the way in which the parameter will be represented in the Composition Editor. In this case, we are using a text input field with auto-completion features. The auto-completion feature is provided by a dedicated service operation that can be reached at the address specified in the {\tt url} option. Finally, we can see the presence of the {\tt configTemplate} tag, which is just used to set the order in which the parameters must be presented in the component representation in the Composition Editor.

\medskip

The other main artifact that constitutes a ResEval Mash component is its \emphbf{implementation}. As already discussed above, a component can be implemented in two different ways: through client-side JavaScript code (client component) or through a server-side web service (server component).
The choice of having a client-side or a server-side implementation depends mainly on the type of component to be created, which may be a UI component (i.e., a component the user can interact with at runtime through a graphical interface) or a service component (i.e., a component that runs a specific business logic but does not have any UI). UI components (e.g., the Bar Chart of our scenario) are always implemented through client-side JavaScript files since they must directly interact with the browser to create and manage the graphical user interface. Service components (e.g., the Microsoft Academic Publications of our scenario), instead, can be implemented in both ways, depending on their characteristics. In the research evaluation domain, since they typically deal with large amounts of data, service components are commonly implemented through server-side web services. In such a way, they do not have the computational power constraints present at the client-side and, moreover, they can exploit the platform features offered at the server-side, like the Shared Memory mechanism, which, e.g., permit to efficiently deal with data-intensive processes. In other cases, where we do not have particular computational requirements, a service component can be implemented via client-side JavaScript, which runs directly in the browser. The JavaScript implementation, both in case of UI and service components, must include the functions implementing the component's business logic.

\begin{figure}[t]
  \centering
   \includegraphics[width=0.90\columnwidth]{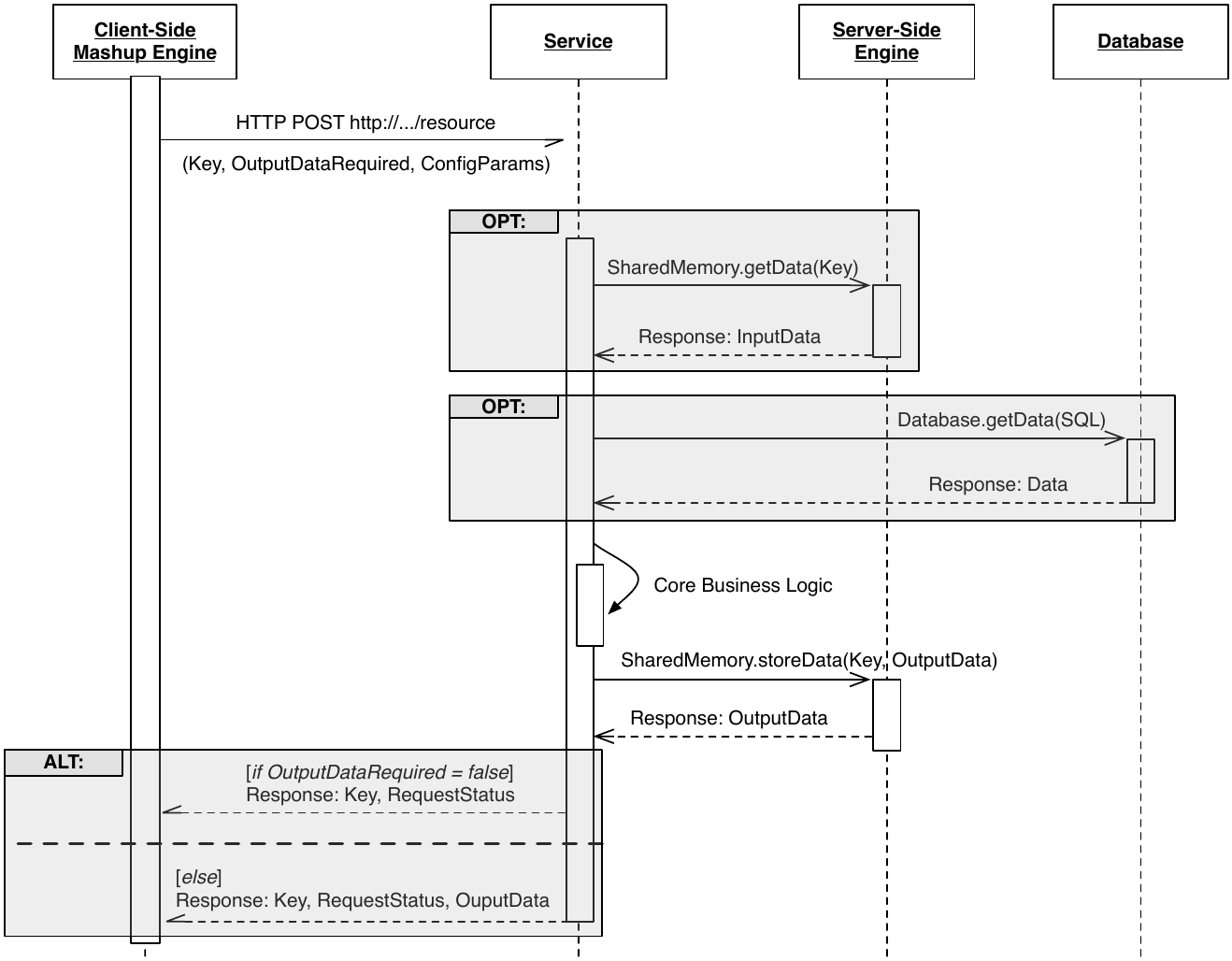}
  \caption{Platform-specific interaction protocol each service must comply with}
  \label{fig:service-server-interaction}
\end{figure}

For example, our Italian Researchers  service component is implemented at server-side since it has to deal with large amounts of data (i.e., thousands of researchers), so it belongs to the server components category (introduced in Section \ref{sec:component-service-interaction}). 
%Hence we propose for such components to be implemented 
%Our Italian Researchers example component, since it has to deal with (extract and filter) data related to thousands of researchers, falls among the service components implemented at the server-side. 
This type of components, to correctly work within our domain-specific platform, must be implemented as Java RESTful web service following specific implementation guidelines. In particular, the service must be able to properly communicate with the other parts of the system and, thus, it must be aware of the data passing patterns discussed before and the shared memory. Figure \ref{fig:service-server-interaction} shows the interaction protocol with the other components of the platform the service must comply with.

The service is invoked through an HTTP POST request by the client-side Mashup Engine, performed through an asynchronous Ajax invocation (the half arrowheads in the figure represent asynchronous calls). The need to expose all the operations through HTTP POST comes from the fact that in many cases it must be possible to send complex objects as parameters to the service, which would not be possible in general using a GET request. For instance, in our example, the operation is invoked through a POST request at the URL \url{http://.../resevalmash-api/resources/italianSource/researchers} and the component's configuration parameters (e.g., selected university or department) are posted in the request body. Besides the parameters, the body also includes control data, that is the {\tt key} and the {\tt OutputDataRequired} flag. 
%The first is used as reference to read the input data from or write the output data in the Shared Memory. The second is set depending on the composition structure, as discussed in Section \ref{sec:component-service-interaction}, and it has the following semantic: when set to {\tt true}, it is used by Mashup Engine to ask the service to include in the service response, beside the control data, also the actual data produced as output by the service operation, when it is set to {\tt false}, the actual data must only be stored in the Shared Memory but not sent back to the client-side Mashup Engine.

Once the request coming from the Mahup Engine is received by the service, the service code must process it following the sequence diagram shown in Figure \ref{fig:service-server-interaction}. If the service is designed to accept input data, first it will get the data from the Shared Memory through the API provided by the Server-Side Engine, using the received key as parameter. 

Then, the service may need to have access to other data for executing its core business logic. The services developed and deployed by us (as platform owners) can use the system database to persistently store their data, as shown in the second optional box. This is, for instance, the case of our Italian Researchers component that retrieves the researchers from the system database, where the whole Italian researchers data source has been pre-loaded for efficiency reasons. Third-party services, instead, do not have access to the system database but they can use external data sources as external databases or online services available on the Web. Clearly, the usage of the system database guarantees higher performances and avoids possible network bottlenecks.

Once the service has retrieved all the necessary data, it starts executing its core business logic (for our example component, it consists in the filtering of the researchers of interest based on the configuration parameters). The business logic execution results are then stored in the Shared Memory. Typically, all the services will produce some output data, although, possibly, there could be exceptions like, for instance, a service that is only designed to send emails. Finally, the service must send a response back to the Mashup Engine. The response content depends on the {\tt OutputDataRequest} flag value. If it is set to false, as shown in the upper part of the alternative box in the figure, the response will contain the {\tt Key} and the {\tt RequestStatus} of the service (success or error). If the flag is set to true, in addition to those control data, the response will also contain the actual {\tt OutputData} produced by the service logic.

So far, all components and services for ResEval Mash have been implemented by ourselves, yet the idea is to open the platform also to external developers for the development of \emph{custom components}. In order to ease component development, e.g., the setup of the connection with the Shared Memory and the processing of the individual control data items, we will provide a dedicated Java interface that can be extended with the custom logic. The description, registration, and deployment of custom components is then possible via the dedicated Component Registration Interface briefly described in Section \ref{sec:architecture}.

\section{ResEval Mash in Action: Various Mashup Compositions}
This section presents various mashup compositions that are developed using the ResEval Mash tool. The first two compositions implement the scenarios described in chapter \ref{sec:scenarios}. 

\subsection{UniTN Department Evaluation Scenario}
The department evaluation procedure that is used by the University of Trento (UniTN) is described in the section \ref{sec:unitn-scenario}. According to its description, we need to fetch UniTN and Italian researchers those belong to the same discipline as of UniTN ones. For both, the UniTN and the Italian researchers the publications have to be retrieved from a publication data source, which are then ranked based on the UniTN venue ranking scheme. Finally ranked publications are used to compute impact percentile using negative binomial distribution. 

\begin{figure}[t] 
  \centering
   \includegraphics[width=0.90\columnwidth]{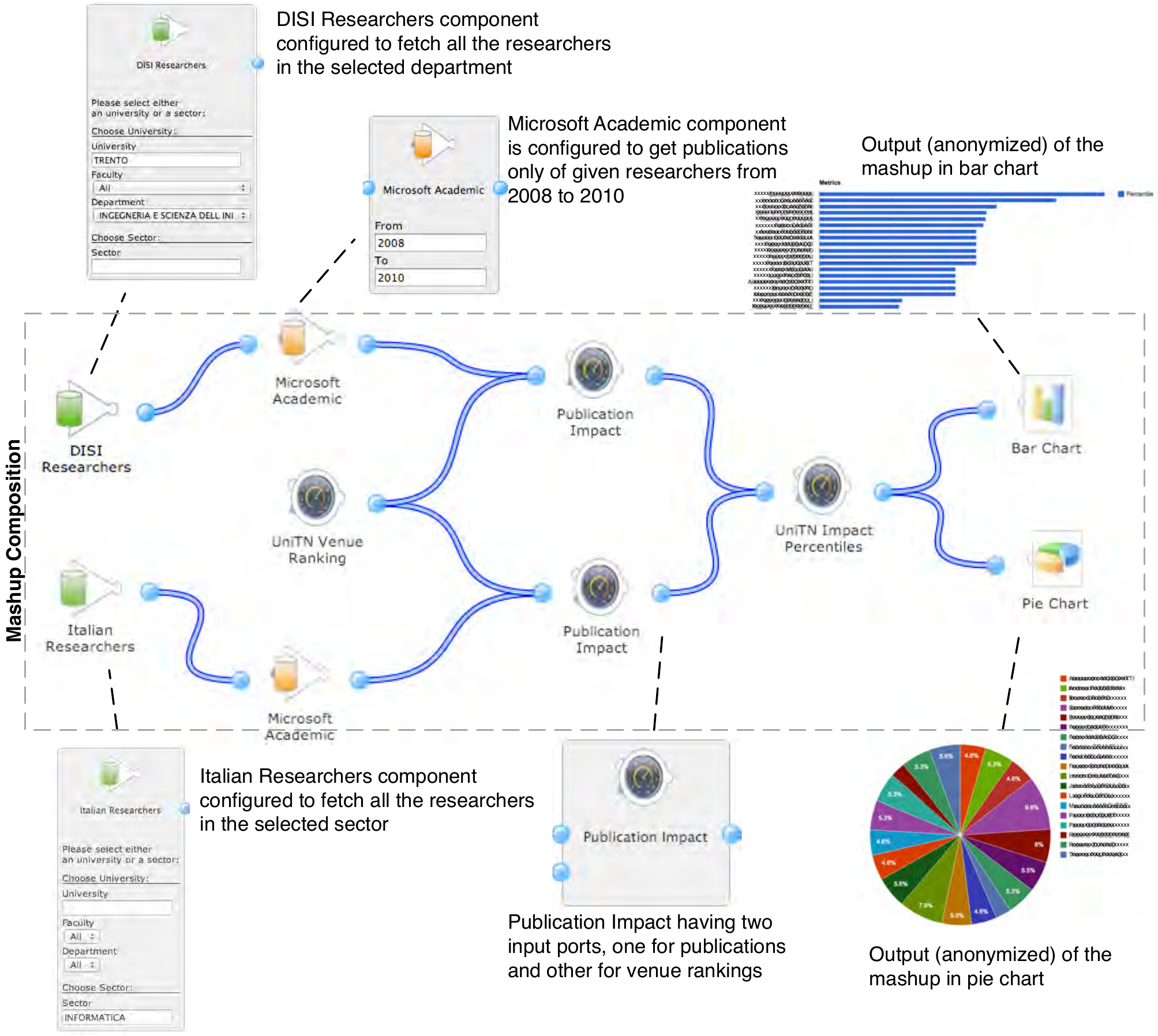}
  \caption{UniTN Dept. Evaluation Mashup Composition: showing components config panels and output (anonymized) with detail description}
  \label{fig:unitn-mashup-imp}
\end{figure}

Figure \ref{fig:unitn-mashup-imp} depicts the implemented version of UniTN department evaluation scenario using ResEval Mash tool. The figure shows in the center (in dotted border) the original mashup, along with configuration panels of a few important components and the final output of the mashup. In total ten components are used to compose this procedure in which seven are distinct and other three components are instances of some of these components (e.g., DISI researchers, Microsoft Academic, Publication Impact etc). The composition starts with two parallel flows: one computing the weighted publication number (the \emph{impact} metric in the specific scenario) for all Italian researchers in a selected discipline sector (e.g., Computer Science). The other computes the same "impact" metric for the researchers belonging to the UniTN computer science department. The former branch defines the distribution of the Italian researchers in the Computer Science discipline sector, the latter is used to compute the impact percentile of the UniTn's researchers and to determine their individual percentile, which are finally visualized in a bar and a pie chart. 

Most of the components that are used in this composition are of server-side type, that is, the actual computation is performed on the server, except two components (i.e., bar chart and pie chart). As described earlier in this chapter that server-side implementation (i.e., using a web service) is preferable for components those manipulate big data.

\subsection{Italian Professorship Selection Scenario}
In section \ref{sec:it-prof.scenario}, we elaborated the evaluation procedure used by the National Agency for the Evaluation of Universities and Research Institutes (ANVUR) for hiring and promoting professors. The procedure states that metrics (e.g., contemporary h-index, number of articles, number of citations) used for the evaluation must be normalized prior to perform comparison with the provided thresholds values.  These values have been fixed by ANVUR as a research quality threshold for a specific area. 

\begin{figure}[t] 
  \centering
   \includegraphics[width=\columnwidth]{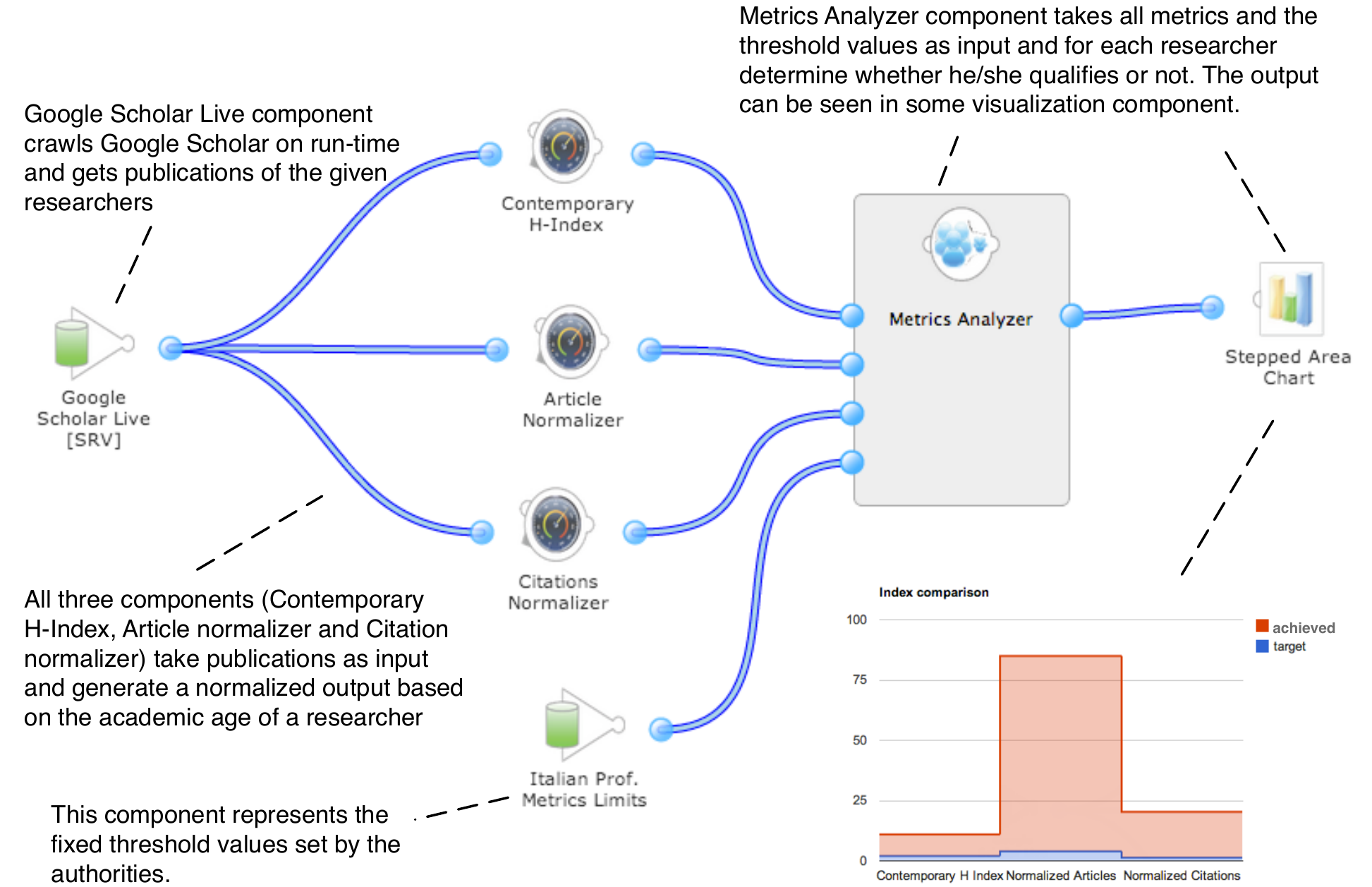}
  \caption{Italian Professorship Selection Mashup Composition: showing components and output with detail description}
  \label{fig:italian-mashup-imp}
\end{figure}  

Figure \ref{fig:italian-mashup-imp} depicts the implementation of the evaluation procedure, which has been developed using ResEval Mash. In the mashup composition we use seven components in total. The composition starts from the Google Scholar Live component, which takes one or more researchers' names as input and crawls Google Scholar web site on run-time to get their publications. Retrieved publication list for each researcher is then given to three components (i.e., contemporary h-index, article normalizer and citation normalizer) to compute normalized metric according to the defined procedure in the original evaluation document. The metric analyzer component, however, takes input of all required metrics and the thresholds to determine for each researcher that he/she qualifies or not. The results of this component can be displayed in a visualization component, as in our case we show the results in a stepped area chart component.

\begin{figure}[t] 
  \centering
   \includegraphics[width=0.80\columnwidth]{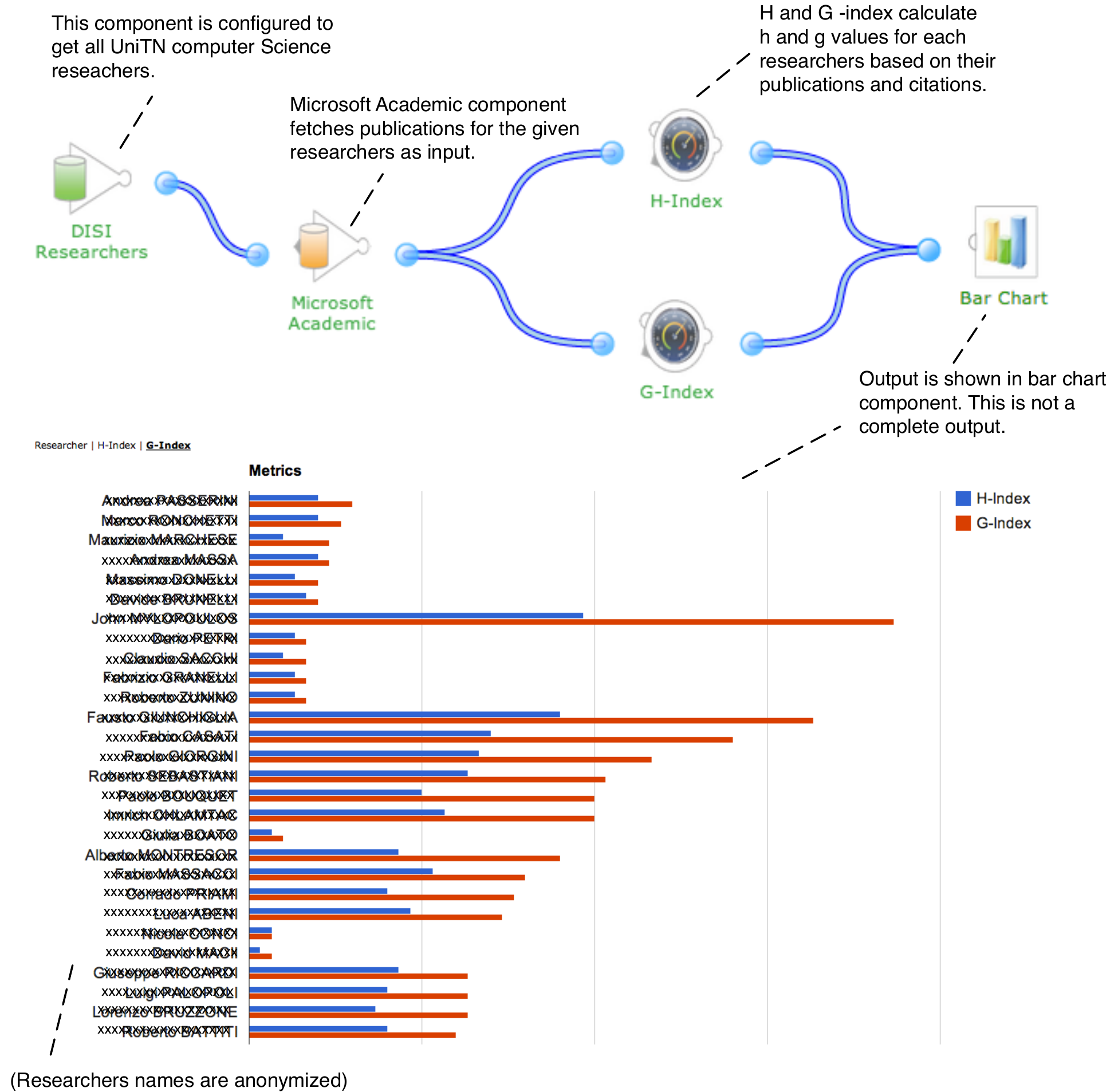}
  \caption{Mashup composition showing H and G -index values of DISI researchers (anonymized names)}
  \label{fig:disi-hg-index}
\end{figure}  

\subsection{Computing and Comparing H and G -Index Values of Researchers}
To show how ResEval Mash can be used to compute various metrics, we compose a mashup that calculates the H and G -Index values of researchers who belong to the University of Trento Computer Science Department. Figure \ref{fig:disi-hg-index} (top) depicts the mashup composition that is developed in ResEval Mash. 

The composition starts with DISI Researchers component, which is configured to retrieve (i.e., from a local repository) all the researchers of Computer Science department of UniTN. The list of researchers is then passed to the next component, which is in this case "Microsoft Academic" component. This component takes as input the researchers list and retrieves publications for each. Next, the output of the Microsoft Academic component is consumed by two components, which are H-Index and G-Index. These two components compute the H and G values, which is then visualized in a bar chart as shown in the figure \ref{fig:disi-hg-index} (bottom).

\begin{figure}[t] 
  \centering
   \includegraphics[width=\columnwidth]{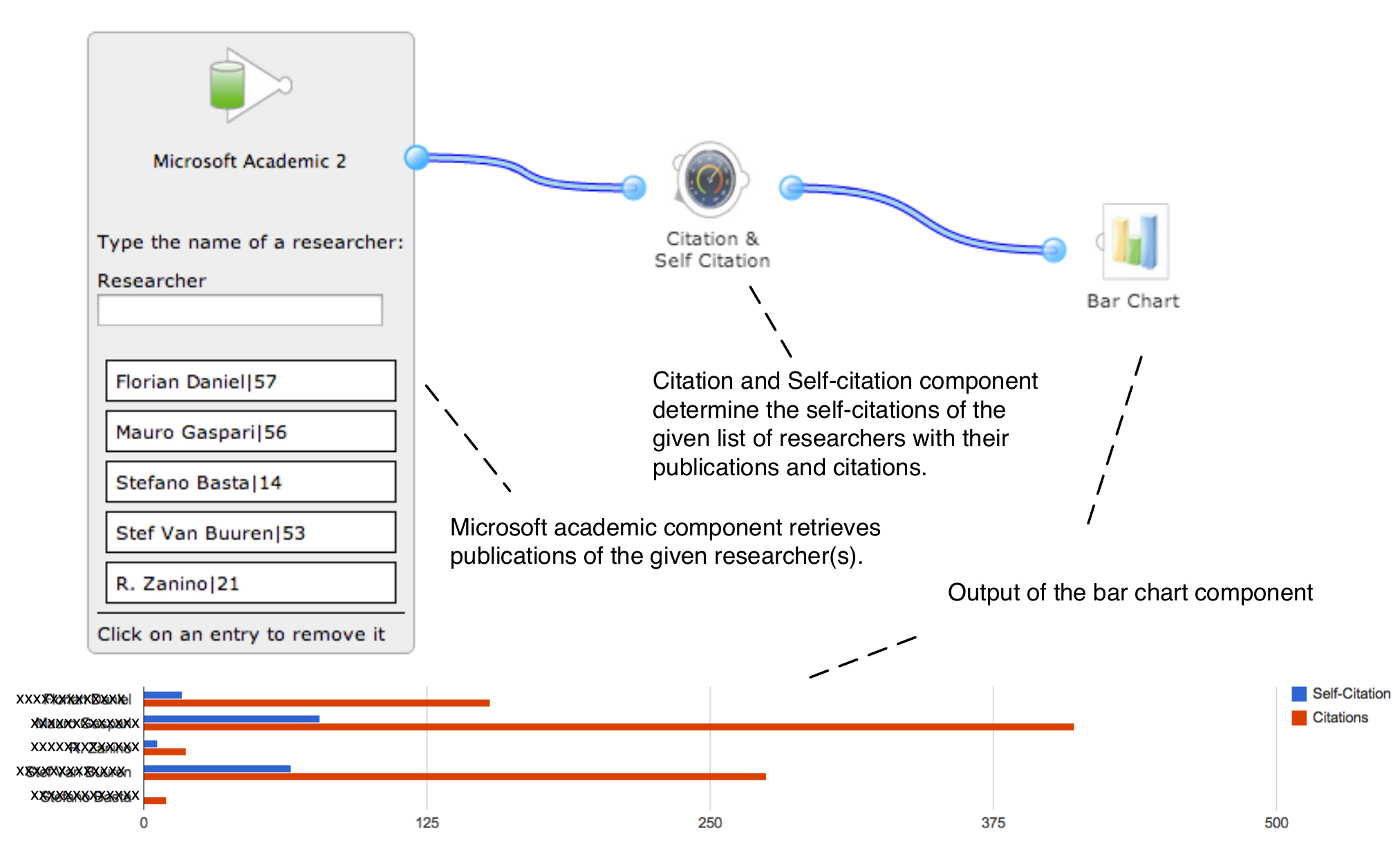}
  \caption{Mashup composition showing citation and self-citation comparison for a given list of researchers (names anonymized)}
  \label{fig:self-citation}
\end{figure}
\subsection{Comparison of Citations and Self-Citations}

Figure \ref{fig:self-citation} depicts a mashup composition, which can be used to compare the citation versus self-citation of one or more researchers. The task is achieved using three components in ResEval Mash. The first component (i.e., Microsoft Academic), given one or more researchers, retrieves publications from Microsoft Academic source. The next component, which is Citation \& Self Citation, takes publications as input and determines self-citation count for each researcher. The self-citation count is determined parsing all the publications of a researcher and checking if the researcher (being investigated) appears in any publication that cites his/her publication. Finally, we use bar chart component to show the results. 

This section presented a few example mashup compositions which are used for different evaluation tasks. The ResEval Mash\footnote{http://open.reseval.org} tool is capable to do more, as now all the effort depends on the availability of new components that users can use according to their broad wisdom. In the next chapter we present user studies that we have conducted to test whether the tool really useful for the non-technical users or not. The studies also investigate various other associated aspects of a mashup tool usability.

\newpage

% Chapter 7: Experimental Evaluation of the thesis' proposal.
% !Tex root = ../phd-thesis-arxiv.tex

%%% Chapter start. %%%
\chapter{User Studies and Evaluation}
\label{sec-experiments}

\section{Overview}
To evaluate different aspects associated with the work presented in this thesis, like whether being domain-specific is preferable or not for end-users; and if yes, how much expressive a tool should be, that is, what level of flexibility a mashup tool should offer so that the corresponding complexity stays within the boundaries of non-technical users. For this purpose, we conducted a few user studies. The first study, which was mainly focused on the usability evaluation of our mashup tool also partially used for the comparative analysis between ours and other mashup based tool. To this end, to understand the users' preference over domain-specific versus generic mashup tools, we used Yahoo! Pipes as a generic tool example. Moreover, to determine what level of complexity a non-technical domain-expert can deal with in case of a domain-specific mashup tool preference, we built four different prototypes. Each prototype encompasses different level complexity which surly depends on the flexibility and customization features that these tools provide. 

After the first user study, we addressed and incorporated the suggestions, feedbacks and new requirements gathered during the first user study. The improvements and changes made our mashup tool more usable and useful. Again, to validate the usability of the ResEval Mash tool, we conducted the second user study to test relatively advance features specifically related to the usability of the ResEval Mash tool. The next section elaborate on the first user study and the section \ref{sec:us-2} presents the details of our second user study. Finally we conclude this chapter with analysis and discussion.

\section{Comparative \& Usability Evaluation: User Study-1}
Being generic versus specific is fundamentally different from their roots, that is, the former covers a broader level in comparison to the latter, which stays specific and can be a specialized form of what former offers. For instance, in this context, simply asking someone to make a computer application and to its opposite asking to make an accounting application is different. Likewise, for building generic or specific application depends what constructive constructs are provided to a developer that is we believe especially for non-technical users those constructs would be more valuable if they are aligned to their level of domain-expertise. Moreover, what a development environment conveys or understand from an end-user point of view is different, like whether the language that an application speaks understandable by the end-users or not. In order to assess all these aspects, we designed our first user evaluation experiment as described in the next section. 

\subsection{Task Design}
To evaluate the different aspects related to the usability of our domain-specific tool as well as to determine users' preference, we performed contextual interviews of 28 users. Among them, 7 were \emph{professors}, 5 \emph{administrative people}, 1 \emph{post-doc}, 3 \emph{PhD students} and 12 \emph{master's level students}. These participants were having different levels of technical skills. The technical skills of the participants were determined asking the following questions:

\begin{itemize}
\item What is the user skill level with tools such as MS Excel, MS Word etc?
\item Is the user aware of the meaning of web service?
\item Is the user able to draw/understand a (simple) process following a given graphical notation (e.g., flowchart)?
\end{itemize}

In addition to the above mentioned questions, the participants were also asked to write (in a text box) if they program computer applications, or involved in programming tasks and other details related to their technical as well as domain skills. From those recorded answers, we can say that all of the users were domain-experts (i.e., they know research evaluation and were involved in some kind of evaluation tasks), excepts the master's level students, which were having low domain-expertise than the former group. Among all the user, 5 administrative were highly domain-expert and most of them were directly involved in the research evaluation task that we used during the study. Table \ref{tab:User-study-1-details} presents the details of all the participants with their technical and domain skills. %We used the scenario described in section \ref{sec:scenario}, which was implemented by all the four prototype having different flexibility (described below).

\begin{table}[h]
\centering
\begin{tabular}{ | l | l | l | l |}
\hline
\textbf{Users} & \textbf{Position} & \textbf{Technical Skills} & \textbf{Domain Skills} \\ \hline
7 & professor & 4 (good skilled), 3 (moderate) & very skilled\\ \hline
5 & administrator & 4 (moderate), 1 (very skilled) & very skilled\\ \hline
1 & post-doc & good skilled & very skilled \\ \hline
3 & phd  & good skilled & good skilled \\ \hline
12 & ms student & moderate & moderate\\ \hline

\end{tabular}
\caption{User Details}
\label{tab:User-study-1-details}
\end{table}

We use one of the famous mashup tool Yahoo! Pipes as a generic tool example and our ResEval mashup tool as a domain-specific one. To better understand the appropriate level of expressive power, flexibility, and difficulty that our tool (i.e., ResEval Mash) offers, we developed four separate prototypes of the tool based on our selected scenario, each prototype offered a different level of flexibility and, consequently, complexity. %These four prototypes implemented the same scenario with different level of flexibility and expressiveness.

\begin{figure}
  \centering
    \includegraphics[width=0.90\columnwidth]{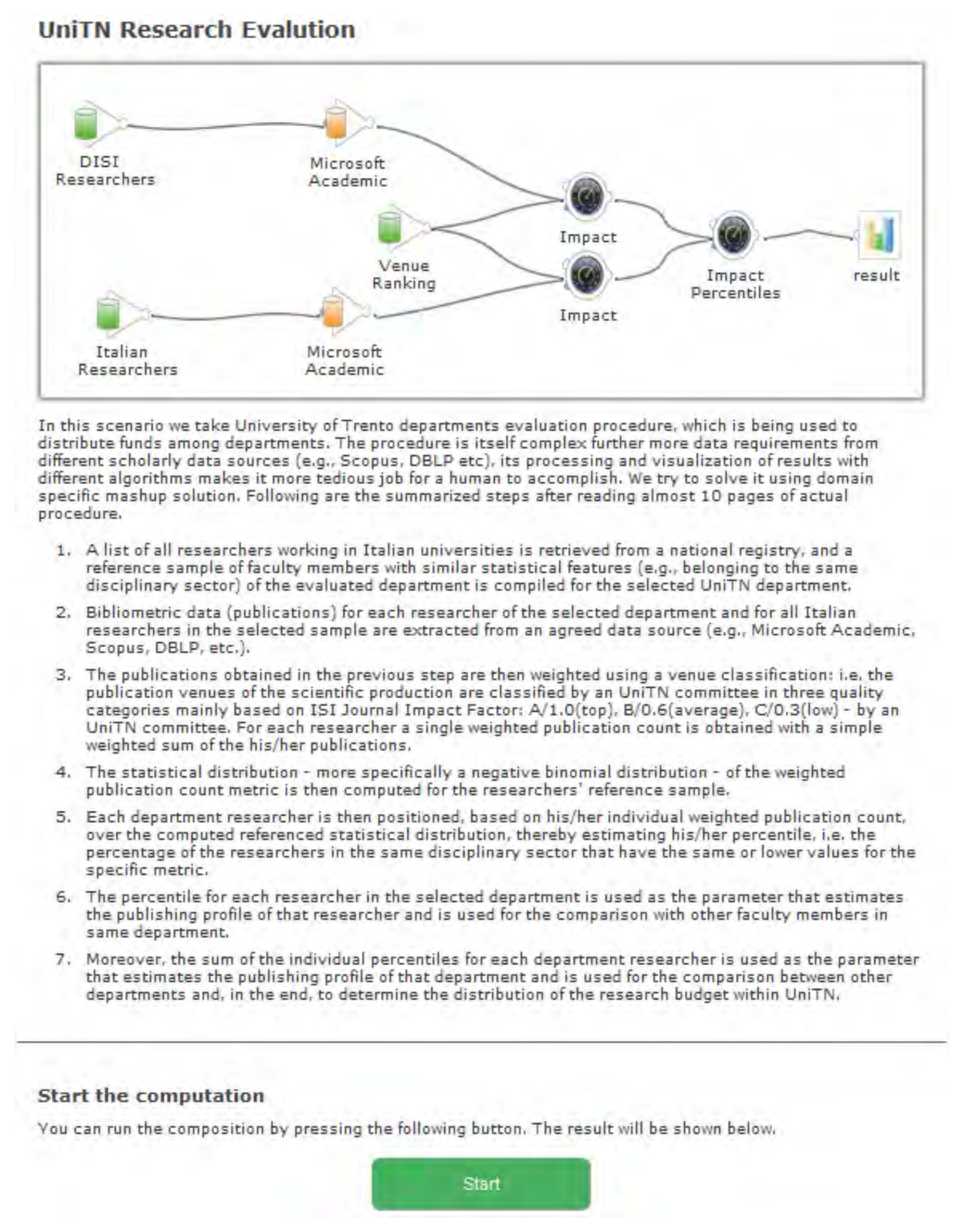}
  \caption{Prototype-1: fixed components with fixed configuration options}
  \label{fig:tool-1}
%\vspace{-3mm}
\end{figure}

\subsection{Evaluation Procedure}
\label{sec:study1-procedure}
\begin{figure}[ht]
  \centering
    \includegraphics[width=0.80\columnwidth]{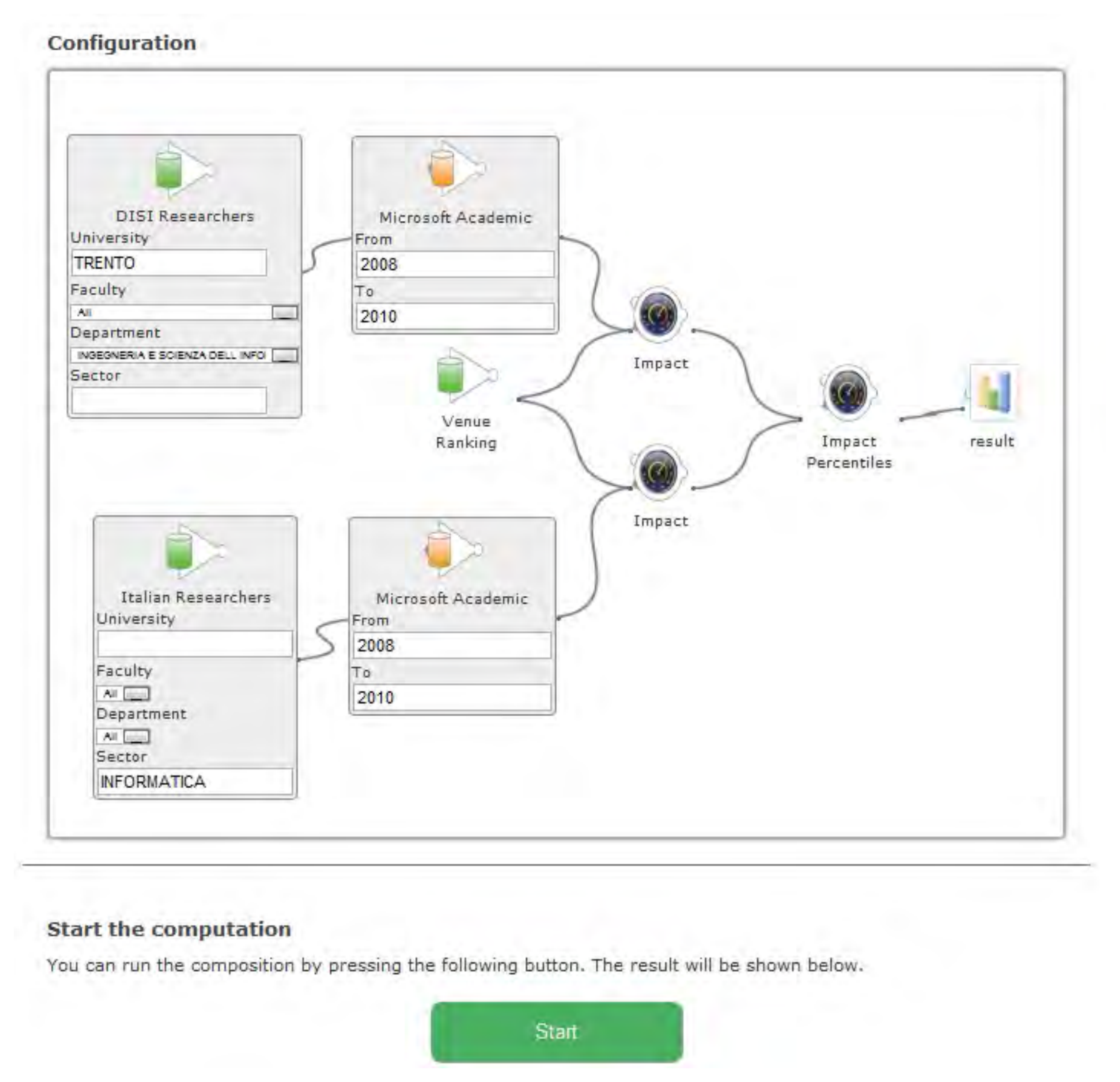}
  \caption{Prototype-2: Showing a more customizable approach, where user allowed to configure the components}
  \label{fig:tool-2}

\end{figure}

\begin{figure}[ht]
  \centering
    \includegraphics[width=0.85\columnwidth]{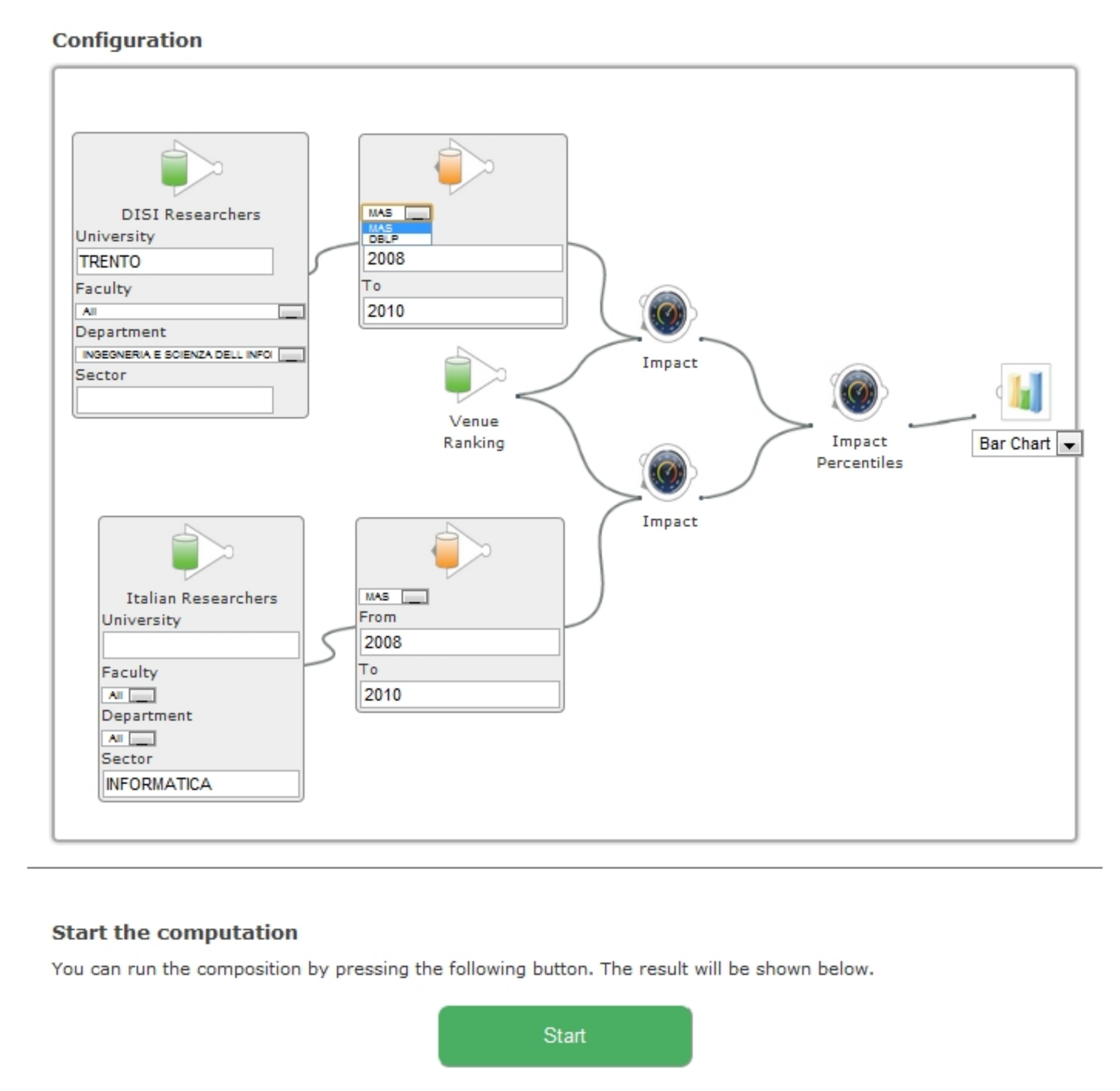}
  \caption{Prototype-3: showing a more flexible and customizable tool to the users}
  \label{fig:tool-3}
\end{figure}

First, the participants were asked about their technical and domain -skills (to assign them to the appropriate user category as mentioned in the previous section), then we instructed them to perform the following tasks in steps, providing help only upon explicit request:

\begin{enumerate}

\item In the first step, we introduced prototype 1 as depicted in Figure \ref{fig:tool-1}. This prototype consists of an explanation of the scenario and a pictorial form of it in which different components are connected making a composition. On the same page we provide a button to start the execution of the process. That is, the mashup is pre-built (by us; acted as a developer) and can only be executed by the participants. The participants were only allowed to click the start button in order to execute the process and the final results shown to them on the same page. The provided composition did not allow the participants to make any type of interaction with the components thus restricting them only with the control over process execution (i.e., start or stop).
	
\item In the next step, prototype 2 was presented, as depicted in Figure \ref{fig:tool-2}. This time the participants were presented the same scenario with configurable components, as it can be seen in the figure that the components' configuration panels are open. Thus allowing participants to configure components through components' configuration panels, that is various parameters (e.g., filtering options, date ranges etc.) values can be of user-defined. Once the configured components are ready, the participants can start the process execution using a \emph{start} button.
	
\item In this step, prototype 3 was presented to the participants as depicted in Figure \ref{fig:tool-3}. The participants are now allowed to change components in the mashup model by choosing among different implementations of the same component class, again through configuration panels. For example, a Microsoft Academic Search (MAS) data source can be replaced with a DBLP data source. The possibility to change the configuration parameters and to substitute the components, provides more flexibility to the participants in order to tune the scenario according to their needs.
	
\item During this step, a fully functional mashup composition environment (i.e., ResEval Mash) was presented to the participants. 
%Figure \ref{fig:tool-4} depicts the screenshot of the tool. 
This tool provides the possibility to drag-and-drop components onto a composition canvas, to fill their configuration parameters, to connect them together and to execute composition, hence giving maximum flexibility to the participants so they can be as expressive as they want. The participants during this step were allowed to use all the features of the ResEval Mash tool, as during this step they were asked to compose a mashup composition based on the scenario they have been experiencing during the previous steps.

\item Finally, we presented to the participants the Yahoo! Pipes tool, which is a popular generic mashup tool. An example pipe, i.e., a composition, is shown to the participants as a short tutorial to introduce the tool. Then, participants were asked to imagine how they would implement our specific scenario in Pipes and were asked to implement it to whatever level they can reach.
	
\end{enumerate}

\subsection{Questionnaires}
\label{sec:us1-questions}
After each step of the procedure, which are mentioned in the previous section, the participants were presented with a set of questionnaires to answer. The questions related to various aspects like, what difficulties they encountered, their understandability level, their suggestions for the improvements etc. were asked. In following the detail of these questions is presented, and the metaphors used for recording their answers are mentioned in (parentheses):

\begin{enumerate}

\item What is your opinion about the difficulty level of this task? (a set of six radio-option buttons ranging from extremely difficult to extremely easy were presented)

\item What are the main difficulties you encountered? and why? (a multi-line text box)

\item What are the advantages of this step as compared to the manual/previous approach? (a multi-line text box)

\item What do you think are the disadvantages of this step as compared to the manual/previous step approach? (a multi-line text box)

\item Do you think the increased flexibility of the tool with respect to the previous step would be useful for you to adapt the process to your specific needs? (ten radio-option buttons ranging from extremely efficient to extremely inefficient)

\item Do you understand how the process executed behind the scene? (ten radio-option buttons ranging from easily understandable to not understandable at all)

\item Do you feel comfortable having such control over the process execution or you would like to have a clearer idea of what is going on? (ten radio-option buttons ranging from extremely comfortable to extremely uncomfortable)

\item Based on your experience during this step, would you prefer to do this task by yourself using a similar tool or you would prefer explaining and asking a technician to implement it for you? (two options were presented: (i) I'd like to do by myself (ii) I'd like to ask technician)

\item Do you have specific suggestions/requirements to improve usability/usefulness of this tool? (a multi-line text box)

\item Are you happy about the flexibility given by the current tool or you would like to be able to change something to adapt the tool to be used for solving other similar problem? (a multi-line text box)

\item In your opinion what is the difference between previous and this step? (a multi-line text box)

\end{enumerate}

The participants completed the above mentioned questionnaires by themselves after the completion of each step, help was provided if asked by any participants. Finally, after the 5th step that is described in the section \ref{sec:study1-procedure}, in which we also presented the Yahoo! Pipes tool, we asked participants a set of general questions in order to constitute an overall consensus among all the tools with respect to the flexibility, usefulness and complexity they offer. These questions are presented below, answering options are in parentheses. Throughout the phase of answering questions the participants were allowed to ask assistance if they have difficulty in understand a question. Final questions were as follows:

\begin{enumerate}

\item What do you think what is the complexity of computing a research evaluation metric manually? (ten radio-option buttons ranging from extremely complex to extremely easy)

\item Now, you have seen all the different tools (steps), how would you judge them? (for all the five steps, we presented three radio-option buttons with \emph{flexible}, \emph{useful} and \emph{complex} as options.)

\item Which step/tool would you consider closer to your needs considering both simplicity and flexibility among 5 different tools and why? (a multi-line text box)

\item Would you use this tool is your real life? (yes/no)
\end{enumerate}

%\begin{figure}[ht]
%  \centering
%    \includegraphics[width=\columnwidth]{tool-4}
%  \caption{Prototype-4: ResEval Mash a fully functional mashup tool giving full freedom to users}
%  \label{fig:tool-4}
%\end{figure}

\subsection{Results}
As the objective of the study was to collect feedback about two main questions. First, if the participants are indeed more comfortable with domain-specific mashup tools compared to general purpose tools like Yahoo Pipes. Second, in the case domain-specific tools are preferred, which is the right tradeoff among flexibility and complexity, i.e., which of the tools 1-4 is most effective. 

\begin{figure}[ht]
  \centering
    \includegraphics[width=\columnwidth]{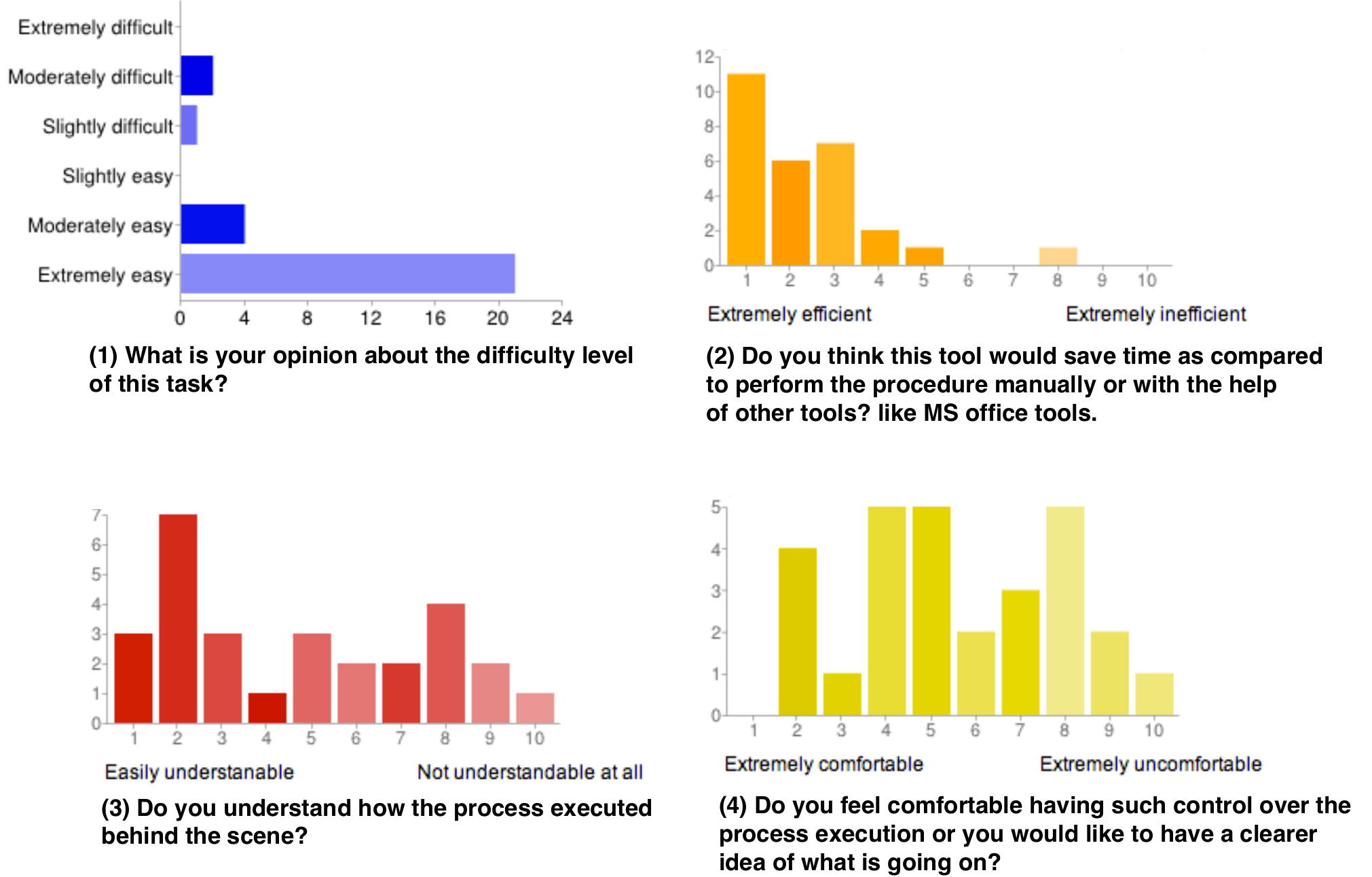}
  \caption{Results of user study-1, prototype-1}
  \label{fig:us1-step-1}
\end{figure}

\subsubsection{Results of the first step}
As for the first step in which we presented the prototype that allows the participants only to the execute process, leaving configurations and component modification options fixed (i.e., hard coded). The results of the quantifiable questions (question number 1, 5, 6 and 7) whose answers can be presented in a chart are depicted in Figure \ref{fig:us1-step-1}. The figure shows four charts in which it can be noticed that in chart-1, most of the participants found the first step extremely easy to perform, as the prototype used in this step had lowest complexity that is to only execute the process with a button click. As chart-2 depicts, most of the participants consider the approach as an efficient upon asking whether they feel the approach is time saving as compared to the other tools or manual effort. However, the chart-3 shows, a half-half division of the opinions when the participants were asked whether they understand how the process executed behind the scene or not. As no execution status was conveyed, nor a progress bar was shown, and also they were having no idea how data is flowing, so many participants remained in dark showing they did not understand it. The chart-4 shows most of the participants feel uncomfortable with the level of control that the tool offers, which implies that they certainly need more control over the execution of the process. 

Regarding the remaining questions (i.e., 2, 3, 4, 8, 9, 10 and 11; as presented in the section \ref{sec:us1-questions}) whose answers were collected in textual format, we show here some important responses. For the second question, main responses were \emph{"not flexible, but simple"}, \emph{"not difficult"}, \emph{"I can't use for my daily job, as I need to deal with loads of variations"}. For the third question typical answers were, \emph{"easier, fast, efficient", "not difficult at all", "much faster than the manual approach"} etc. Regarding the fourth question, user respond like, \emph{"you can not check whats going on behind the scene", "unskilling people", "all data are registered with more accuracy and speed"} etc. For the eighth question 90\% of the participants said "I'd like to do by myself", whereas 10\% said "I'd like to ask technician". In response to the ninth question, which was about users' suggestions, they answered as, \emph{"having an interface to change the parameters that you need to look for", "need much more customization", "I'd like to make more choices"} etc. were the main ones. Mostly, similar to the ninth, in response to the tenth question mainly participants asked for \emph{"I'd like to change configurations", "need more flexibility", "there is no flexibility"} etc.

\subsubsection{Results of the second step}

\begin{figure}[t]
  \centering
    \includegraphics[width=\columnwidth]{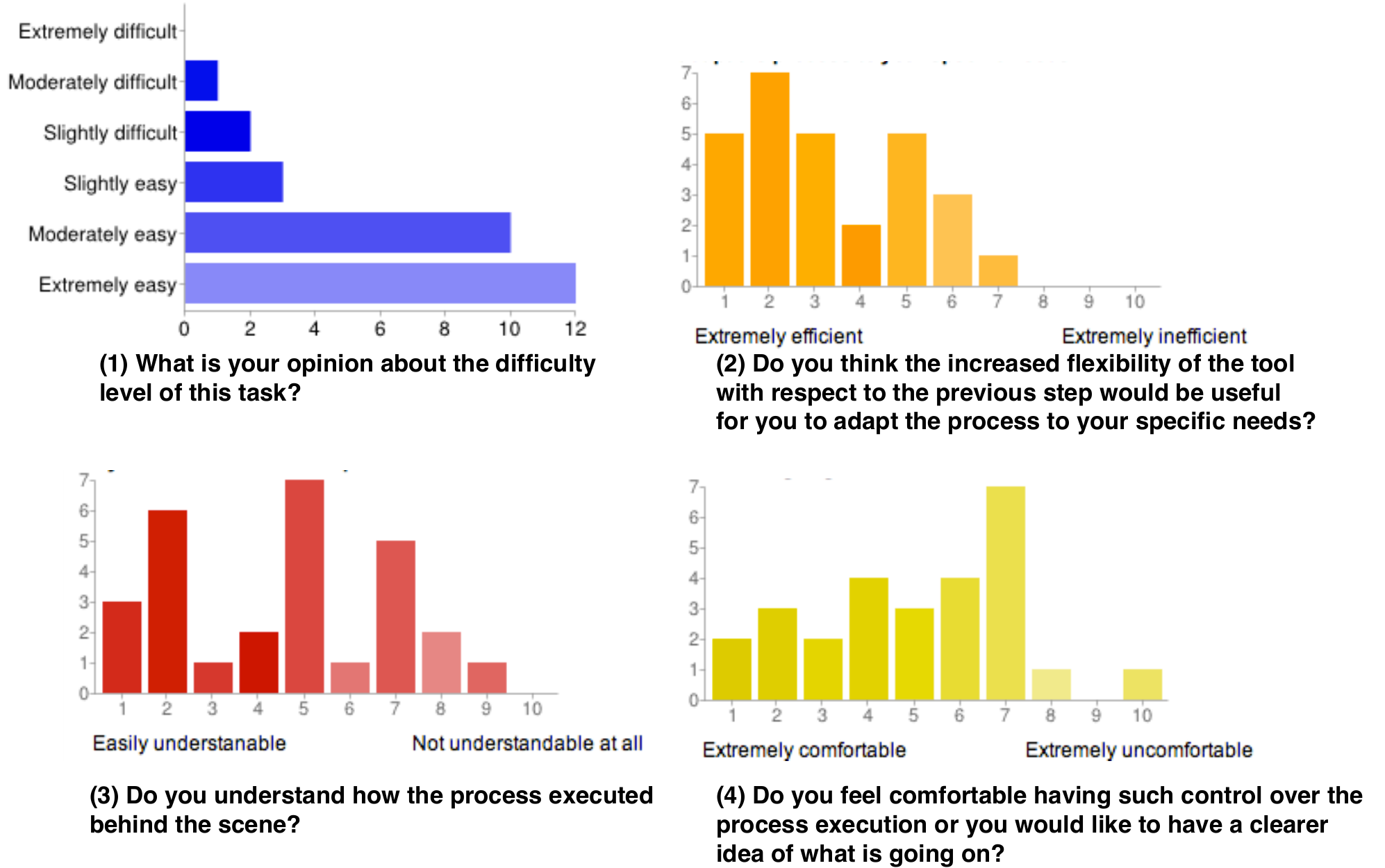}
  \caption{Results of user study-1, prototype-2}
  \label{fig:us1-step-2}
\end{figure}

In response to the second step, which is mentioned in the section \ref{sec:study1-procedure}, Figure \ref{fig:us1-step-2} depicts the charts of the four questions that are 1, 5, 6 and 7. The chart-1 shows that again the participants found step-2 task (i.e., prototype-2) easy to use. As shown in the chart-2, most of the participants liked the increased flexibility of the tool. The third chart shows that almost an equal division of the perception of the understanding of the execution of the process. However, most of the participants still not comfortable using this tool and thus demanded for more control over the process, as depicted in the chart-4. Regarding questions (i.e., 2, 3, 4, 8, 9, 10 and 11; as presented in the section \ref{sec:us1-questions}) whose answers are in text format, mainly the users' responses to the second question were \emph{"more flexible than the previous but not flexible to change components", "understanding configuration parameters"} etc. For the third question responses were \emph{"parameteric approach is better than the previous one", "its flexible", "you can choose different parameters", "auto-completion is perfect for me"} etc. The fourth question received responses like \emph{"still no idea what's behind the process", "still not able to change the execution flow", "not always reliable"} etc. 

In response to the eighth question, 90\% said \emph{"I'd like to do by myself"} and 10\% said \emph{"I'd like to ask technician"}. The ninth question where participants gave suggestions as \emph{"if you provide me right choice of component then I can do myself", "it should provide the ability to check input parameters", "details about what kind of data are being used", "better than one button"} etc. The tenth question received like \emph{"more flexibility would be good for me", "I'd like to change the flow of execution", "more options are needed"} etc. In response to the eleventh question, which was not the part of the first step, the participants respond as \emph{"more customization of the search", "i have more flexibility", "more flexibility"} etc.

\subsubsection{Results of the third step}
\begin{figure}[t]
  \centering
    \includegraphics[width=\columnwidth]{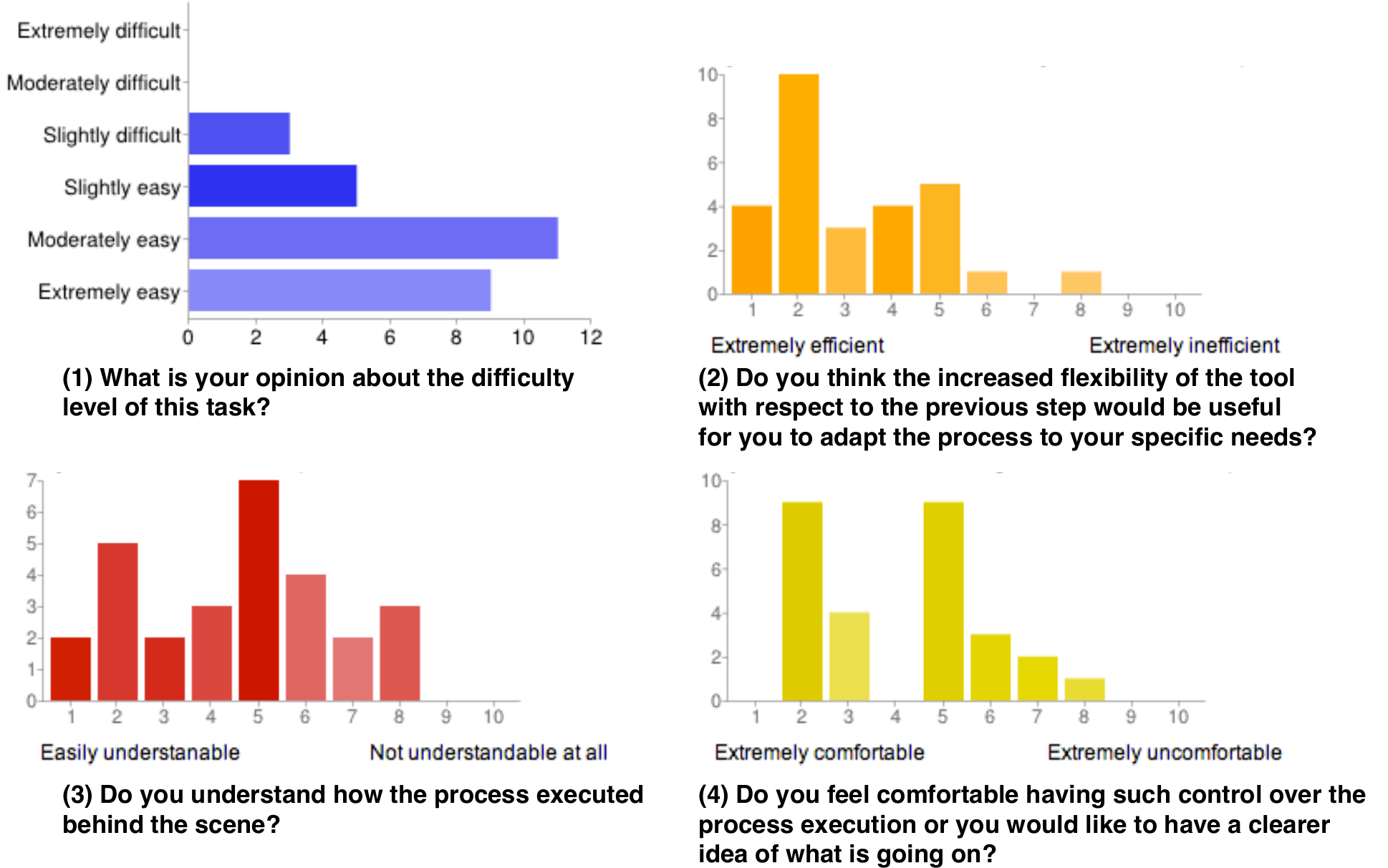}
  \caption{Results of user study-1, prototype-3}
  \label{fig:us1-step-3}
\end{figure}

In response to the third step, which is mentioned in the section \ref{sec:study1-procedure}, Figure \ref{fig:us1-step-3} depicts the results of the four questions (i.e., 1, 5, 6 and 7; presented in the section \label{sec:us1-questions}). As shown in the chart-1, majority of the participants still feel that the difficulty of the presented step is manageable and hence easy to handle. Likewise, most of the participants considered the increased flexibility still an efficient approach, as shown in the chart-2. However, the chart-3, which conveys the understanding level of the execution of the process, still plot that largely the process execution was less understandable for many. The chart-4 shows many of the participants remain uncomfortable, which means they still demand for more control over the process. On the other side, from textual answers, for the second question main responses were \emph{"no difficulties", "more details about the sources would be good", "more documentation", "none"} etc. For the third question \emph{"more clear, more accurate", "more flexible than the previous one", "presentation styles changing is good"} etc. were the main responses and for the fourth question \emph{"provide more choices", "its more error prone", "none"} etc. were the main responses. In response to the eighth question, 93\% said \emph{I'd like to do by myself} and 7\% said \emph{"I'd like to ask technician'}.

As for the ninth question, the participants suggested as: \emph{"having more visual charts would be more useful", "having more sources would be good", "check configuration parameter validity"} etc. Regarding the tenth question, main responses were \emph{"yes I'm happy, but if you provide me more flexibility I will be more happy", "I'd like to alter the execution flow", "I'd like to change flow", "yes"} etc. For the eleventh question, the participants differentiated this tool with the previous as \emph{"more flexibility", "this is better", "you can choose other databases"} etc.

\subsubsection{Results of the fourth step}
\begin{figure}[t]
  \centering
    \includegraphics[width=\columnwidth]{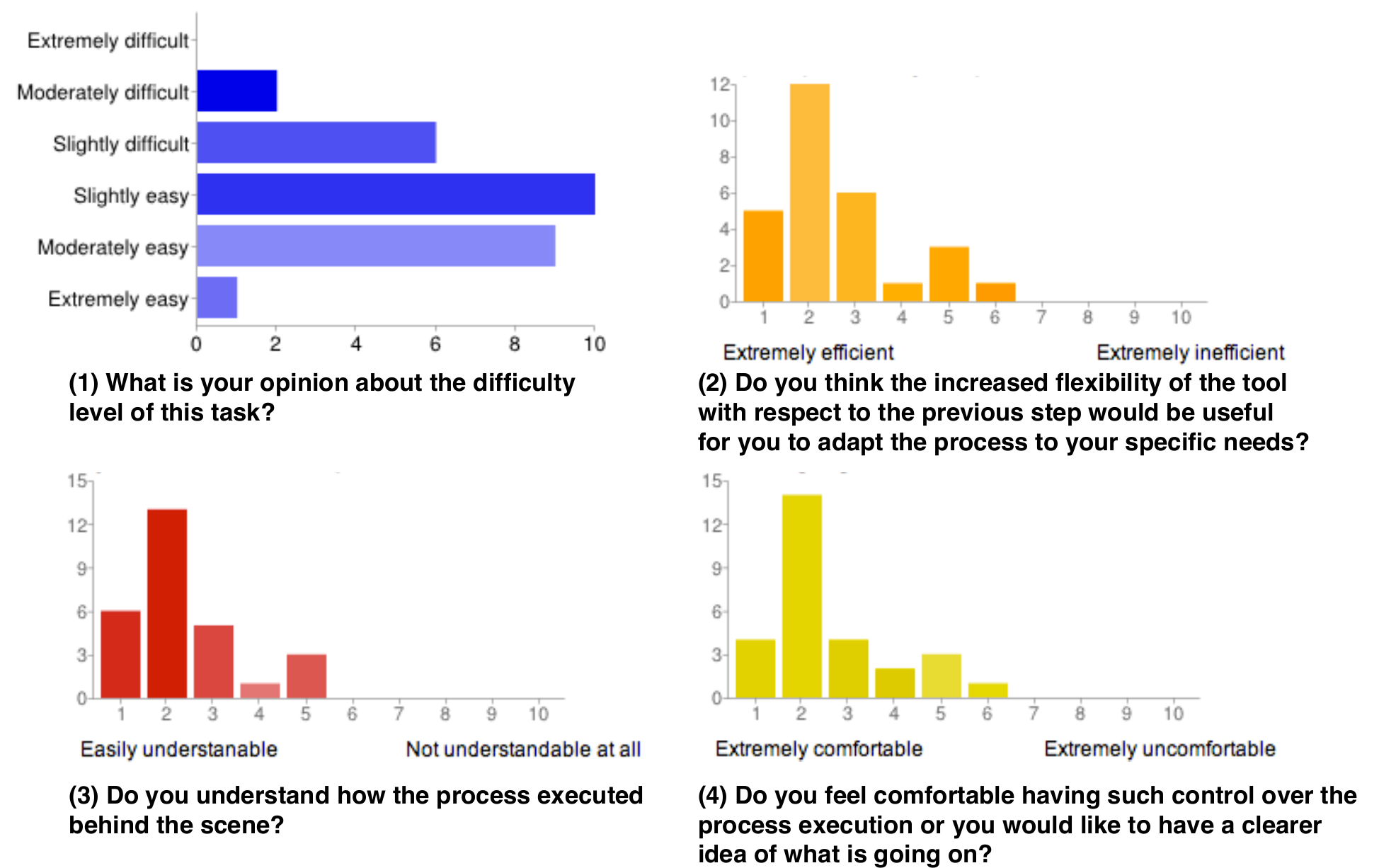}
  \caption{Results of user study-1, prototype-4}
  \label{fig:us1-step-4}
\end{figure}

As stated in the section \ref{sec:study1-procedure} that we presented to the participants the fully functional tool (i.e., ResEval Mash) during the fourth step. The responses regarding the questions (1, 5, 6 and 7) are depicted in the Figure \ref{fig:us1-step-4}. In the chart-1, it is clearly shown that this prototype slightly increased the difficulty level but still majority voted it as easy to use. On the other hand, the increased flexibility of the tool was mainly perceived positively and the majority felt it as an efficient approach, as depicted in the chart-2. However, this prototype was extremely understandable by the participants as compared to all the previous ones, as depicted in the chart-3. Moreover, as depicted in the chart-4, majority of the participants felt comfortable with the control that the prototype-4 provided.

Regarding the responses against the questions (2, 3, 4, 8, 9, 10 and 11; as presented in the section \ref{sec:us1-questions}), for the second question main responses were \emph{"no", "the more you are free in changing configuring component, the more you are risky", "need to know components", "documentation of the components required"} etc. For the third question mainly the users' responses were \emph{"Its all advantages and not disadvantages", "the more you can personalize the more you feel comfortable", "i can customize", "I can better adopt my needs", "I can reuse compositions, components'} etc. And for the fourth question, \emph{"more degree of freedom so there could be more chances of error", "more knowledge is required", "takes time to understand"} etc. In response to the eighth question, 95\% of the participants said \emph{"I'd like to do by myself"}, whereas 5\% of the users said \emph{"I'd like to ask technician"}. Main responses for the question number nine were \emph{"some training course for the user would be good", "add more components", "give suggestion/assistance to the user during composition", "check intermediate results"} etc. The tenth question was received as \emph{"I am fine, not more than this", "I think its enough", "I don't need more details than this because then we go into the programmers world", "absolutely happy", "I'm happy with the approach", "yes, happy"} etc. And finally for the eleventh question, the participants' responses were like \emph{"interesting", "more configurable and more useful", "need more skills", "more creative and more options"}.

\begin{figure}[ht]
  \centering
    \includegraphics[width=\columnwidth]{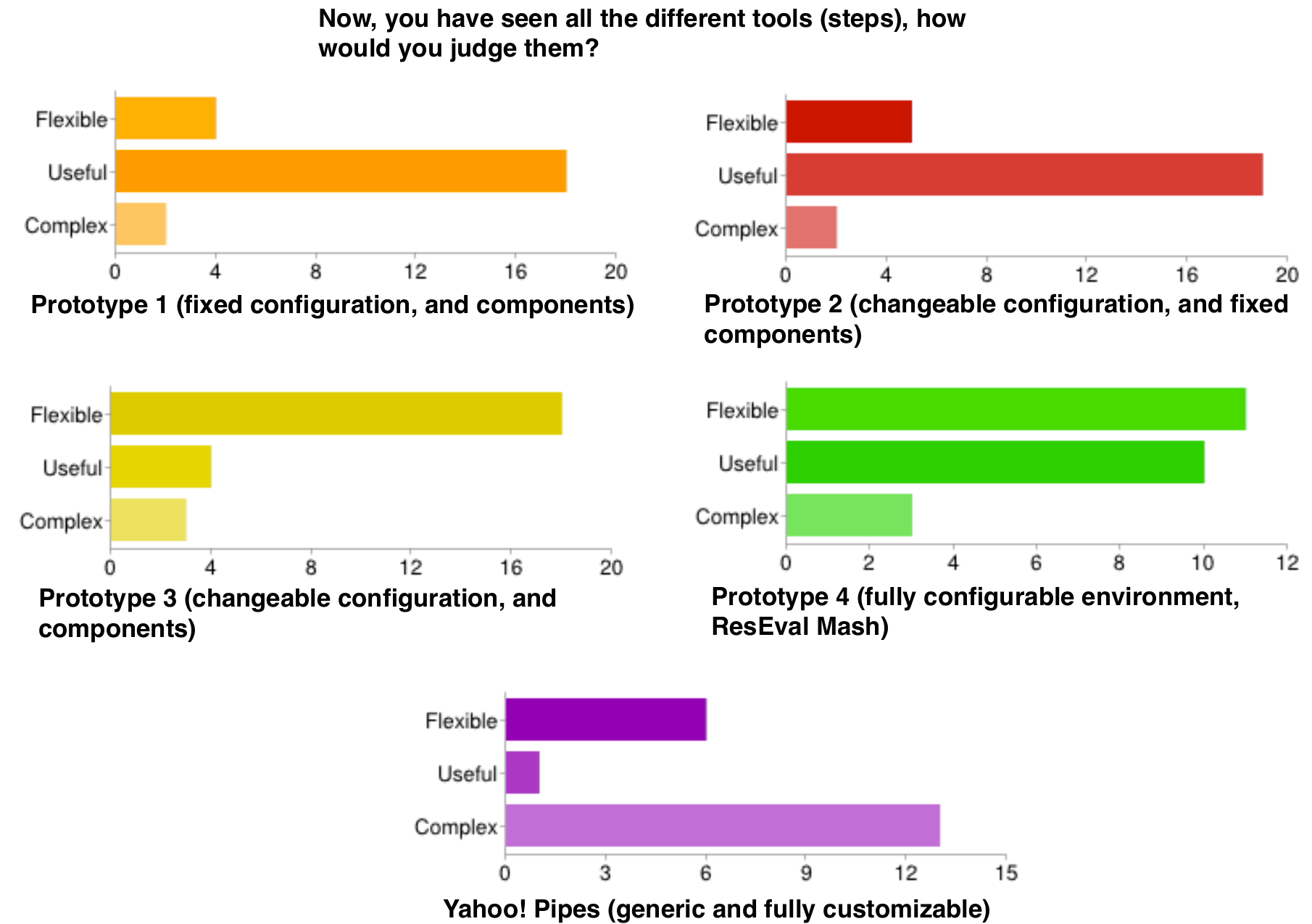}
  \caption{Results of user study-1, general results}
  \label{fig:us1-step-final}
\end{figure}

\subsubsection{Results of the fifth step}
As mentioned in the section \ref{sec:study1-procedure} that at the end of the study we asked a set of general questions to the participants. Figure \ref{fig:us1-step-final} depicts the results of the important questions, that is how users judge all different tools in terms of flexibility, usefulness and complexity. As it can be noticed that prototype-1 is little flexible, more useful and little complex, whereas prototype-2 is a bit more flexible, highly useful and little complex too. The prototype-3 is considered more flexible than its usefulness and still considered as little complex. However, it is clearly shown that in case of the prototype 4, which is the full mashup environment (i.e., ResEval Mash), the flexibility is largely increased along with its usefulness. The participants found it more useful and flexible as the tool gives freedom to drive their requirements as they want. As compared to the previous three prototypes, the complexity of prototype 4 increased and that is normal. Because most of the participants stated in their remarks that they would need an introductory training to fully utilize its advantages. In the same figure the results on Yahoo! Pipes tool can also be seen (the last chart). As anticipated it has a high complexity, very low usefulness for the users, however, it is fairly flexible as it offers more options to play with, but mainly suitable for programmers only. 

\subsection{Evaluation Analysis \& Discussion}
In the previous section we have presented the results, which reflect the exact representation of the participants' responses. However, this section provides an analysis of the overall study in which we analyze inter as well as intra -steps variabilities and patterns particularly focusing on the participants' skills (e.g., technical, non-technical). Mainly, the technical expertise of the participants mentioned in the table \ref{tab:User-study-1-details}, which shows 19 out of 28 participants have moderate/low technical and 9 have good technical skills. As mentioned earlier that users with good technical expertise are familiar with the programming languages and they were involved in some sort of programming. On the other hand, users with low technical expertise are not programmers. Based on this, we can divide these users into two groups, that is, technical group (i.e., those who know web services, programmings, etc.) and non-technical group (i.e., those who know MS world, Excel etc. but do not know programming). 

\begin{figure}[t]
  \centering
    \includegraphics[width=\columnwidth]{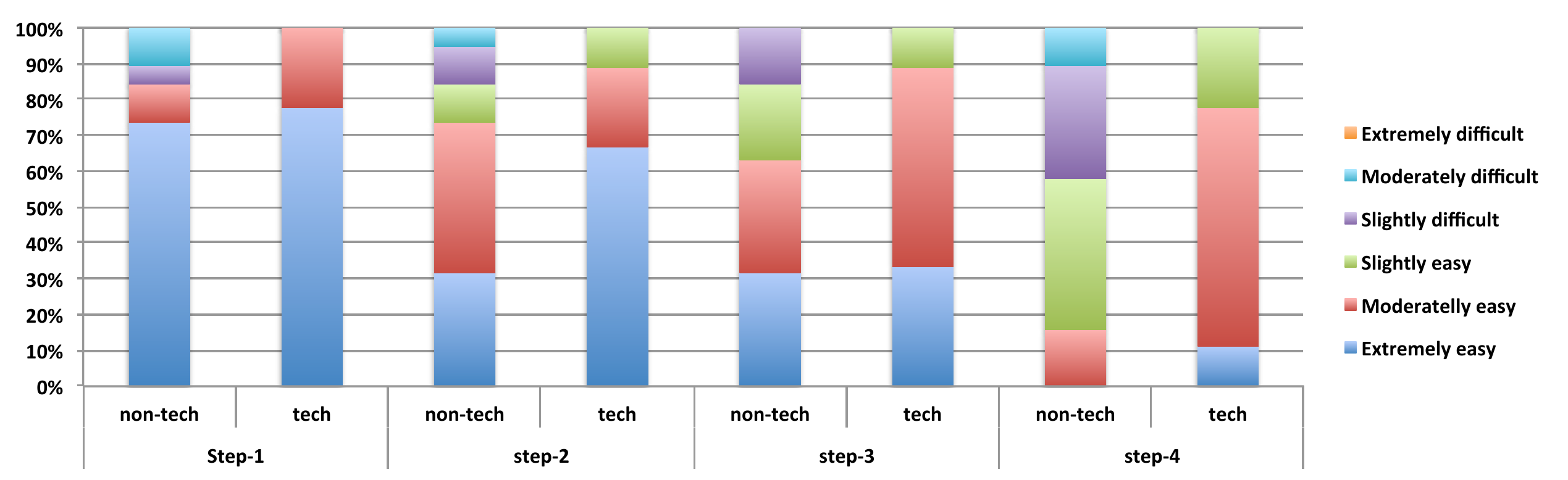}
  \caption{For both tech and non-tech groups the difficulty level of steps (1-4)}
  \label{fig:us1-step1-4-q1}
\end{figure}

Regarding the question number 1, as presented in the section \ref{sec:us1-questions}, for all the four steps, as described in the section \ref{sec:study1-procedure}, the difficulty level slightly increased for the non-technical group as compared to the technical one. Figure \ref{fig:us1-step1-4-q1} depicts the distribution of the both groups along various difficulty levels. Non-technical group faced difficulties as the process started providing more flexibility and customization (see non-tech column of step 3 \& 4). However, one can notice that even during the step-4 in which the prototype-4 was presented, the majority of the non-technical participants are still within the boundaries that they consider it easy to use, and a few considered it "slightly difficulty". In response to the very next question (i.e., question-2, where participants were asked to provide textual answers about what difficulties they faced) during the step-4, most of the participants demand for more training and tutorial prior to the use of the tool to effectively deal with difficulty.

\begin{figure}[t]
  \centering
    \includegraphics[width=\columnwidth]{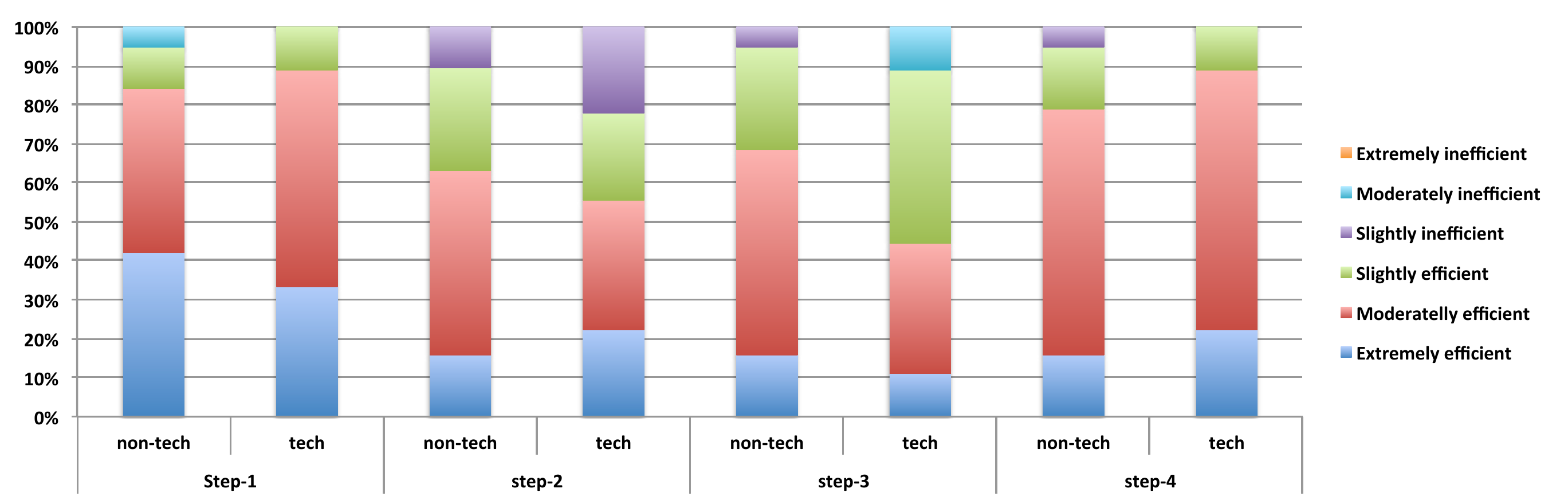}
  \caption{For both tech and non-tech groups, how increased flexibility perceived for all steps (1-4)}
  \label{fig:us1-step1-4-q5}
\end{figure}

\begin{figure}[t]
  \centering
    \includegraphics[width=\columnwidth]{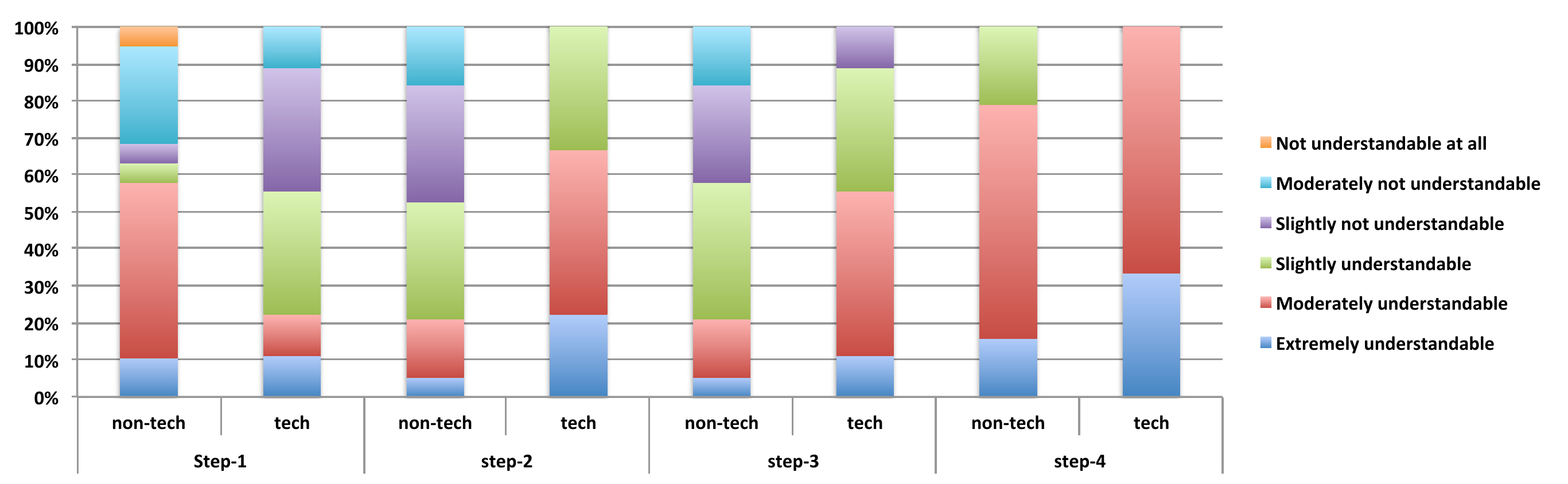}
  \caption{For both tech and non-tech groups, process execution understandability for all steps (1-4)}
  \label{fig:us1-step1-4-q6}
\end{figure}

In figure \ref{fig:us1-step1-4-q5}, the distribution regarding both, the technical and non-technical groups in terms of process adaptation with respect to increased flexibility for all the four steps is depicted. Clearly for all the steps the "moderately efficient" pattern is consistent that is to some extent increased for the fourth step. The technical group considered the prototype-4 more efficient than the non-technical. Even then the majority of the non-technical participants voted for the efficient option except one participant who considered it slightly inefficient. Figure \ref{fig:us1-step1-4-q6} depicts the level of understandability of the both groups. A very low understandability level can be seen in the step number 1 \& 2 and a slight increase is detected in the step-3 but still majority could not easily understand how the process execution is performed. However, for the step number four both non-technical and technical groups shown a good understanding of the process execution. Obviously technical users have advantage than non-technical users with their technical skills, that is the reason the process execution during the step-4 were easily understandable for the former group. 

\begin{figure}[t]
  \centering
    \includegraphics[width=\columnwidth]{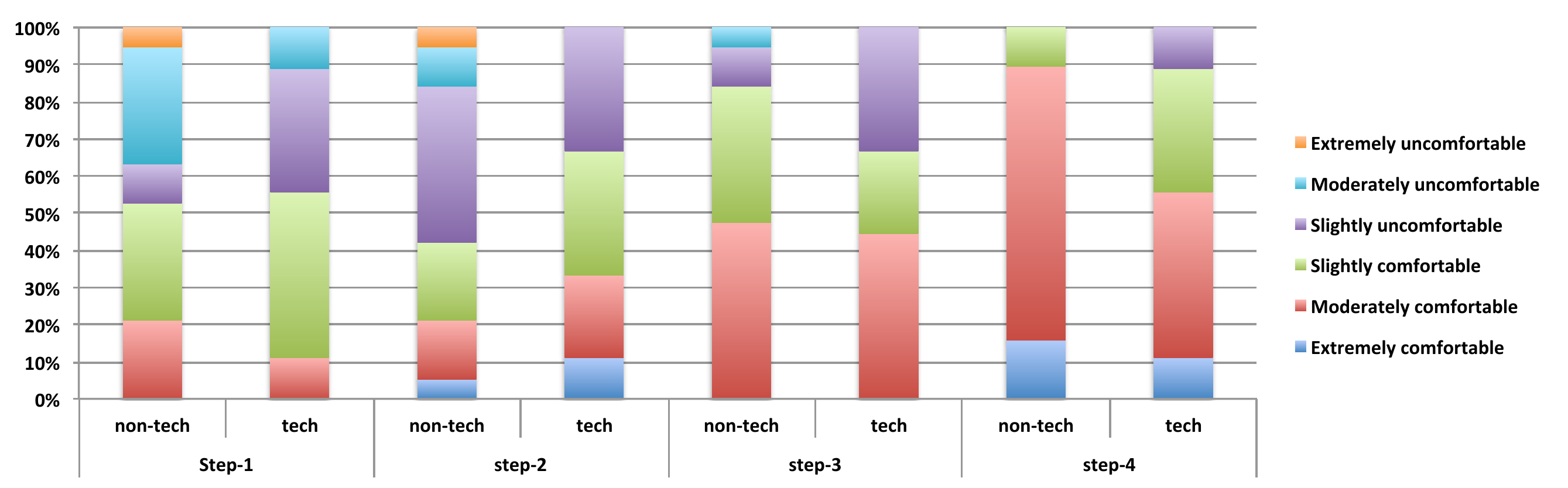}
  \caption{For both tech and non-tech groups, control over process results for all steps (1-4)}
  \label{fig:us1-step1-4-q7}
\end{figure}

The accumulated results of the both groups (i.e., technical and non-technical) on question number 7, which represents participants' comfortability about the given control over the process, are depicted in Figure \ref{fig:us1-step1-4-q7}. Clearly the demands for more control over the process have emerged till the step number 3. However, one can notice that during the step-4 majority of the participants of both groups feels comfortable with the given control, that is they now feel they can tailor it as they want up-to the level of their expertise. The demand for more control mainly asked by the technical participants, as a small number of participants still want to go beyond the flexibility that the prototype-4 provides, whereas non-technical participants largely consider prototype-4 as a boundary line for them, or otherwise the complexity will increase, responded many non-technical participants. 

\begin{figure}[t]
  \centering
    \includegraphics[width=\columnwidth]{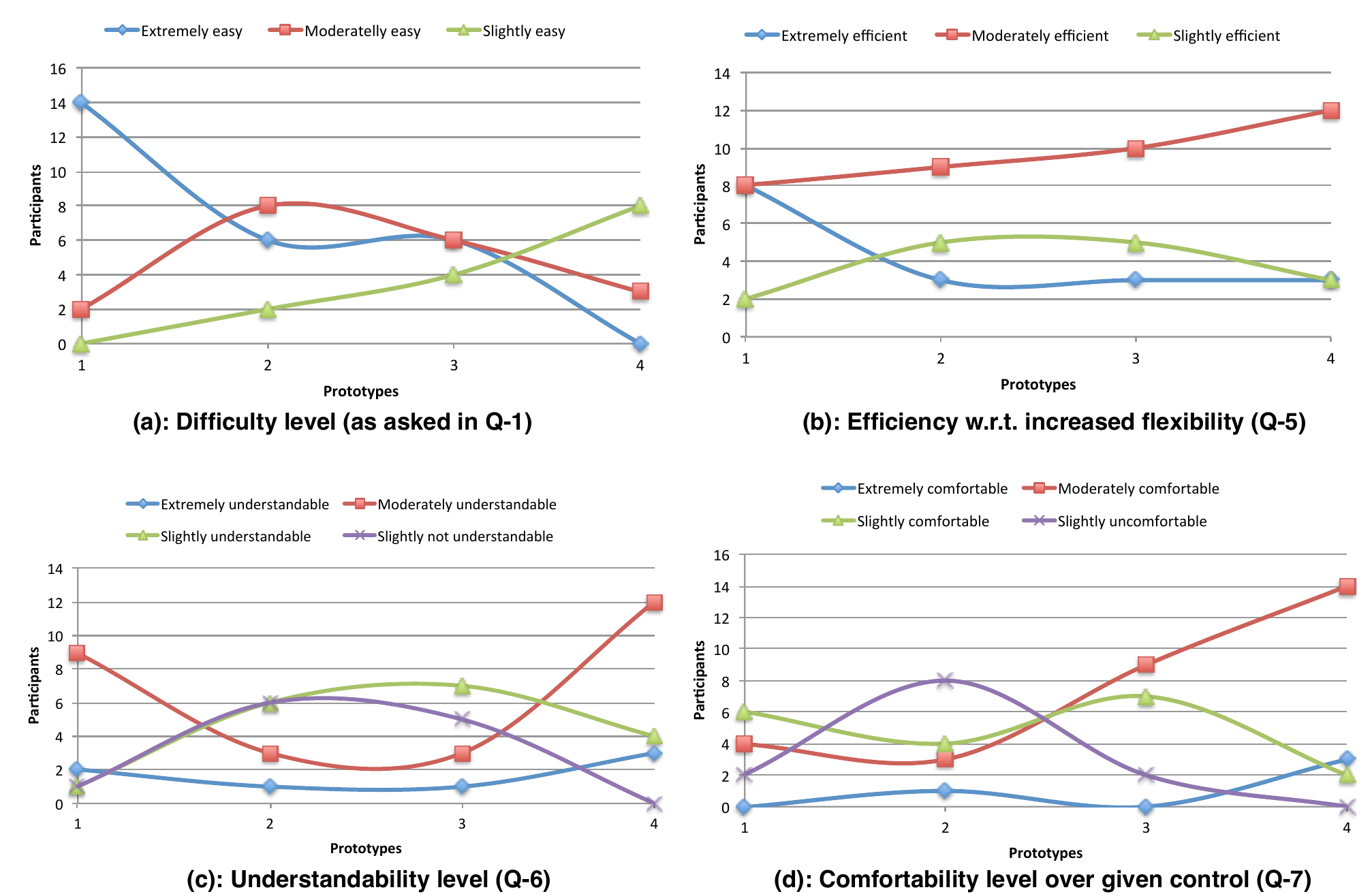}
  \caption{Various results of non-technical participants for all four prototypes against questions (1, 5, 6, \& 7)}
  \label{fig:us1-step1-4-vr}
\end{figure}

As an overall, for non-technical participants the difficulty level increased against each richer prototype, however, most of the non-technical participants still found prototype-4 easy to use, as shown in chart (a) in figure \ref{fig:us1-step1-4-vr}. The demands for more flexible tool emerged from both technical and non-technical groups and the increased flexibility still within the range of users, as chart (b) depicts in figure \ref{fig:us1-step1-4-vr}. Relatively low but some participants, during the first three steps, with high technical background did not even understand how the process executed behind the scene, whereas participants having low technical skills did not understand at all as chart (c) depicts in figure \ref{fig:us1-step1-4-vr}. Largely all types of participants were uncomfortable with the give control over the process for the first three steps, however, a clear satisfaction on such control can be seen in the fourth prototype which was presented in the step-4, especially for non-technical users as depicted in chart (d) in figure \ref{fig:us1-step1-4-vr}. 

Likewise, during the final step in which the generic mashup tool (i.e., Yahoo! Pipes) was presented, the non-technical participants said the tool is too complex for them and that they are not able to understand most of the component and parameter names. However, the technical participants were not scared by the programming-oriented functionalities and terminology of the tool, but they identified a problem with the level of abstraction of the tool. They think the tool could be potentially able to allow the design of a process similar to our reference scenario. However, a complete implementation of the scenario in Pipes would be too complex and time consuming, since the level of abstraction of the provided functionalities is low and not focused on the domain constructs. All participants in this test declared to prefer the domain-specific tools, like ResEval Mash.

Regarding the tool they consider best among the domain-specific ones -- considering both their needs and abilities -- the majority of the participants stated that their preference go to the fully functional mashup composition environment of step-4 (i.e., prototype-4). This was motivated by the fact that the tools up to step 3 are too rigid and cannot adapt to different scenarios. Indeed, in the feedbacks after each step a need for more flexibility is evident until step 3, while in step 4 most of the participants declared that they think the flexibility level is enough to make the tool actually useful. Most of the participants also think they are able to cope with the increased complexity. 

Moreover, the majority of the participants was overall satisfied with the prototype-4 (i.e., ResEval Mash), they also pointed out some shortcomings. In particular, some participants would have liked some assistance during the design phase, e.g., in the form of contextual information about the components' functionality and input and output data they use. Other users were instead more concerned about the availability of components providing all the necessary functionality and allowing for high customization (e.g., specification of custom filters, custom impact functions). Another common concern was related to the trust in the final result. The transparency provided by the tool at runtime was useful to establish trust, but there is still room for the improvements.

In summary, the results of this user study show that there is a clear preference for domain-specific tools and that users can indeed imagine to use such tools to build their own processes (only 1 user said that, instead of using tool 4 by himself, he would prefer to ask a technician). Both classes of the participants judge the complexity of the tool-4 as acceptable and think the benefits in terms of time and effort savings are in balance with the effort it takes to learn its proper and effective use.

\section{Usability Evaluation: User Study - 2}
\label{sec:us-2}
The suggestions, which were gathered during the first evaluation procedure, were addressed and incorporated in the tool. Mainly we focused on the prototype-4 (i.e., ResEval Mash) to increase its usability and intuitiveness, and domain-syntax was also introduced in components' list, process transparency enhanced by conveying more intuitive execution status. Moreover, an important and much requested feature, which is to give suggestions while composing compositions, is added. Now the tool provides suggestions on the possible connections that a component can make with other components. This feature would surly enhance composition problems that non-technical users could face. Now the tool also shows helping tips regarding various things that we think would help users to effectively adjust to the user experience that the tool provides.

A summative evaluation was conducted to analyze the user experience with improved ResEval Mash. The results reported in this section concentrate on usability, with an emphasis on the role of prior experience on learning. Lessons learned from the first evaluation study, this time we give an introduction of the mashup tool to participants before they use it. Prior experience was differentiated in two categories which are fundamental in our approach to mashup design: domain knowledge and computing skills. Domain knowledge was controlled by selecting all users with expertise in research evaluation, computing skills varied in the sample from people with no programming knowledge at all, to expert programmers.

The study applied a concurrent talk-aloud protocol, a technique requiring users to verbalise all their thoughts and opinions while performing a set of tasks. Verbalisation capture techniques have been found to be particularly effective when conducting experimental investigations, which provide an opportunity to study communication between products, designers and users \cite{jarke1998scenario}\cite{rouse1986looking}. The responses given during task completion are considered more representative of the behavior and problems users have during assessment \cite{hands2001recency} and concurrent talk-aloud protocols have been shown to encourage participants to go into greater detail, to provide more in-depth evaluation, and help pin-point usability problems and places where their expectations fail to be met \cite{teague2001concurrent}. 

%\begin{table}[h]
%\centering
%\begin{tabular}{ | l | l |}
%\hline
%\textbf{IT Skills} & \textbf{Position} \\ \hline
%High Computing Skills	& {PhD Students (3), Post-doc (1), Senior Faculty Member (1)} \\ \hline
%No Computing Skills	& {Administrative People	 (3), PhD student (1), Senior Faculty Member (1)} \\ \hline
%\end{tabular}
%\caption{User categories}
%\label{tab:UserCategories}
%\end{table}

\begin{table}[h]
\centering
\begin{tabular}{ | l | l | l |}
\hline
\textbf{Users} & \textbf{Position} & \textbf{Computing Skills}\\ \hline
3 & Administrative People & No Computing Skills \\ \hline
1 & PhD student & No Computing Skills \\ \hline
1 & Professor &No Computing Skills \\ \hline
3 & PhD Students & High Computing Skills \\ \hline
1 & Post-Doc & High Computing Skills \\ \hline
1 & Professor & High Computing Skills \\ \hline
\end{tabular}
\caption{User categories}
\label{tab:us2-UserCategories}
\end{table}

\subsection{Evaluation Procedure}
Ten participants covering a broad range of academic and technical expertise were invited to use ResEval Mash. At the beginning of the study, they signed a consent form presenting ResEval Mash as a tool for allowing non-programmers to develop their own computing applications. Then, they were asked to fill in a questionnaire reporting their computing skills and knowledge about research evaluation alongside some basic demographic information (e.g., age and job position). Specifically, participants were asked to estimate their skills with the use of software similar to the Microsoft Office Suite tools, programming languages, flowcharts and mashup tools, on a 4-point scale, ranging from very skilled to no skilled at all. They were also presented with a list of 21 concepts related to research evaluation and asked to indicate for each of them whether they were aware of these and able to understand their meaning, on a 2 point scale (yes vs. no). 

\begin{figure}
  \centering
    \includegraphics[width=\columnwidth]{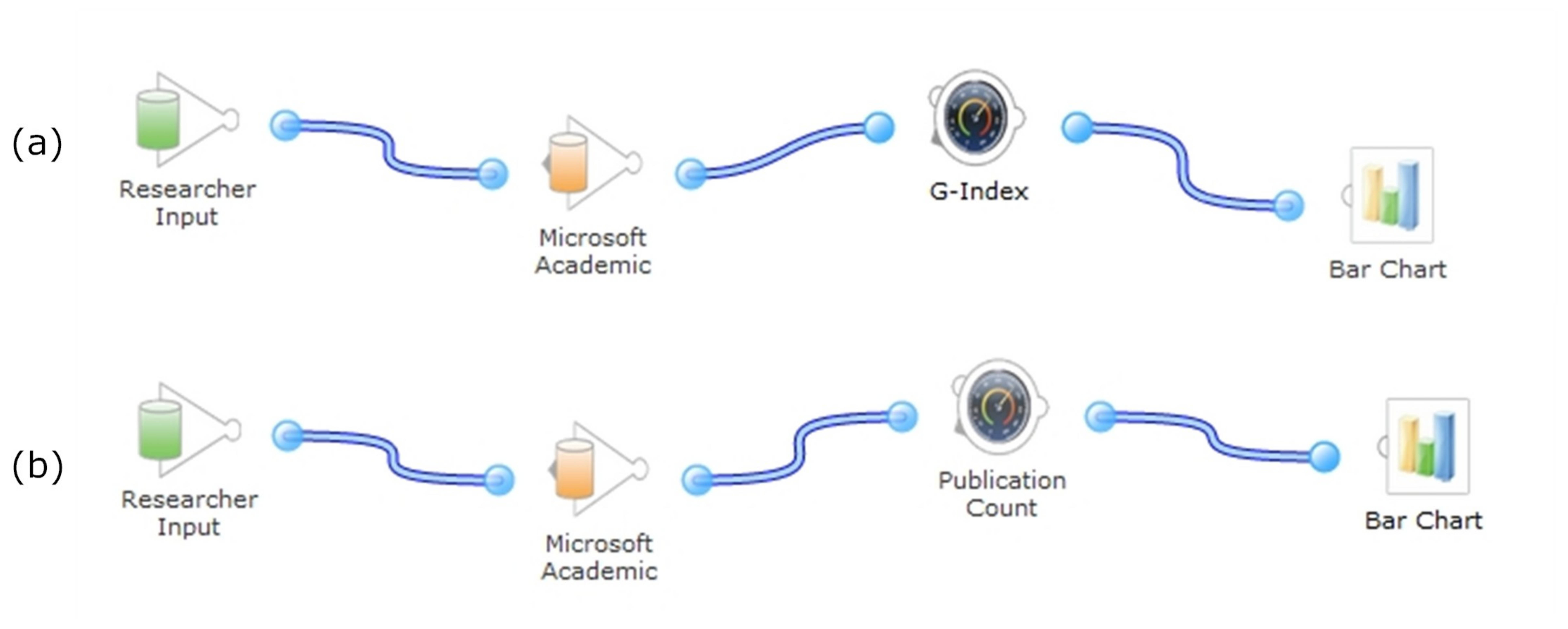}
  \caption{Mashup compositions to compute G-Index (a) and publication count (b)}
  \label{fig:gindex}
\end{figure}

After the questionnaire, participants watched a video tutorial (lasting approximately 10 minutes) that instructed them how to operate ResEval Mash. The video introduced the basic functionalities of the tool, quickly explaining the concept of components, configuration parameters, and data compatibility. It then showed how to create a simple mashup of 4 components to display the H-index of the researchers of the Department of Computer Science and Engineering of the University of Trento on a bar chart according to the Microsoft Academics publication source. Finally, the video presented another mashup example used to summarize and reinforce the concepts shown up to this point, where 4 components were connected to visualize on a bar chart the G-index of a researcher (Figure \ref{fig:gindex}.a).

After training, participants were asked to use the system and to perform a series of tasks as described follow: 
\begin{enumerate}

\item The first task asked people to start from the first composition presented in the video tutorial and to modify the year parameter of the Microsoft Academic component, to select a different department from the Italian Researchers component and finally to replace the publication source component currently used in the composition with the Google Scholar component.

\item The second task required them to design a composition to compute the participant's own publication count and visualize it on a chart. The correct solution required linking together 4 components, as highlighted in Figure \ref{fig:gindex}.b.

\end{enumerate}

Whilst completing these two tasks, participants were asked to "talk aloud" regarding their thoughts and actions. This interaction was filmed, as was the interview that followed task completion. The interview focused on interactional difficulties experienced, the evolution of participants' conceptual understanding over time, and a detailed usability evaluation stressing a feature based assessment reporting which features were considered to be beneficial to interaction, which were understood, and what participants, as users, would like to see in the system. Finally, participants asked to answer a set of questions as listed below and answering metaphor are in braces:

\begin{enumerate}
\item What are the main difficulties you encountered? Why? (a multi-line text box was given to get users' comments)

\item Did you understand the concept of component? (a multi-line text box)

\item Did you encounter difficulty configuration the components? What? (a multi-line text box)

\item How usable did you find the tool? (choice options as: (1) I feel very comfortable and I can use it. (2) I could use it with some more training. (3) It is very unlikely I would be able to use it, even with much more training. (4) I would not be able to use it at all.)

\item Why do you think the tool is useful? (a multi-line text box)
\end{enumerate}

\subsection{Participants Description}
\begin{figure}
  \centering
    \includegraphics[width=0.90\columnwidth]{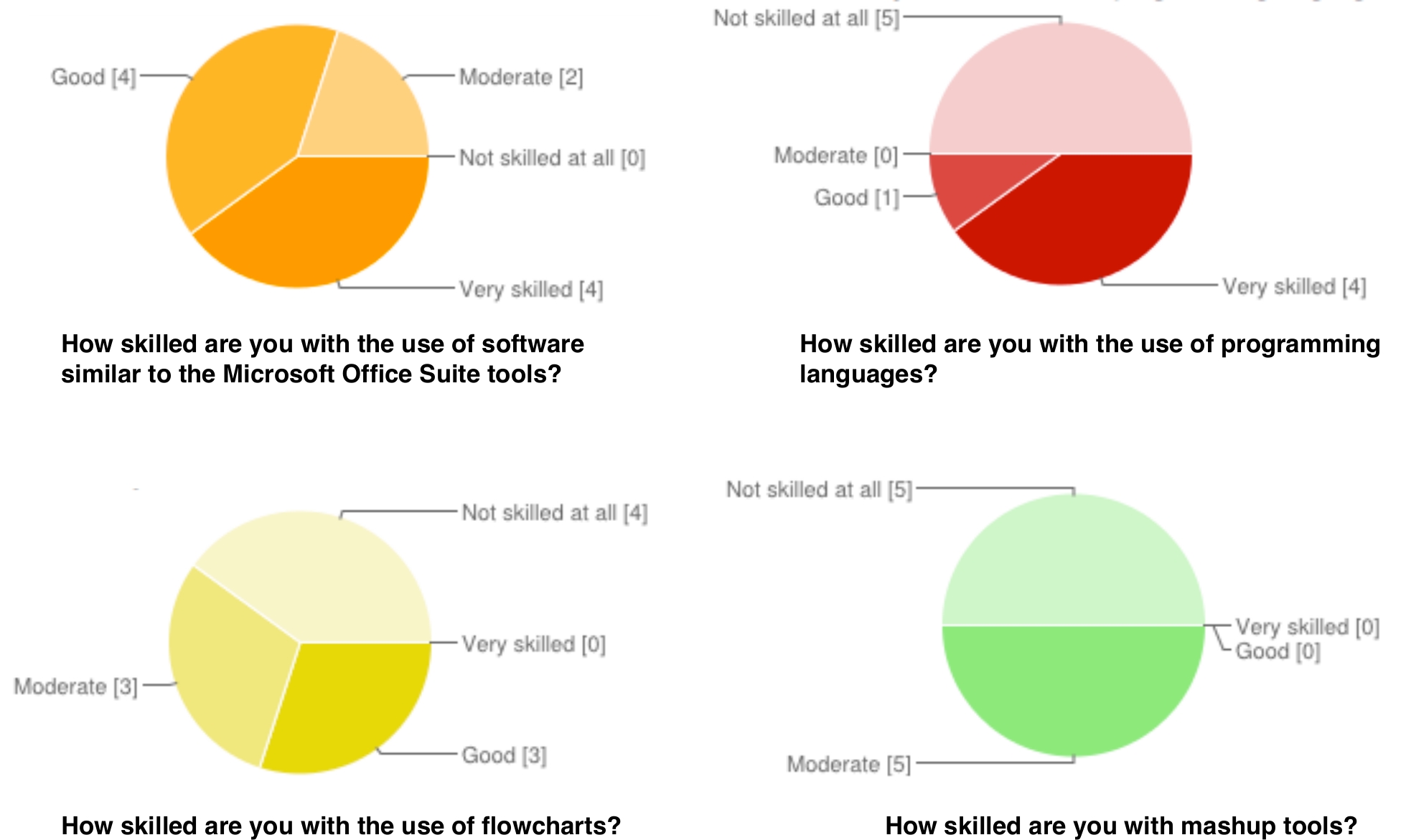}
  \caption{Participants technical skills breakdown}
  \label{fig:us2-tech-skills}
\end{figure}

The sample covered a broad range of job positions and technical skills. Half of it was composed of people who reported not being skilled in programming languages, the other half reported being very skilled or good in relation to programming languages. All the participants possessed moderate to no experience with mashup tools. The breakdown of participants according to Position is reported in Table \ref{tab:us2-UserCategories} and a detailed breakdown of the participants based on their technical skills is depicted in Figure \ref{fig:us2-tech-skills}.

On average as a group, participants had a good understanding of the domain. They possessed experience of 80\% of the 21 domain specific conceptual components listed during the pre-interaction assessment. This value ranged from a minimum of 48\% to a maximum of 100\%.  Table \ref{tab:us2-UserCategories} shows the details of the participants. 

\subsection{Analysis: Usability Evaluation}

The video-capture and talk aloud protocols were used to establish strengths and weaknesses in design and conceptual understanding. A subsequent usability assessment was used to identify the difficulties participants reported experiencing and their understanding of the key features of mashup tool interaction. %Anonimized data related to the initial and final questionnairs are available on the ResEval Mash project website\footnote{\url{http://open.reseval.org/project-updates/user-studies/}}. Videos recording user interactions with the tool can not be provided for privacy reasons.

Overall, the tool was deemed as usable and something with which participants were comfortable. Independently of their level of computing knowledge, all participants were able to accomplish the tasks with minimal or no help at all. The only visible difference reflected a variable level of confidence in task execution. The IT expert users reflected less before performing their actions and appeared to be more confident during the test. Overall, among the users with lower computing skills there was agreement that more training in the use of the tool would be beneficial, whereas this requirement did not emerge from the more skilled sample. It is worth noticing however that the people reporting this need also indicated a lower level of domain knowledge as compared to the other users.

All participants understood the concept of "component" and had no specific issues in terms of configuring or connecting components. However, the post-doc researcher suggested that it might be beneficial for the system to indicate clearly when a proposed connection was inappropriate or illegal by using color to differentiate the states of legality or appropriateness. Another participant suggested the possibility of disabling the illegal components from the selection panel when a component was selected in the composition canvas. Selection of components was highlighted as a potential problem, as identification of the right component required some time to be performed. During the study, this did not appear to be a major problem, as only a selected number of components (N= 8) were tested. Yet, it is reasonable to assume that this problem will increase as the number of available components grows. One participant suggested a search feature, to complement the current menu selection interaction mode. Regarding the question number 4, which asked "how usable did you find the tool?", 80\% of the participants responded as "I feel very comfortable and I can use it", and 20\% said "I could use it with some more training".

The task requiring tailoring an existing mashup was generally performed better than the task requiring creating a new mashup. In the latter case, a problem emerged with the selection of the first component (i.e., Researcher Input), as several participants selected the Italian Researchers component expecting to be capable to personalise their query there. Saving of configurations was also a source of uncertainty for some participants. The configuration parameters only needed to be filled in by the users and no other action from them was required. This was not clear to the users that in many cases expected an explicit saving action to be performed (e.g., through a "Save configuration" button) and that also expected a feedback to be returned on configuration completion. Several people used the "Close" button after updating the configuration, leading to deletion of the component. 

Furthermore, most participants reported some difficulty interacting with the tool due to the physical interaction of double-clicking on the component image in order to open it and been capable to configure its parameters. This constraint was referenced as taking time to learn.

\section{General Analysis \& Discussion }
 
Our studies indicate real potential for the domain-specific mashup approach to allow people with no computing skills to create their own applications. The comparison between the two groups of users highlighted good performance independently of participants computing skills. The request for higher training emerging from a few less expert users appeared to be rather linked to a weaker domain knowledge than to their computing capabilities. Further research could explore the relative role of these two factors by a full factorial experimental study on a larger sample. However, these studies suggested that ResEval Mash is a successful tool appealing both to expert programmers and end-users with no computing skills. 
 
All participants reported a good level of understanding of the basic concepts implemented in ResEval Mash, although some suggestions for improvement were collected, mainly related to verbal labels used to denote components. Most usability issues evinced from behavioral observations can be easily solved. For instance, the uncertainty experienced by several users with saving the configuration parameters can be counteracted by adding an explicit saving option in the interface of the components. A more serious issue was highlighted as regards the selection of components, which was found to be an error prone and time demanding task. This problem is likely to increase exponentially with the availability of more components, but it can be partially counteracted by a smart advice system decreasing the number of items available for selection based on a comparison between the current application context and previous successful implementations, as presented in \cite{DeAngeli11}. For instance, illegal components could be automatically disabled and the one used most often made salient.
 
Overall, the studies provided some interesting results and highlighted the important role of user evaluation in the design of interactive systems. A major finding is related to the ease with which our sample (independently of their technical skills) understood that components had to be linked together so that information could flow between different services. This is a well-acknowledged problem evinced in several user studies of EUD tools (e.g., the ServFace Builder, \cite{NamounICWEWS2010}), which surprisingly did not occur at all in the current study. The mismatch can be due to a different level of complexity of the evaluation tasks, but also to an important design difference. Indeed, ResEval Mash only requires users to connect the components as holistic concepts, whereas other tools, such as the ServFace builder required the user to perform complex connections between individual fields of user interfaces. %More research is needed to understand the boundaries of ResEval Mash, testing it with more complex development scenarios.

\newpage

%chapter on name disambiguation
%\input{chapters/ch9-name-disam.tex}
%\newpage

% Chapter 8: Conclusions and future work.
% !Tex root = ../phd-thesis.tex

%%% Chapter start. %%%
\chapter{Conclusions and Future work}
\label{sec-conclusions}

\section{Overview}
Many crucial decisions such as research funds distribution, faculty recruitment, promotions, PhDs selection, award of grants require measuring quality, productivity, and impact of researchers. The research impact is determined based on the scientific research outputs of a researcher, which is measured either through traditional approaches that are based on published papers, citation records, journal prestige (where papers were published) or through advanced bibliometric approaches such as h-index, g-index, ch-index. Over the last few years scientific production has increased to a large extent, researchers are growing in number and also making collaborations, producing and disseminating scientific results. However, primarily the growth is positive, but it takes research organizations, funding bodies under pressure who want to maximize the wider impact of their investment in research. Evaluating research work, determining research impact is a notoriously challenging problem which so far has no well accepted solution. Bibliometrics approaches, which are typically comprised of citation and content analysis methods, have largely well perceived by many communities. However, a general consensus is that the problem remains unsolved, which is mainly due to its multi-dimensionality nature. Moreover, the proliferation of the scientific data sources (e.g., DBLP, Google Scholar, Microsoft Academic etc.), and highly customized bibliometric indices, and locally developed evaluation procedures are the three highly diverse aspects of this field, which make the overall approach highly subjective. Moreover, people involved in such evaluation processes, most of the time, are not IT experts, and not capable of building appropriate software for crawling data sources, automatically parsing relevant information, merging data and computing the needed personalized metrics. Therefore, in order to empower the interested people, we need to design an appropriate and possibly easy-to-use IT platforms, which could make life easier of those domain-experts who do not expert in IT.

The ever increasing number of computer users, especially those non-technical users (as also in our case), who are not computer programmers, use computer applications to fulfill their daily life situational requirements. A large proportion of such users are teachers, doctors, researchers, administrative persons etc. Those are more expert in their domains than in computing skills. In the past,  several enabling approaches have been proposed that  aimed at facilitating non-technical end-users. Despite many efforts, it is still a challenging endeavor for users to develop applications that support or fulfill their goals. This is because, generally proposed technologies require expertise in programming languages and their complicated user experience poses difficulties for the users.

End-user development is a way to solve this problem with an aim to empower non-technical end-users in such a way that they can effectively participate in development processes. Specifically through the end-user development a set of techniques, methods and tools collectively enhance user experiences that then can be easily utilized by the non-technical users. To this end, several approaches have been proposed which we have explained in the chapter \ref{ch:eud-soa}. Mainly, these approaches whose initial goal was to enable non-technical users to design and develop applications with little or no help from developers, we are still in a situation in which these solutions can only be used by specifically trained developers.

On the other hand, the recent emergence of \emph{mashup tools} has refueled research on end-user development. We see that \emph{mashups} are simple applications that rather than being developed from scratch by developers, are composed by integrating and reusing available services, data, functionalities or user interfaces. Likewise, \emph{mashup tools} provide enabling environments for mashups development, ambitiously aim at enabling non-technical users to develop their own on-demand ad hoc applications.

We believe that doing so is even harder than enabling non-technical users to develop applications because developing full applications is simply complex. The reason is the mashup platforms developed so far come with so many functionalities and too many technicalities that are only suitable for programmers. Yet, being amenable to non-technical users is extremely important as the availability of a wider range of online applications, services and data raised the need for situational, short-lifespan applications that cannot be anticipated and developed through traditional software development processes.

However, we believe that it is impractical to design tools that are generic enough to cover a wide range of application domains, powerful enough to enable the specification of non-trivial logic, and simple enough to be actually suitable for non-programmers. Instead, in our view, we need to give up something and that is a generality since reducing expressive power would mean supporting only the development of simple applications, which is useless, while simplicity is our major aim. That means, giving up generality in practice requires  narrowing the focus of a design tool to a domain and tailoring the tool's development paradigm, models, language, and components to the specific needs of that domain only.

\section{Contributions Summary of the Thesis}
This thesis presented a novel approach for an effective end-user development. The target end-users in this context considered non-programmers, less-technical but domain-experts. The approach, which leverages mashups philosophy, can effectively involve end-users in development tasks. Opposite to the existing mashup based solutions, this thesis presented a novel idea of domain-specific mashups, that is,  we specifically focused, during the development of the mashup platform, to a well-defined domain. So that, a tool whose design and development is based on a domain can speak the language of the users, a key aspect that existing approaches lack and consequently failed in aiding end-user development. 

To this end, we first introduced the notion of \emph{domain-specific mashups} and described what they are composed of, how they can be developed and how non-programmers can effectively get benefits. We described which design artifacts are necessary for a domain-specific mashup tool's development. In this regard, we presented the \emph{mashup meta-model} and the \emph{domain concept model}, the \emph{domain syntax model} and showed how these can be merged into a \emph{domain-specific mashup meta-model}, which provides a consolidate basis and expressive power to a mashup tool whose development follows it.

Based on the above mentioned design and modeling artifacts, the thesis proposed a systematic \emph{approach} and \emph{methodology} (presented in chapter \ref{chp-method} ), which consists of a number of steps. The proposed methodology can be used for both, the development of an end-user oriented mashup platform and the development of a domain-specific mashup tool. The developed platform initially stays empty (i.e., in terms of domain-specific concepts) and generic (i.e., ready to be tailored for other similar domains) so then it can be tailored to make a domain-specific mashup tool. In this thesis we followed the same strategy, that is, we first developed a mashup platform (presented in chapter \ref{sec-contrib03}) whose capabilities are as defined in the \emph{mashup meta-model}, second, we developed ResEval Mash (presented in chapter \ref{chp-reseval}), which is a domain-specific mashup tool for the research evaluation domain. The development of ResEval Mash was also driven by the requirements that are of end-user specific, which we gathered during the analysis of various domain-specific research evaluation procedures (presented in chapter \ref{sec:scenarios}). 

Moreover, this thesis introduced a novel and an efficient approach for data-intensive web applications, specifically applications which follow mashups philosophy. The approach is suitable for those applications, which deal with large amounts of data that travel between client and server. We observed that web-service based mashups applications extensively communicate data not only between client server but also between web services too. In this case, particularly, if these web services are deployed on a web server, then data-intensive communication can be reduced to a large extent with the help of proposed approach (presented in chapter \ref{chp-reseval}, section \ref{sec:component-service-interaction}).

For the evaluation of our proposed approach and to check the usability of our mashup tool, we performed two separate user studies. In the first user study, which was focused on the usability and comparative evaluation aspects, participants were presented four different mashup based prototypes comprising different level of expressive power and flexibility. Moreover, participants were presented Yahoo! Pipes as a generic mashup tool example. The results presented in the chapter \ref{sec-experiments} clearly show users' preference to domain-specific mashup approach and most of the participants preferred ResEval Mash tool among the other prototypes. The participants found ResEval Mash offering right balance in terms of expressive power and complexity that a non-programmer can easily deal with. Many useful comments, feedbacks, and requests for improvements were gathered during this study, which were applied to the tool for further improvements.

After incorporating improvements, the second user study was performed, which was mainly focused on the advance usability evaluation aspects of the improved ResEval Mash tool (presented in chapter \ref{sec-experiments}). The results of this study show that the real potential of our domain-specific mashup approach to allow people with no programming skills to create their own applications. Domain experts found very attractive the no-data mapping approach offered by the ResEval Mash, as asking non-programmers or less-technical users to perform complex data-mapping always creates difficulties for them. As an overall, both studies suggested that ResEval Mash is a successful tool appealing both to expert programmers and end-users with no programming skills.

\section{Discussion and Lessons Learned}
The work presented in this thesis mainly focused on the research evaluation domain, however, throughout the presentation of methodological approach, and the development of the mashup platform we keep separate aspects those are of domain-specific type from the ones of generic type. This helped us to learn from both, the technology and the domain, ends about demands (from domain) and support (from technology) both in the context of less skilled end-users. As we have seen in the reference domain and this is also the case for other domain, that the ever increasing number of end-users, especially non-programmers, those use computers in their daily life routine for different tasks. For instance, considering only those users which belong to a particular field such as medical field (doctors), research field (scientists), teaching (teachers) and banking field (bankers) are only a few examples of non-programmers. The nature of work they usually involved in vary on a daily basis, which requires more easy-to-use, flexible, and intuitive software solutions which can provide enabling environments to support such rapid development. Despite many efforts, little is achieved in this direction. 

However, in this thesis we learned in detail what these kinds of specific fields (i.e., domains) and their users requirements from technology. We presented that traditional software development practices largely failed to achieve their needs. This is due to many reasons, as also listed below a few important ones:

\begin{itemize}

\item From the traditional software development point of view, anticipation of rapidly changing requirements is an aspect which is almost impossible to achieve. Simply traditional requirements elicitation approaches cannot gather and thus software developer cannot anticipate what a specific type of requirement a user can face or think of. This is the reason, we focused in our work on giving more expressive power to users keeping simplicity and complexity within a user's expertise domain. 

\item Existing enabling environments (e.g., those based on mashups approaches) do not effectively communicate with the users. That is, they do not speak the language of the users, which creates a communication gap between users and technology (e.g., Yahoo Pipes, Taverna and many other DSLs). We believe that not only these technologies must be simply to use but also the interaction medium between technology and its users must be the same (i.e., communication language should be same).  

% Software solutions built on the current development approaches do not effectively communicate with the users. That is, the current approaches do not speak the language of the users, which creates a communication gap between the users and the technologies. This is also the case of existing mashup based environments.

\item Most customizable and tailorable mashup based tools provide enabling environment to these users to develop their own applications. However, often these tools offer components of generic types, as they cover a broader set of functionalities, in result not suitable for many non-programmers. In our opinion, the compositional constructs on one side should reflect domain processes or activities and on the other side they should wrapped by domain-syntax.

\item Most mashup based tools require complex data mappings in order to develop mashup compositions. Non-technical users always find this aspect very hard to grasp, and also they don't show interest in learning these complex issues. A tool, which is specific to a domain, can get rid of complex data mappings, as also in our case.

\end{itemize}

As an overall, we believe that less technical users (i.e., non-programmers) can be involved in development tasks provided if they are given with appropriate technology support. This thesis presented a comprehensive analysis of all end-user development related research directions, and found that either these approaches require high technical skills or even in some cases where less-skilled users were targeted they still not able to achieve their objectives due to the generic nature of compositional constructs. We believe a successful end-user oriented development environment on one side must provide an intuitive user experience and on the other side the development constructs should be within the users' domain expertise. Moreover, an enabling environment having these two characteristics must maintain a right balance between flexibility and the expressive power it offers, as these aspects consequently change the corresponding complexity of a tool. 

%For this purpose, we get inspiration from the mashups, and proposed a complete methodology that is used to build mashup platforms, that is then the platform can be tailored for a specific domain for specific users. We applied the proposed methodology for the creation of a mashup platform, which we have presented in chapter \ref{sec-contrib03}. Based on this generic mashup platform, we tailored it to the domain of research evaluation, which we presented in the chapter \ref{chp-reseval}, and performed two user studies. Overall, the studies suggested that ResEval Mash is a successful mashup tool appealing both to expert programmers and end-users with no computing skills.

\section{Future Work}
During this thesis work we observed a number of interesting directions those if applied can strengthen the overall platform. In the following sub-sections we provide details on these future directions.

\subsection{Persistent Cache Support}
The platform presented in this thesis provides cache support for mashups to efficiently process huge data. Server-side components (i.e., those components which use web services) take advantage of server-side cache, a feature which prevents heavy communication between client and server. The current design of our platform uses server's physical memory (RAM) as cache, thus under server's memory limitations. Although, we have not observed a noticeable drawback of using server's physical memory as cache purposes, however, we believe that the support of a persistent cache (e.g., like database which uses persistent means to store data) can act like an effective backup plan. Most database management solutions such as Oracle, Cassandra provide a very robust mechanism to read \& write data to and from disks. In some cases, for example Oracle, even provide a physical memory based shared area (much like our cache) that intelligently  uses RAM to provide high hit-ratio and efficiency.  

To this end, an extension for a persistent cache requires just an implementation of the same interfaces, which are currently being used in the case of physical memory, for the persistence solution. That is, the schema, which we used to generate implementation classes, has to be used for this purpose too and in case of a database-based persistence this schema can be used to create database-dependent schema. The rest of the task is then to communicate with the persistence solution from our CDM manager. A more efficient approach in this case would be to use both caches (i.e., RAM and disk).

\subsection{Third Party Services Registration \& Deployment}
In addition to the client-side components support, which can be implemented using JavaScript language, the platform provides the support for server-side components so that components can use web services. These web services can be deployed and used from anywhere, through a publicly accessible application server. However, if one wants to deploy web services on our server, in that case the current implementation of the platform does not provide an interface for third party services for the registration and deployment purposes. However, we, as a platform provider deploy these services manually in case of a third party service deployment request. A significant improvement can be achieved by providing an interface for third party web services to register and deploy automatically via a web interface. Moreover, these web services should allow to be set as public or private, much like the concept we use for components and compositions. 

\subsection{Component-Mappers for Third Party Components}
Components, which are the basic building blocks, follow \emph{component definition language} and a representational format of our choice. However, the platform does not restrict components representational format specific to ours, but components defined and having different formats can also be plugged-in. As described in the chapter \ref{sec-contrib03} that the components mappers perform conversions from a format to the one understandable by our platform. We provide default mappers that are compatible to our specified format, however, new mappers can be written which then can be used to convert different representations into the platform specific one. 

\subsection{Recommendation Support for Mashup Compositions Development}
Recommender systems such as Amazon.com\footnote{http://www.amazon.com/}, Pandora Radio\footnote{http://www.pandora.com/}, Netflix\footnote{http://www.netflix.com/} have become very successful in recent years. A recommender system makes predictions for users based on content-based or collaborative filtering based approaches. This is also inline to support the reuse of development knowledge from more expert users in an automated fashion. In our case, intelligent recommendation during mashup composition surely can greatly aid end-users in their development tasks. For instance, component recommendations, composition recommendations would greatly increase the overall experience of the platform. 
\newpage

%%% Bibliography. %%%
\clearemptydoublepage\thispagestyle{empty}\footnotesize

% Add the bibliography page to the table of contents.
\addcontentsline{toc}{chapter}{Bibliography}

% Use named style for the citations. kbib, amsalpha, plainnat
\bibliographystyle{ieeetr}

% Build the bibliography page from a BibTeX file (bibliography.bib).
\bibliography{Bibliography}

% End of document.
\end{CJK*}
\end{document}